\documentclass[aps,prd,twocolumn,amssymb,showpacs,superscriptaddress,nofootinbib,10pt,longbibliography]{revtex4-1}

\usepackage{amsfonts}
\usepackage{latexsym}
\usepackage{yfonts}
\usepackage{amsmath}
\usepackage{amssymb}
\usepackage{upgreek}
\usepackage{bm}
\usepackage{dcolumn}
\usepackage{epsfig}
\usepackage{graphicx}
\usepackage[latin1]{inputenc}
\usepackage{latexsym}
\usepackage{rotating}
\usepackage{hyperref}
\usepackage{color}

\hypersetup{
    pdfauthor={Santos-Oliv\'an, Daniel and Sopuerta, Carlos F.},
    pdftitle={Moving Closer to the Collapse of a Massless Scalar Field in Spherically-Symmetric Anti-de Sitter Spacetimes} }

\newcommand{\PD}[2]{\frac{\partial {#1}}{\partial {#2}}}

\newcommand{\be}{\begin{equation}}
\newcommand{\ee}{\end{equation}}
\newcommand{\bea}{\begin{eqnarray}}
\newcommand{\eea}{\end{eqnarray}}
\newcommand{\onehalf}{\textstyle{\frac{1}{2}}}
\newcommand{\met}{\mbox{g}}

\begin{document}

\title{Moving Closer to the Collapse of a Massless Scalar Field in Spherically Symmetric Anti-de Sitter Spacetimes}

\author{Daniel Santos-Oliv\'an}
\affiliation{Institut de Ci\`encies de l'Espai (CSIC-IEEC), 
Campus UAB, Carrer de Can Magrans s/n, 08193 Cerdanyola del Vall\`es, Spain.}

\author{Carlos F.~Sopuerta}
\affiliation{Institut de Ci\`encies de l'Espai (CSIC-IEEC), 
Campus UAB, Carrer de Can Magrans s/n, 08193 Cerdanyola del Vall\`es, Spain.}

\date{\today}


\begin{abstract} 
We present a new hybrid Cauchy-characteristic evolution method that is particularly suited for the study of gravitational collapse in spherically-symmetric asymptotically (global) Anti-de Sitter (AdS) spacetimes. The Cauchy evolution allows us to track the scalar field through the different bounces off the AdS boundary while the characteristic method can bring us very close to the point of formation of an apparent horizon. Here, we describe all the details of the method, including the transition between the two evolution schemes and the details of the numerical implementation for the case of massless scalar fields.  We use this scheme to provide more numerical evidence for a recent conjecture on the power-law scaling of the apparent horizon mass resulting from the collapse of subcritical configurations.  We also compute the critical exponents and echoing periods for a number of critical points and confirm the expectation that their values should be the same as in the asymptotically-flat case.
\end{abstract}


\maketitle


\section{Introduction}\label{introduction}

One of the main results in General Relativity (GR) is the discovery of critical phenomena in gravitational collapse by Choptuik~\cite{Choptuik:1992jv}.  Using numerical relativity computations Choptuik investigated a suggestion by Christodoulou~\cite{Christodoulou:1986zr} (see also~\cite{Gundlach:1995kd,Gundlach:1996eg} and~\cite{Gundlach:2002sx,Gundlach:2007gc} for reviews) on whether gravitational collapse of a massless scalar field in an asymptotically-flat spacetime can form a black hole (BH) with arbitrarily small mass.  This study ignores quantum effects that may create stable stationary states that do not collapse.  Indeed, in the case of degenerate stars, quantum properties of matter establish a lower limit for the mass of an astrophysical BH~\cite{Chandrasekhar:1931sc}. In the original Choptuik computations~\cite{Choptuik:1992jv} there are only two possible final states: (i) The formation of a (Schwarzschild) BH.  (ii) Dispersion of the scalar field leading to flat space as the final state of the evolution.  Choptuik found that near the threshold between these two possibilities, but within BH formation, the behavior of the system exhibits critical phenomena with the BH mass depending on the deviations of a single initial-data parameter from a critical value in a universal way (the same for any family of initial configurations and for any initial-data parameter). Previous to Choptuik's work, Goldwirth and Piran~\cite{Goldwirth:1987nu} found out that collapse to a BH takes place for initial scalar field configurations with large enough amplitudes.  The work of Choptuik is based on a Cauchy-type initial-value problem formulation and relies on the use of Adaptive Mesh Refinement (AMR). Soon after Choptuik's work, Garfinkle~\cite{Garfinkle:1994jb} reproduced the main results by using a characteristic formulation of the problem that did not use AMR techniques.  These important results also constitute the first great triumph of the field of numerical relativity.  Other important results of numerical relativity in asymptotically-flat spacetimes are motivated by the developments in gravitational wave astronomy (see~\cite{Lehner:2001wq,Bona:2009bo,Baumgarte:2010bs,Lehner:2014asa,Sperhake:2014wpa} for reviews).

There are three maximally-symmetric families of spacetimes: Minkowski (Mink) spacetime, which has vanishing Riemann curvature tensor; de Sitter (dS) spacetime, which has vanishing positive and constant scalar curvature; and Anti-de Sitter (AdS) spacetime, which has negative and constant scalar curvature. These three maximally-symmetric spacetimes, (Mink, dS, AdS), are vacuum solutions of the Einstein Field Equations (EFEs) for a null, positive, and negative cosmological constant respectively. Both Mink and dS spacetimes have been extensively studied. The first one due to its relevance not only in GR but also in Special Relativity and in Quantum Field Theory, and the second for its importance in Cosmology. However, the AdS spacetime was forgotten for many years and there were only a few classical books (see, e.g.~\cite{Hawking:1973uf}) where it was mentioned. However, recently, AdS and asymptotically-AdS (AAdS) spacetimes have been extensively studied in connection with the AdS/Conformal Field Theory (AdS/CFT) correspondence~\cite{Maldacena:1997re} (see, e.g.~\cite{Witten:1998qj}), also known as the Gauge/Gravity duality. This conjecture establishes a relationship between a purely gravitational theory in an AAdS spacetime and a CFT that lives in its boundary. This duality has found diverse applications to subjects like the thermalization of quark gluon plasma~\cite{Chesler:2013lia}, superconductors~\cite{Horowitz:2014awa}, and many others in condensed matter physics~\cite{Hartnoll:2009sz}.  This activity has motivated a number of developments in numerical relativity to deal with AAdS spacetimes (see~\cite{Cardoso:2012qm,Sperhake:2013wva}).

A natural question that arises is the stability of the (global) AdS spacetime. Mink~\cite{Christodoulou:1994ck,Lindblad_theglobal} and dS~\cite{Friedrich:1986cm} spacetimes have been proven to be stable under {\em small}  finite perturbations.  We also know that AdS is stable under linear perturbations~\cite{Ishibashi:2004wx}, but the problem of the stability of AdS is still under debate~\cite{Friedrich:2014raa}.  The first numerical studies of gravitational collapse in AdS spacetime appeared in~\cite{Husain:2000vm,Pretorius:2000yu} for the 2+1 dimensional case (see also~\cite{Garfinkle:2000br}). Recent abundant activity on this subject has shown an instability in AAdS spacetimes endowed with a real massless scalar field~\cite{Bizon:2011gg,Jalmuzna:2011qw} and also for a complex massless scalar field~\cite{Buchel:2012uh}.  What has been shown is that certain families of initial configurations lead to the formation of an apparent horizon (AH) independently of their total energy.  This is due to the AdS causal structure that allows light-like signals to reach the AdS boundary in a finite time.  Then, a scalar field packet, no matter how small its energy, can bounce repeatedly off the AdS boundary while the non-linearity of Einstein's equations transfers power from long wavelength components to short wavelength ones until the scalar field packet gets compact enough to form an AH.  This instability, in reference to the cascade towards shortwave modes typical in fluids, is called the {\em turbulent} instability of AdS~\cite{Bizon:2011gg,Maliborski:2013via}. In this sense, the AdS boundary provides a box-like structure to space that is fundamental for the instability.  Indeed, studies~\cite{Maliborski:2012gx,Okawa:2013jba,Okawa:2014nea,Cardoso:2016wcr,Cai:2016yxd} in asymptotically-flat spacetimes with an artificial boundary at a finite distance find the same type of instability. A previous study of gravitational collapse in AdS~\cite{Husain:2002nk} showed that the cosmological constant does not change the properties of the collapse with respect to the asymptotically-flat case, in particular the critical exponent for direct collapse does not change. However, the {\em turbulent} instability was not identified. 

There are some open questions about this instability, in particular whether or not it is generic.  Recent studies~\cite{Buchel:2013uba,Maliborski:2013ula} indicate that some initial configurations never form an AH, but numerical studies can only provide evidence for these islands of stability.  The properties and characteristics of these islands are being currently investigated combining non-linear perturbation methods and numerical simulations~\cite{Fodor:2013lza,Maliborski:2014rma,Balasubramanian:2014cja,Craps:2014vaa,Dimitrakopoulos:2014ada,Bizon:2014bya,Craps:2014jwa,Buchel:2014xwa,Bizon:2015pfa,Balasubramanian:2015uua,Dimitrakopoulos:2015pwa,Green:2015dsa,Deppe:2015qsa,Craps:2015iia,Craps:2015xya,Menon:2015oda}, and the structure of the spectrum of AdS {\em perturbations} appears to play an important role.  Applications of these studies to the AdS/CFT correspondence can be found in~\cite{Abajo-Arrastia:2014fma,daSilva:2014zva}. New mathematical studies~\cite{Ishibashi:2012xk,Friedrich:2014rpa,Holzegel:2015swa,Holzegel:2015bna} have been motivated by the rich structure of AAdS spacetimes and the turbulent instability has been recently studied (in spherical symmetry) for massive scalar fields~\cite{Kim:2014ida,Okawa:2015xma}, self-interaction potentials~\cite{Yang:2015jha,Basu:2015efa,Evnin:2015gma,Cai:2015jbs}, and even for modified theories of gravity like Einstein-Gauss-Bonnet gravity~\cite{Deppe:2014oua,Menon:2015oda}, where the outcome of the evolution is very different from the Einstein gravity case, as it already happens in the asymptotically-flat case~\cite{Golod:2012yt}.  There are also studies of the turbulent instability outside spherical symmetry~\cite{Dias:2011ss,Dias:2012tq,Horowitz:2014hja}.

In this paper, instead of looking at the long-term evolution we focus on the details of the collapse, i.e. we look at the dynamics of AH formation. This is motivated by the fact that the landscape of gravitational collapse in AAdS spacetimes is much richer than in the case of asymptotically-flat spacetimes~\cite{Bizon:2011gg,Bizon:2013gxa}. This  problem is quite challenging from the numerical point of view because the scalar field keeps bouncing between the origin and the AdS boundary while it develops sharper and sharper profiles that we need to evolve all the way to the AdS boundary and back near the origin until an AH is finally formed. Most of the numerical computations done until now to evolve a massless scalar field in spherically-symmetric AAdS spacetimes use a Cauchy-type evolution like in Choptuik's work~\cite{Choptuik:1992jv}, which allows us to follow the different bounces off the AdS boundary.  In contrast, a characteristic approach like the one proposed by Goldwirth and Piran~\cite{Goldwirth:1987nu}, although it can potentially get much closer to the formation of an AH than a Cauchy scheme, cannot follow the bounces because the grid covers only a part of the spacetime.  In order to combine the best of the two worlds, in this paper we introduce a new scheme that combines them and has three essential ingredients: (i) A Cauchy-based evolution scheme numerically implemented using pseudospectral methods and a compactified radial coordinate in order to follow the scalar field through all the bounces off the AdS boundary. (ii) Construction of initial data on a null slide for the characteristic evolution from the results of the Cauchy evolution. (iii) A characteristic evolution scheme to follow the scalar field all the way to AH formation.  This scheme has already been successfully used in~\cite{SantosOlivan:2015fmy}, where it was found that, in AdS spacetime, the AH mass of BHs formed following the evolution of subcritical solutions associated with the $n$-th critical point ($n=0,1,\ldots$) follow a power law of the form
\begin{equation}
M_{\mathrm{AH}}-M^{n+1}_{g} \propto (p_{n}-p)^{\xi}\,,
\end{equation}
where $p_n$ is the critical value of the initial-data parameter $p$ for the $n$-th branch, $M^{n+1}_g$ is the corresponding mass gap (minimum mass from subcritical configurations), and $\xi\simeq 0.7$ is the exponent.  Moreover, $\xi$ appears to be universal, independent of the initial data family and initial-data parameter and the same for all the critical points. In this paper, we give all the details about this hybrid Cauchy-Characteristic scheme and its numerical implementation (including validation, convergence properties, and other details about evolution in AAdS spacetimes), and provide new results that make the findings of~\cite{SantosOlivan:2015fmy} more robust and reliable. In addition, we compute the scaling exponents of the supercritical configurations confirming the expected result~\cite{Bizon:2011gg} that they are the same as in the asymptotically-flat case, i.e. $\gamma \simeq 0.374 $.

The plan of this paper is: In Sec.~\ref{equations} we introduce the Einstein-Klein-Gordon (EKG) system of equations and, for the spherical-symmetric case, we adapt them to a Cauchy-type formulation in Sec.~\ref{equations_cauchy} and to a characteristic formulation in Sec.~\ref{analytics_characteristic}.  In Sec.~\ref{cauchy-to-characteristic} we describe  the transition from the Cauchy-type evolution to the characteristic one. In Sec.~\ref{numerics} we describe the ingredients of the numerical implementation. In Sec.~\ref{code-validation} we show numerical results for the validation of the code, including convergence and some special features arising in AAdS spacetimes. Finally, in Sec.~\ref{results}, using this hybrid Cauchy-Characteristic code, we present new numerical results on the dynamics of AH formation. 

We use units in which $c=1$ and $8\pi G_{d} = d-1$, where $G_{d}$ is the $(d+1)-$dimensional Newton's gravitational constant and $d$ is the number of spatial dimensions. A semicolon denotes covariant differentiation with respect to the canonical connection; a dot denotes differentiation with respect to the time coordinate $t$, $\dot{\phi} \equiv \partial\phi/\partial t$; and differentiation with respect to the compactified radial coordinate $x$ is denoted by a prime,  $\phi'\equiv\partial \phi/\partial x$. For other partial derivatives we use the notation: $\partial^{}_{y}f\equiv \partial f/\partial y \equiv f_{,y}$. We use small-case Greek letters $\alpha,\beta,\ldots = 0,\ldots,d$ for spacetime indices.


\section{Field Equations}\label{equations}

The field equations for a self-gravitating, real massless scalar field in an AAdS spacetime are the EKG system of equations for the metric $\met_{\mu\nu}$ and scalar field $\phi$:
\begin{eqnarray} 
&& G^{}_{\mu\nu} + \Lambda \met^{}_{\mu\nu} = (d-1) \left( \phi^{}_{;\mu}\phi^{}_{;\nu} - \onehalf 
\met^{}_{\mu\nu}\phi^{}_{;\alpha}\phi^{;\alpha}  \right) \,, \label{efes} \\
&& \met^{\mu\nu} \phi^{}_{;\mu\nu} = 0\,, \label{kg_eq}
\end{eqnarray}
where $G^{}_{\mu\nu}$ is the $(d+1)$-dimensional Einstein tensor and $\Lambda$ is the (negative) cosmological constant.

We restrict our study to spherically-symmetric configurations.  This assumption simplifies the structure of the spacetime metric and the field equations. Spherically symmetric spacetimes have a warped geometry, which means that their metric tensor can be written in the form $ds^{2} = \met_{AB}(x^{C})dx^{A}dx^{B} + r^{2}(x^{C})\gamma_{ab}dx^{a}dx^{b}$ ($A,B,\ldots=0,1$ and $a,b,\ldots=2,\ldots,d$), where $\met_{AB}$ is a Lorentzian metric (with associated manifold $M^{2}$), $\gamma_{ab}$ is the unit curvature metric on the $(d-1)$-sphere (with associated manifold $S^{d-1}$), and $r=r(x^{A})$ is the radial area coordinate.  The fact that $r^{2}(x^{C})\gamma_{ab}$ is not a true metric on $S^{d-1}$ is what prevents the spacetime manifold $M^{d+1}$ to be a true product of the two manifolds $M^{2}$ and $S^{d-1}$.  Instead, it is said that the spacetime manifold is the warped product of $M^{2}$ and $S^{d-1}$ and this is sometimes denoted in the literature as $M^{d+1} = M^{2} \times_{r} S^{d-1}$.

We can freely choose the coordinates in the Lorentzian manifold $M^{2}$.  In this work we will consider two different choices according to the type of spacetime slicing that they induce: (i) Timelike slicing: We will consider coordinates $(x^{A})=(t,x)$ so that the spacetime is sliced in spacelike (with timelike normal) hypersurfaces $\{t=const.\}$.  In addition, we take $x$ to be a radial coordinate that compactifies the radial direction so that it is in the range $x\in [0,\pi/2]$, where $x=0$ corresponds to the {\em center} of the radial coordinate system and $x=\pi/2$ corresponds to the AdS boundary.  Using these coordinates we can set up a Cauchy-type system of evolution equations with some constraints.  Given the causal structure of AAdS spacetimes, the scalar field can propagate to reach the AdS boundary in a finite time. Previous works on this problem (see, e.g.~\cite{Bizon:2011gg,Buchel:2012uh}) showed that we can expect the field to bounce off the AdS boundary a number of times and eventually collapse near the center $x=0$. It is for this reason that we use the compactified coordinate $x$ in order to track the field up to the AdS boundary. The equations are given in Sec.~\ref{equations_cauchy}.  (ii) Lightlike slicing:  We will consider coordinates $(x^{A})=(u,r)$ so that the spacetime is foliated by outgoing null slices (composed by outgoing null rays) $\{u=const.\}$.  The radial coordinate $r$ is not a compactified radial coordinate as in the previous case, in the sense that the AdS boundary is located at $r\rightarrow\infty$. This system of coordinates allows us to set up a characteristic-type system of evolution of equations.  In contrast to the coordinates $(t,x)$, the coordinates $(u,r)$ do not allow us to follow the field up to the AdS boundary.  Instead, we want to use them in order to track the evolution of the field near collapse, that is, near the center $r=x=0$.  The fact that the $\{u=const.\}$ slides are outgoing means that as we evolve in $u$ we approach faster the collapse than in the case of Cauchy evolution.  As we will see, the characteristic evolution allows us to get much closer to the formation of an AH than the Cauchy evolution.  The equations for this case are given in Sec.~\ref{analytics_characteristic}.


\subsection{Cauchy-type Evolution of the EKG System}\label{equations_cauchy}

Following~\cite{Bizon:2011gg,Jalmuzna:2011qw} and our previous discussion, the metric of an spherically-symmetric AAdS spacetime in $d+1$ dimensions can be written as
\be
ds^{2} = \frac{\ell^{2}}{\cos^{2}x}\left( - A {e}^{-2\delta}\,dt^{2} 
+\frac{dx^{2}}{A} + \sin^{2}x\, d\Omega^{2}_{d-1} \right)\,, \label{aadstx}
\ee
where $d\Omega^{2}_{d-1}$ is the metric of the unit ($d-1$)-sphere ($S^{d-1}$); $A$ and $\delta$ are the two metric functions that completely determine the metric and depend only on $(t,x)$; and $\ell$ is the AdS length scale, which is related to the cosmological constant $\Lambda$ by the expression: $\ell^2 = -{d(d-1)}/{2\,\Lambda}$.  The time coordinate $t$ has an infinite range, i.e. $t\in$ ($-\infty$,$\infty$), whereas $x$ is a radial compactified that goes from $x=0$ (center) to $\pi/2$ (AdS boundary). We can recover AdS spacetime by setting $A = 1$ and $\delta=0$.

\begin{figure}[t]
\begin{center}
\resizebox{.45\textwidth}{!}{ \includegraphics{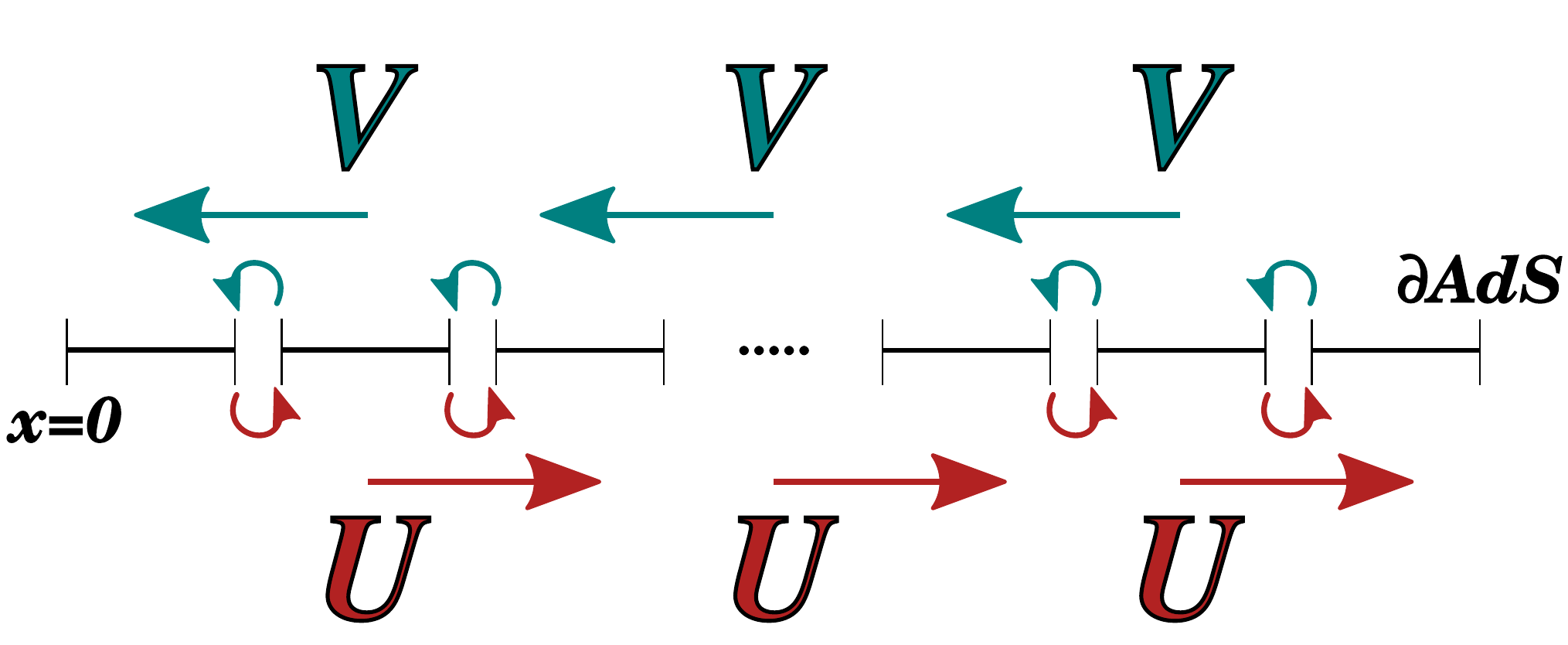}}
\caption{Diagram of the multidomain structure. The evolution variables need to be communicated between domains using the characteristic variables, $U$ and $V$, which have a well defined direction of propagation. The communication is done by copying the boundary values in the direction indicated by the arrows. \label{plot_diagram_U_V}}
\end{center}
\end{figure}

From the field equations~\eqref{efes} and~\eqref{kg_eq} we can derive partial differential equations (PDEs) for $A$, $\delta$, and $\psi$ (see, e.g.~\cite{Bizon:2011gg}).  Here, due to the type of numerical implementation we use for the Cauchy evolution, based on the pseudospectral multidomain method (see Appendix~\ref{app_psc} for a summary of the main ingredients), we are interested in a first-order formulation of the equations based on the characteristic variables associated with our dynamics (see~\cite{Courant:1989aa} for a definition of the characteristic variables of a hyperbolic system of PDEs and Appendix~\ref{app_char_var} for a derivation of the hyperbolic structure of our equations). In our case, the only true hyperbolic sector is the one corresponding to the Klein-Gordon equation~\eqref{kg_eq}.  In Appendix~\ref{app_char_var} we find the characteristic variables of our system, $U$ and $V$ [see Eq.~\eqref{characteristic-variables-u-v}].  Variable $U$ corresponds to the scalar field mode that propagates always with positive velocity and $V$ to the mode with negative velocity (see Fig.~\ref{plot_diagram_U_V}). In terms of the scalar field they are:
\be
U = \frac{1}{\cos^{d-2}x} \left( \phi' - \frac{e^{\delta}}{A}\dot \phi  \right) \,, 
\label{Udef}
\ee
\be
V = \frac{1}{\cos^{d-2}x} \left( \phi' + \frac{e^{\delta}}{A}\dot \phi  \right) \,. 
\label{Vdef}
\ee
It is also convenient to introduce the following normalized variable associated with the scalar field:
\be
\psi = \frac{\phi}{\cos^{d-1}x} \,. \label{psi-def}
\ee
Then, from the field equations~\eqref{efes} and~\eqref{kg_eq} we derive the PDEs for $(\psi,U,V,A,\delta)$: (i) Evolution equations:
\begin{eqnarray}
\dot{\psi} & = & \frac{A e^{-\delta}}{2\cos x} \left( V - U\right)  \,,  
\label{evol_psi}
\end{eqnarray}
\begin{eqnarray}
\dot{U} & = & - A e^{-\delta} U^{}_{,x} - \frac{(d-2 \cos^2 x)}{\sin x \cos x}  U \,e^{-\delta}\,(1-A) \nonumber \\
        & - & \onehalf \frac{(d-1) A e^{-\delta} }{\sin x \cos x} \left( U+V \right)  + \left( d-2\right) \frac{\sin x}{\cos x}U\,Ae^{-\delta} \,,   
\label{evol_U} 
\end{eqnarray}
\begin{eqnarray}
\dot{V}    & = & A e^{-\delta} V^{}_{,x} + \frac{(d-2 \cos^2 x)}{\sin x \cos x}  V \,e^{-\delta}\,(1-A) \nonumber \\
           & + & \onehalf \frac{(d-1) A e^{-\delta} }{\sin x \cos x} \left( U+V \right) - \left( d-2\right) \frac{\sin x}{\cos x}V\,Ae^{-\delta} \,,  
\label{evol_V}
\end{eqnarray}
\begin{eqnarray}
\dot{A}    & = & - \onehalf A^2 e^{-\delta} \sin x \;\cos^{2d-3} x \,(V^2 - U^2)\,, 
\label{evol_A}
\end{eqnarray}
(ii) Constraint equations\footnote{The distinction between evolution and constraint equations we make here is not in correspondence with the evolution and constraint equations of the 3+1 ADM formalism~\cite{adm:1962ok}.}:
\begin{eqnarray}
A'  & = &  \frac{d-2+2\sin^2 x}{\sin x\cos x}(1-A) \nonumber \\
    & - & \frac{A}{2}\sin x\cos^{2d-3} x\left( V^2 + U^2\right) \,,   
\label{A_prime}
\end{eqnarray}
\begin{eqnarray}
\delta' & = & - \onehalf \sin x \cos^{2d-3} x  \left( V^2 + U^2 \right)\,. 
\label{delta_prime}
\end{eqnarray}
It is interesting to note that we have both an evolution and a constraint equation for $A$.  As we have already indicated, only the scalar field sector have a distinctive hyperbolic structure, Eqs.~\eqref{evol_U} and~\eqref{evol_V}, while the evolution of $A$ does not contain any gradients of the variables.  In practice, we can solve for $A$ either by evolving it using Eq.~\eqref{evol_A} or by solving the {\em elliptic} equation~\eqref{A_prime}. On the other hand, from the definition of $U$ and $V$, Eqs.~\eqref{Udef} and~\eqref{Vdef}, we can find also a  constraint equation for the scalar field variable $\phi$: 
\begin{eqnarray}
\phi' = \frac{1}{2}\cos^{d-2}x \left( U+V\right) \,, 
\label{phi_prime}
\end{eqnarray}
and therefore also for the normalized scalar field $\psi$:
\begin{eqnarray}
\psi' =  \psi \,\frac{\sin x}{\cos x} (d-1) + \frac{1}{2}\, \frac{U + V}{\cos x} \,.
\label{psi_prime}
\end{eqnarray}
Then, in the same way as with A, we can solve for $\psi$ either by evolving Eq.~\eqref{evol_psi} or by solving this constraint equation.

To solve these equations we need boundary conditions at the center $x=0$ and at the AdS boundary $x=\pi/2$. Near $x=0$ we find that the scalar field variables admit the following power expansion:
\begin{eqnarray}
\psi & = & \psi^{}_0 + \psi^{}_2\, x^2 + O (x^4) \,, \\
U    & = & U^{}_0 + U^{}_1\, x + U^{}_2\, x^2 + O (x^3) \,, \\
V    & = & - U^{}_0 + U^{}_1\, x - U^{}_2\, x^2 + O (x^3) \,,
\end{eqnarray}
and the metric coefficients admit the expansions:
\begin{eqnarray}
A      & = & 1 + A^{}_2 x^2 + O (x^4)\,, \\
\delta & = & \delta^{}_0 + \delta^{}_2 x^2 + O (x^4)\,,
\end{eqnarray}
where $\delta^{}_0$ is a time-dependent quantity always greater than zero. 

We can also obtain a power expansion near the AdS boundary by just introducing a coordinate change in the radial direction: $\rho = \pi/2 - x$. Then, the expansions for the normalized scalar field $\psi$ and the characteristic variables $U$ and $V$ are:
\begin{eqnarray}
\psi   & = & \psi^{}_1 \rho + O(\rho^{3})\,, \\
 U     & = & U_1 \rho  + U_2 \rho^{2} + O (\rho^{3})\,, \\
 V     & = & U_1 \rho  - U_2 \rho^{2} + O (\rho^{3})\,,
\end{eqnarray}
and for the metric functions $A$ and $\delta$:
\begin{eqnarray}
A      & = & 1 + O(\rho^d)\,, \\
\delta & = & O (\rho^{2d}) \,.
\end{eqnarray}
%


\subsection{Characteristic-type Evolution of the EKG System}\label{analytics_characteristic}

For the characteristic evolution, we adapt the scheme first introduced by Christodoulou~\cite{Christodoulou:1986zr} and later used in~\cite{Goldwirth:1987nu,Garfinkle:1994jb} to the case of AAdS spacetimes with spherical symmetry.   The form of the metric is:
\be
ds^{2} = -g\bar{g}\,du^{2} - 2 g \,dudr + r^{2}\, d\Omega^{2}_{d-1}\,, 
\label{charmet}
\ee
where $u$ is an outgoing null coordinate ($u=const.$ is an outgoing null geodesic) and $r$ is the radial area coordinate.  The coordinate range for $(u,r)$ is: $u\in(-\infty,\infty)$ and $r\in(0,+\infty)$ (although the range of $r$ depends on whether gravitational collapse takes place).  The AdS boundary corresponds to $r\rightarrow\infty$. The functions $g=g(u,r)$ and $\bar g=\bar g(u,r)$ are always greater that some normalization value at the origin that we choose to be $1$. The AdS limit is: $g\rightarrow 1$ and $\bar g \rightarrow 1 + {r^2}/{\ell^2} $.  The coordinates $(u,r)$ have dimensions of length and, throughout this paper, the numerical values that we quote are in units of $\ell$.

To write down the field equations~\eqref{efes} and~\eqref{kg_eq} in the coordinates of Eq.~\eqref{charmet} we introduce two variables associated with the scalar field $\phi$:
\be
\bar{h} = \phi\,, 
\label{h-def}
\ee
and
\bea
\frac{d-1}{2}r^{\frac{d-3}{2}}\,h = \left( r^{\frac{d-1}{2}}\,\bar{h}\right)_{,r} \,. 
\label{slicedq1}
\eea
Then, we can recover $\bar{h}$ from $h$ as follows
\begin{equation}
\bar{h}(u,r) = \frac{d-1}{2} r^{\frac{1-d}{2}} \int^{r}_{0}  {r'}^{\frac{d-3}{2}}   h(u,r') \, d r'  \,. 
\label{slicedq1bis}
\end{equation}
Moreover, from the $(r,r)$ and $(u,r)$ components of the Einstein equations~\eqref{efes} for the metric of Eq.~\eqref{charmet}, we get:
\begin{align}
g_{,r} &= r g ({\bar h}_{,r})^2\,, \label{ee_gr} \\
\left( r^{d-2} \, \bar g \right)_{,r} &= \left(  d -2 + d \frac{r^2}{l^2}\right) r^{d-3} g\,, \label{ee_gbr} 
\end{align}
and from here, we can solve for the metric variables $(g,\bar{g})$ in terms of scalar field variables $(h,\bar{h})$ as:
\begin{align}
g(u,r) & =  \exp\left\{ \frac{(d-1)^{2}}{4} \int^{r}_{0}dr' \frac{\left( h(u,r')-\bar{h}(u,r')\right)^{2}}{r'}\right\}\,, 
\label{slicedq2} \\
\bar{g}(u,r) & =  \frac{1}{r^{d-2}}\int^{r}_{0}dr' \left( d - 2 + d\,\frac{r'^{2}}{\ell^{2}} \right) r'^{d-3}\, g(u,r') \,. 
\label{slicedq3}
\end{align}
That is, as expected we can find all the variables of the problem from $h$.  An important observation about Eq.~\eqref{slicedq3} is that both the numerator and denominator of the right-hand side go to zero as we approach the origin $r=0$, although they do it in a way that the limit is well-defined and finite.  However, this can problematic from the point of view of the convergence of a numerical algorithm.  Then, following~\cite{Bland:2005kk,Bland:2007sg} we can get an alternative form for this equation by using integration by parts. The result is:
\begin{widetext}
\be
\bar{g}(u,r) = \left(1+\frac{r^{2}}{\ell^{2}}\right)g(u,r) - \frac{(d-1)^{2}}{4\,r^{d-2}}\int^{r}_{0}dr' r'^{d-3}\left(1+\frac{r'^{2}}{\ell^{2}}\right)\left(h(u,r')-\bar{h}(u,r')\right)^{2}\,g(u,r') \,,
\label{slicedq3bis}             
\ee
\end{widetext}
where we have used the boundary conditions at the origin and the equation for $g(u,r)$ [Eq.~\eqref{slicedq2}]. Now, the second term in this equation goes to zero as we approach the origin and hence, it is more amenable for numerical computations.  The only remaining equation is the one for $h$, which can be obtained from the Klein-Gordon equation~\eqref{kg_eq} and can be written as:
\begin{equation}
\frac{\partial^{2} \bar{h}}{\partial u \partial r} -\frac{1}{2} \frac{\partial}{\partial r}\left( \bar{g}\frac{\partial \bar{h}}{\partial r} \right)
+ \frac{d-1}{2r}\left( \frac{\partial \bar{h}}{\partial u}-\bar{g}\frac{\partial \bar{h}}{\partial r} \right) = 0 \,.
\end{equation}
Using Eq.~\eqref{slicedq1} we finally get the equation for $h$:
\begin{equation}
\frac{\partial h}{\partial u} -\frac{1}{2} \bar{g}\frac{\partial h}{\partial r} = 
\frac{h - \bar{h}}{2\,r} \left[ \left( d-2 + d\frac{r^{2}}{\ell^{2}}\right) g - \frac{d-1}{2}\bar{g} \right] \,.  
\label{kge-pde}
\end{equation}

\begin{figure}[t]
\begin{center}
\centerline{\resizebox{.50\textwidth}{!}{ \includegraphics{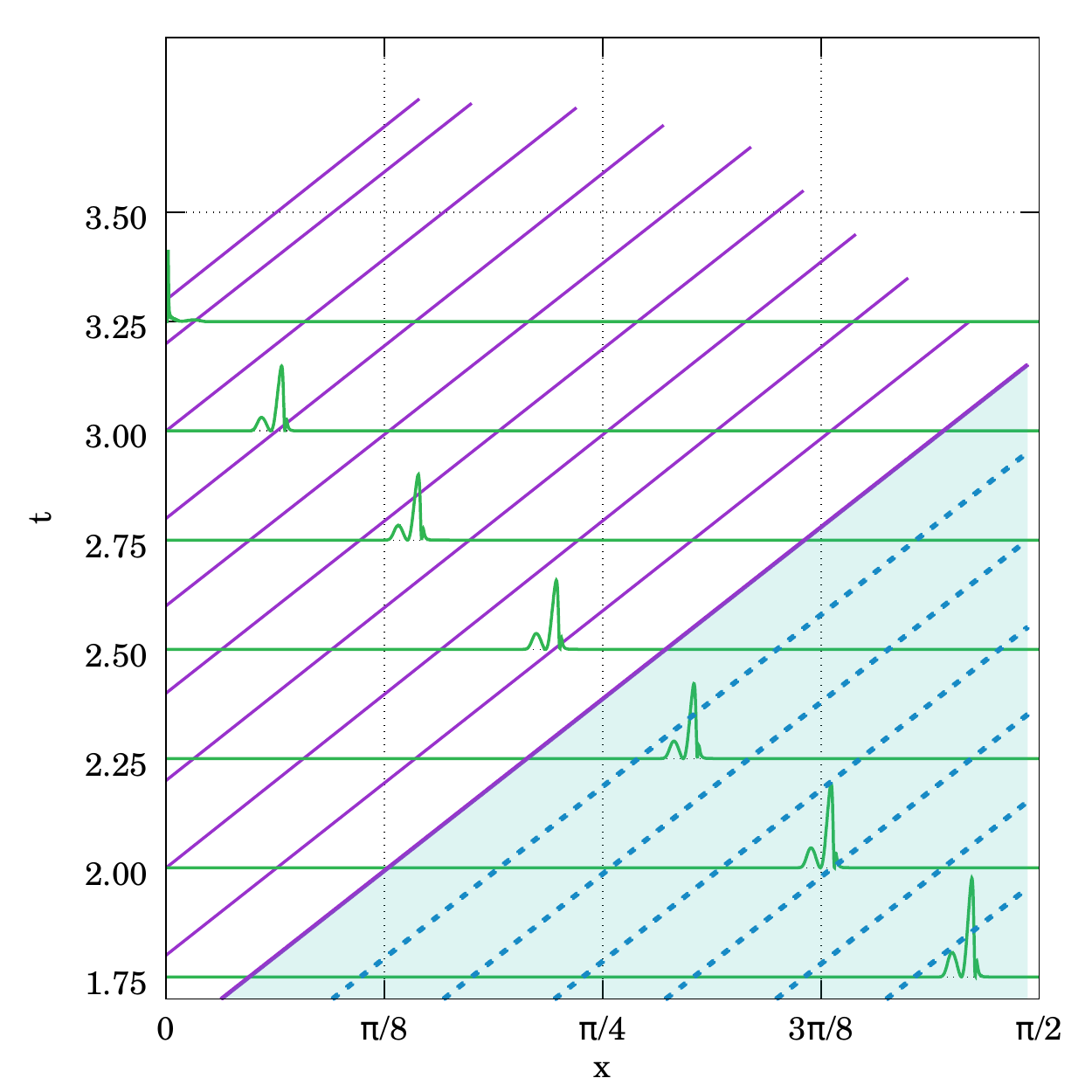}}}
\caption{Characteristic evolution. The green (mostly horizontal) lines represent the energy density as computed using the Cauchy evolution at different times (evolution goes in the vertical time direction). Blue dashed lines are characteristics (outgoing null geodesics) computed through the Cauchy data. Purple lines represent the {\em surfaces} we have evolved using the characteristic evolution using the thickest purple line as a seed.  The characteristic grid moves according to Eq.~\eqref{inullgeo} and therefore the range in $r$ covered decreases over time.
\label{plot_charac_evol}}
\end{center}
\end{figure}

In this work we use the characteristic initial-value problem in the traditional way, in the sense that we integrate the hyperbolic equations along their associated characteristic lines (see, e.g.~\cite{Courant:1989aa,John:1991fj}). Then, we set up initial data on an initial outgoing null slide $\{u = u^{}_{o} = const.\}$ and evolve that data onto the next slide $\{u = u^{}_{o} +\Delta u = const.\}$ through the ingoing null geodesics (the purple lines in Fig.~\ref{plot_charac_evol}), which are given by 
\begin{equation}
\frac{dr}{du} = - \onehalf \bar{g} \,. 
\label{inullgeo}
\end{equation}
Integrating along the ingoing null geodesics allows us to exchange partial derivatives of our variables by total derivatives with respect to $u$. For instance, in the case of the field variable $h$ we have
\be
\frac{dh(u,r(u))}{du} = \left(\frac{\partial h}{\partial u}\right)^{}_{r=r(u)} + 
\left(\frac{\partial h}{\partial r}\right)^{}_{r=r(u)} \frac{dr(u)}{du} \,, 
\label{charevorel}
\ee
where $r(u)$ is an ingoing null geodesic, solution of Eq.~\eqref{inullgeo}.  In this way we can replace Eq.~\eqref{kge-pde} by ordinary differential equations, one for the variable $h$,
\be
\frac{dh}{du} = \frac{h - \bar{h}}{2\,r} \,
\left[ \left( d-2+d\,\frac{r^{2}}{\ell^{2}}\right) g - \frac{d-1}{2}\bar{g} \right] \,, 
\label{charevod}
\ee
and another one for $r(u)$, namely Eq.~\eqref{inullgeo}.  The first one, Eq.~\eqref{charevod}, tells us how to evolve $h$ from a $\{u=$ const.$\}$ slice to the next one.  The second one tells us that the $r$-coordinate of a point in a $\{u=$ const.$\}$ slice changes according to the ingoing null radial geodesic to which that point belongs.

Some of the previous expressions can be problematic at the origin and in order to make perfect sense of them we need to understand their behaviour around $r=0$, from where we can also extract boundary conditions. Assuming the following expansion for the scalar field $\phi=\bar h$:
\begin{align}
\bar h &= \phi_0 + \phi_1 \, r   + \phi_2 \, r^2  + O (r^3)\,,
\end{align}
we obtain the following expansions for the rest of variables:
\begin{align}
h &= \phi_0 + \frac{d+1}{d-1} \phi_1 \,r  + \frac{d+3}{d-1} \phi_2 \, r^{2}  + O (r^3)\,, \\
g &=  1 + \frac{1}{2} \phi_1^2 \, r^2  + O (r^3)\,, \\
\bar g &= 1 + O (r^2)\,.
\end{align}
%

\subsection{Transition from the Cauchy Evolution to the Characteristic Evolution}
\label{cauchy-to-characteristic}

During the Cauchy evolution we can monitor our variables to see when we are approaching the collapse.  Then, we can make the transition from the Cauchy evolution of Sec.~\ref{equations_cauchy} to the characteristic evolution of Sec.~\ref{analytics_characteristic} in order to follow better the dynamics near collapse.   
This transition consists in constructing initial data on an initial null slide $\{u = u^{}_{o} = $ const.$\}$ from the outcome of the Cauchy evolution. It is important to cover a portion of the spacetime that guarantees that the characteristic evolution will cover the formation of the apparent horizon as it is represented in Fig.~\ref{plot_charac_evol}, where we take the first purple line (a null outgoing geodesic) as the initial slice for the characteristic evolution.  As we can see, the initial data on that slice uses the information of the Cauchy evolution for a time $t_{f}-t_{i} \approx \pi/2$, and in this way AH formation is covered.

This transition requires to find the relations between different objects of the Cauchy and characteristic evolutions.  First we need to find the relation between the coordinates $(t,x)$, used for the Cauchy evolution, and the coordinates $(u,r)$ of the characteristic one. The relation between the radial coordinates $x$ and $r$ is quite straightforward considering the factor in front of the metric of the unit $(d-1)$-sphere in Eqs.~\eqref{aadstx} and~\eqref{charmet}, from where we get:
\begin{equation} 
r = \ell\,\tan x \,. 
\label{r-of-x}
\end{equation} 

The second important ingredient is the construction of the initial null slice for the characteristic evolution from the information extracted from the Cauchy evolution.  This can be done by finding the outgoing null geodesics from the Cauchy evolution. From the expression of the metric in Eq.~\eqref{aadstx} we have that the outgoing null geodesics are given by:
\begin{equation}
\frac{dx}{dt} = + A e^{-\delta} \equiv v(t,x)\,, 
\label{outgoing-null-geodesics}
\end{equation}
where the plus sign has been included to emphasize that these radial null geodesics are outgoing (a minus sign would correspond to the ingoing null geodesics).  To integrate this ODE we need to know the metric functions $A(t,x)$ and $\delta(t,x)$, for which we need to have the evolution of the Cauchy initial-value problem over the spacetime region that includes the null geodesics of interest. Some of these geodesics are shown in Fig.~\ref{plot_charac_evol} in a $t-x$ diagram.

The next important ingredient is the construction of initial data for the characteristic evolution on one of the 
null outgoing geodesics that constitute the slicing $\{u = $ const.$\}$.  To begin with, let us apply the coordinate transformation of Eq.~\eqref{r-of-x}, adding the transformation $\tau = \ell\,t$, to the metric in Eq.~\eqref{aadstx}.  This brings this metric to a more familiar form of AAdS spacetime metrics:
\begin{equation}
ds^{2} = -A {e}^{-2\delta}\,\left( 1 + \frac{r^{2}}{\ell^{2}}\right) d\tau^{2} + 
\frac{dr^{2}}{A\,\left( 1 + \frac{r^{2}}{\ell^{2}}\right)} + r^{2}\, d\Omega^{2}_{d-1}\,. 
\label{aadsTR}
\end{equation}
The AdS limit $A\rightarrow 1$ and $\delta\rightarrow 0$ gives us the well-known form of the AdS spacetime metric.  With this in mind, let us perform a general coordinate transformation from the Cauchy-formulation metric to characteristic-formulation metric:
\begin{equation}
\tau  = {\cal F}(u,r)\,, 
\label{coord2}
\end{equation}
which transforms the metric in Eq.~\eqref{aadsTR} to the following metric:
\begin{eqnarray}
ds^{2} & = & -A e^{-2\delta}\,\left( 1 + \frac{r^{2}}{\ell^{2}}\right) {\cal F}^{2}_{u}\, du^{2} \nonumber \\
       & - & 2A e^{-2\delta}\,\left( 1 + \frac{r^{2}}{\ell^{2}}\right) {\cal F}^{}_{u}{\cal F}^{}_{r}\, dudr \nonumber \\
       & + & \left[ 1 - A^{2}e^{-2\delta}\left( 1 + \frac{r^{2}}{\ell^{2}}\right)^{2}{\cal F}^{2}_{r}\right]\frac{dr^{2}}{A\,\left( 1 + \frac{r^{2}}{\ell^{2}}\right)} \nonumber \\
       & + & r^{2}\, d\Omega^{2}_{d-1}\,, 
\label{genmet}
\end{eqnarray}
where ${\cal F}^{}_{u} \equiv \partial{\cal F}/\partial u$ and ${\cal F}^{}_{r} \equiv \partial{\cal F}/\partial r$.  Now, let us impose two conditions on the general coordinate transformation of Eq.~\eqref{coord2}.  The first one comes by comparing with the characteristic metric of Eq.~\eqref{charmet}, namely from the fact that the vector $\partial/\partial r$ is a null vector for this metric.  If we impose the same condition on the metric in Eq.~\eqref{genmet} we are imposing a condition in the transformation between the coordinates $(t,x)$ and $(u,r)$, i.e. on the function ${\cal F}$ in Eq.~\eqref{coord2}, which is the following: $\met^{}_{rr} = 0$. Since the slides $\{u=const.\}$ must be composed of null outgoing radial geodesics we can write
\begin{equation}
{\cal F}^{}_{r} = \frac{1}{\left(1+\frac{r^{2}}{\ell^{2}}\right)A e^{-\delta}} \,.
\label{condition-on-Fr}
\end{equation}
In the case of ingoing geodesics we would have chosen the opposite sign for ${\cal F}^{}_{r}$.  The second condition that we impose on the coordinate change has to do with the freedom in rescaling the coordinate $u$, which is a freedom in the choice of the quantity ${\cal F}^{}_{u}$.  Our choice, motivated by the implementation of the Cauchy-characteristic transition, is: 
\begin{equation}
{\cal F}^{}_{u} = e^{\delta_0}\,,
\label{condition-on-Fu}
\end{equation}
where $\delta_0$ is the value of the metric function $\delta$ at $x=0=r$. Now, by comparing the line element in Eq.~\eqref{genmet} with the one for the characteristic formulation in Eq.~\eqref{charmet}, and using the conditions on the function ${\cal F}$ given in Eqs.~\eqref{condition-on-Fr} and~\eqref{condition-on-Fu}, we find the following relations between $(A,\delta)$ and $(\bar{g},g)$:
\begin{align}
g &= e^{\delta_0 - \delta}\,, 
\label{expression-for-g}\\
\bar{g} &= A e^{\delta_0 - \delta}\, \frac{1}{\cos^2 x}\,, 
\label{expression-for-gbar}\\
A &= \frac{\bar g}{g \left( 1 + \frac{r^2}{\ell^2}\right)}\,. 
\label{Agbarg}
\end{align}
These are key relations to carry out the transition from the Cauchy to the characteristic evolution, both for the construction of the initial slice and for the initial data on that slice. Given that AH formation in the Cauchy evolution is given by the limit $A\rightarrow 0$, from Eq.~\eqref{Agbarg} we have that in the characteristic evolution we can track AH formation by monitoring the right-hand side of this equation and controlling when it approaches unity.  

On the other hand, from our particular choice of coordinate change, Eqs.~\eqref{coord2},~\eqref{condition-on-Fr}, and~\eqref{condition-on-Fu}, we can write the following relation between $\tau$ and $u$ (and $r$):
\begin{eqnarray}
\tau & = & u + \ell \int^{x}_{0} \frac{dx'}{v(t^{}_{+}(x'),x')} \nonumber \\ 
     & = & u + \int^{r}_{0} \frac{dr'}{\left(1+\frac{r'{}^{2}}{\ell^{2}}\right)v(t^{}_{+}(r'),r')} \,.
\label{tau-of-r}
\end{eqnarray}
where $t^{}_{+}(x)$ denotes the solution for the outgoing null geodesics, Eq.~\eqref{outgoing-null-geodesics}, and $v$ is the function of $(t,x)$ defined there.

Finally, we give the relations between the metric and scalar field variables in both formulations. First, $\psi$ and $\bar h$ are, by definition, directly related with the scalar field:
\begin{equation}
\bar{h}  =  \phi = \cos^{d-1}x\,\psi\,.  
\label{hbarexp} 
\end{equation}
The scalar field variable $h$ can be constructed along the outgoing null geodesics in term of the Cauchy variables as follows:
\begin{eqnarray}
h        & = & \bar h + \frac{2}{d-1} r\, \bar h_{,r}  \nonumber \\
         & = & \phi +  \frac{2}{d-1} r(x) \left( \PD{x}{r}\PD{}{x} + \PD{t}{r}\PD{}{t} \right) \phi \nonumber \\
         & = & \cos^{d-1}x\left( \psi + \frac{2}{d-1}  \sin x  \,\,V\right) \,,  
\label{heq}
\end{eqnarray}
where we have used Eqs.~\eqref{Vdef},~\eqref{slicedq1},~\eqref{r-of-x}, and~\eqref{tau-of-r}.  It is remarkable that $h$ depends on the scalar field itself, through the variable $\psi$, and the ingoing (negative speed) characteristic variable $V$ [see Eq.~\eqref{Vdef}], but not on the outgoing (positive speed) characteristic variable $U$ [see Eq.~\eqref{Udef}].  The reason is that we are doing the characteristic evolution using null slides made out of outgoing null geodesics, and hence the evolution from one slide to the next one of $h$ takes place along ingoing null geodesics [see Eq.~\eqref{inullgeo}].

In summary, Eqs.~\eqref{expression-for-g}-\eqref{heq} contain all the information we need to construct the initial null slide during the Cauchy evolution, the associated coordinate change, and the initial data to initiate the characteristic evolution.  This completes the procedure to perform the transition from the Cauchy to the characteristic evolution.


\section{Basics of the Numerical Implementation}\label{numerics}
In this section we describe the basic ingredients for the numerical implementation of the two evolution schemes, and the transition between them, described in Sec.~\ref{equations}.  In the case of the Cauchy evolution we use pseudospectral collocation methods with multiple domains following the line of previous works in the context of the computation of the self-force in black-hole spacetimes~\cite{Canizares:2009ay,Canizares:2010yx,Canizares:2011kw,Jaramillo:2011gu}.  For the characteristic evolution we use the method introduced in~\cite{Goldwirth:1987nu}, consisting in using a null foliation where the points of each slide follow ingoing null geodesics (the {\em characteristic lines}).  Finally, we describe the details of how we store the information from several Cauchy slides in order to construct the initial slide and data for the characteristic evolution.

\begin{figure*}[t]
\centerline{\resizebox{.5\textwidth}{!}{ \includegraphics{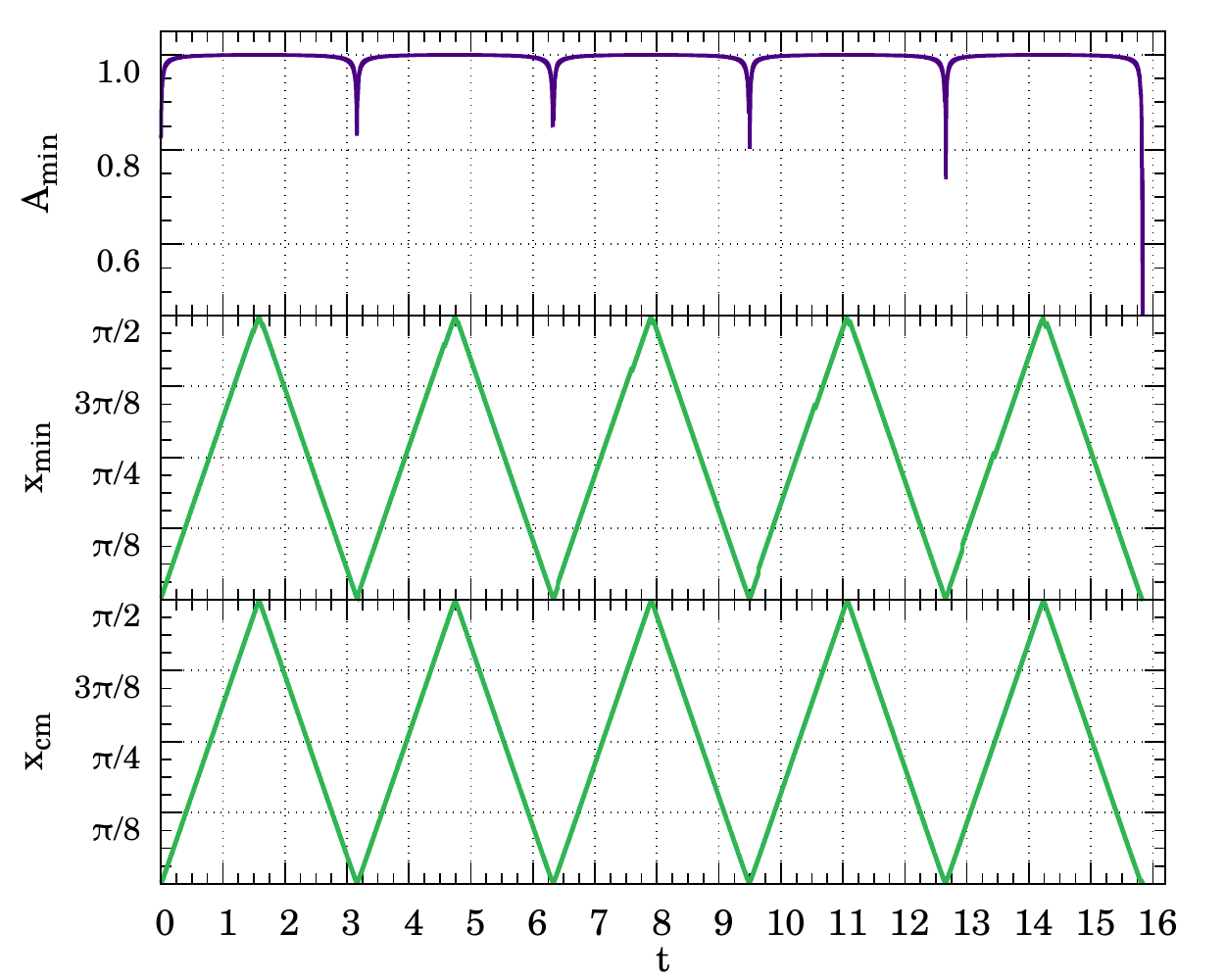}   } 
\resizebox{.5\textwidth}{!}{ \includegraphics{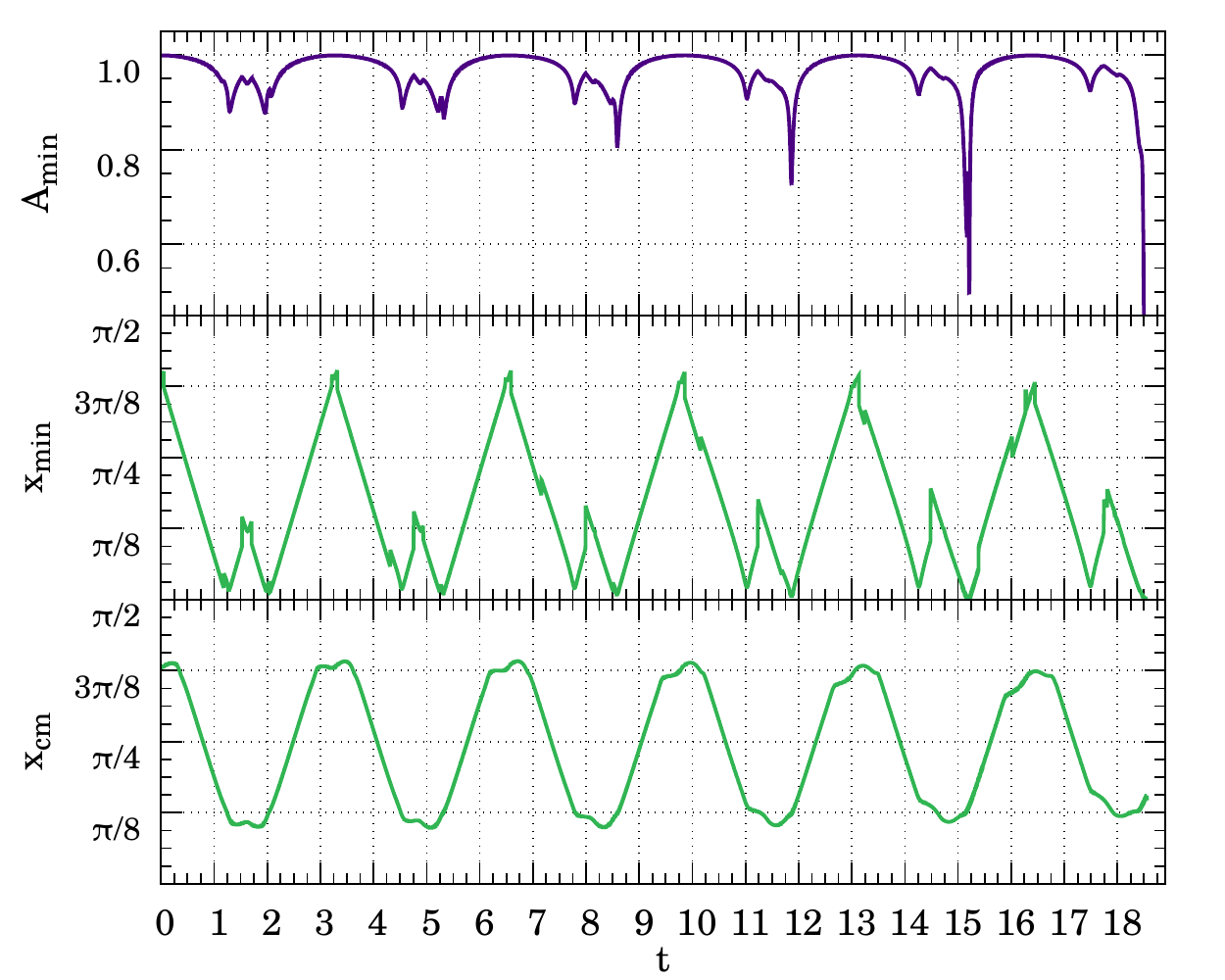}   }  }
\centerline{}
\caption{{\em Center-of-mass} evolution. In the left plot we show the evolution of $A_{\mathrm{min}}$, $x_{\mathrm{min}}$, $x_{\mathrm{cm}}$ for initial conditions in Eq.~\eqref{cauchy-initial-data_Bizon}, that collapse after five bounces. In the right plot we use an initial profile from Eq.~\eqref{cauchy-initial-data-boundary}.}
\label{plot_cdm}
\end{figure*}


\subsection{Numerical Implementation of the Cauchy Evolution \label{numerics_cauchy}}

In order to have a precise numerical evolution we are going to use the PseudoSpectral Collocation (PSC) method (see, e.g.~\cite{Boyd,Fornberg:1996psc,Canutoetal:2006sm1}) for the space discretization, which in our case means in the radial direction, in the compactified radial coordinate $x$ to be more precise. The main tools of the PSC method used in this paper are briefly described in Appendix~\ref{app_psc}. In a standard spectral method the outcome of the spatial discretization of a set of hyperbolic PDEs is a (much larger) set of ODEs for the time-dependent spectral coefficients.  Instead, in the PSC method we obtain a set of ODEs for the time-dependent values of our variables, $\bm{U}=(U,V,...)$ (which variables are evolved in time depends on the choice of equations since some variables, like $A$ and $\psi$, can be found either by time evolution or by radial integration) at the collocation points, $\{\bm{U}_{i}(t)\equiv \bm{U}(t,x_{i})\}$, where the equations are forced to be satisfied exactly.  The number of ODEs that we obtain is equal to the total number of variables ($N_{v}$) times the number of collocation points ($N$), i.e. $N\times N_{v}$.  The numerical evolution of the resulting ODEs for the collocation values $\{\bm{U}_{i}(t)\}$ is performed using a standard Runge-Kutta 4 (RK4) algorithm (see, e.g.~\cite{Butcher:2003jcb,Press:1992nr}).  

The great advantage of the PSC method is that for smooth solutions it provides exponential convergence, i.e. the truncation error of the spectral series, which can be approximated by the last spectral coefficient, $a^{}_{N}$, decays as $e^{-N}$.  In contrast, the cost of most operations like derivatives, computation of non-linear terms, etc. increases as $N^{2}$ with the number of collocation points, unless we use a fast Fourier algorithm to transform from the physical space (the collocation values of our variables) to the spectral space (the coefficients of the spectral series for our variables), in which case the cost increases only as $N\log{N}$. In addition, the Courant-Friedrichs-Lewy (CFL) condition for the stability of the evolution of the PDEs (see, e.g.~\cite{Gustafsson:1995tb}), in the case of the PSC method, is of the form $\Delta t < C\,N^{-2}$ (where $C$ is a certain constant), in contrast with the typical form of standard finite difference schemes for PDEs, where $\Delta t< C'\,N^{-1}$ and $C'$ is another constant. This is due to the structure of the Lobatto-Chebyshev grid that we use (see details in Appendix~\ref{app_psc}), where the points cluster near the boundaries of the domain. As a consequence, the evolution in the PSC method can be significantly more expensive than in the case of finite-difference schemes.  A way to alleviate this is to use refinement via a multidomain PSC method.  The idea is to adapt the size and number of the domains so that different regions in the radial direction with different resolution requirements are covered by an adequate number of collocation points.  We can change the number and size of the different domains along the evolution, following the resolution needs of the problem.  The practical implementation of the AMR is described in Sec.~\ref{refinement-cauchy}.  Most computations are done at each domain in an independent way.  The different domains are connected via the corresponding matching conditions, which depend on the type of equation that each variable satisfies.

The Cauchy evolution allows us to follow the system from its initial conditions to the latest stages, just before the collapse. As we have already mentioned we can expect the scalar field to travel to the AdS boundary ($x= \pi/2$) several times, and in this sense using the compactified radial coordinate $x$ gives us control over the whole space.  On the other hand, when the scalar field is close to collapse, large gradients will be generated in our variables and the AMR is crucial in order to guarantee the high resolution requirements needed to resolve the dynamics near collapse.

In Sec.~\ref{equations_cauchy} we have presented the equations we obtain from Einstein's field equations and from energy-momentum conservation in terms of the Cauchy-type variables, namely $(\psi,U,V,A,\delta)$.  Some variables have two equations, for instance the metric function $A$ can be obtained either by evolving Eq.~\eqref{evol_A} or by integrating Eq.~\eqref{A_prime} and the same happens with the scalar field variable $\psi$ [See Eqs.~\eqref{evol_psi} and~\eqref{psi_prime}].  We have numerically implemented several combinations of equations but in general we obtained the best results and efficiency by evolving in time $U$ and $V$ [with Eqs.~\eqref{evol_U} and~\eqref{evol_V}] and then obtaining $\psi$, $A$, and $\delta$ from radial integration [with Eqs.~\eqref{psi_prime},~\eqref{A_prime}, and~\eqref{delta_prime} respectively].

From Eqs.~\eqref{A_prime} and~\eqref{delta_prime} we can find an integral expression for the metric functions $A$ and $\delta$:
\begin{align}
A(t,x) &- 1 =  \nonumber \\
& - \frac{\cos^d x\;e^{\delta}}{2 \sin^{d-2} x} \int_0^{x}\!\!\! dy\, e^{-\delta}\sin^{d-1}y\cos^{d-3}y (U^2 + V^2) \,,   
\label{ads_integration_for_A}  
\end{align}
\begin{align}
\delta(t,x) &=  \frac{1}{2} \int_x^{\frac{\pi}{2}}\!\!\! dy\, \sin y \cos^{2d-3} y \left( U^2 + V^2 \right)\,,
\end{align}
and using Eq.~\eqref{psi_prime} for the scalar field:
\begin{align}
\psi(t,x) &= - \frac{1}{\cos^{d-1} x} \int_x^{\pi/2}\!\!\! dy\, \cos^{d-2} y \left( U^2+ V^2\right) \,.
\end{align}

On the other hand, we can introduce the energy density
\begin{equation}
\mathcal{E}(t,x) =  e^{-\delta}  \sin^{d-1} y  \cos^{d-3} y \left( \frac{U^2 + V^2}{2}\right)\,,
\end{equation}
and from it we can compute the energy contained inside a sphere of a given radius $x$, which we call the mass function:
\begin{equation}
\mathcal{M}(t,x)  =  e^{\delta} \int_0^{x}\!\!\! dy\,  \mathcal{E}(t,y) \,,
\end{equation}
which is related to the metric function $A$ by
\begin{equation} 
A(t,x) = 1 -  \frac{\cos^d x}{\sin^{d-2} x} \mathcal{M}(t,x) \,.
\end{equation}
Then, the ADM mass is just the limit: $M_{\rm ADM} = \lim^{}_{x\rightarrow\pi/2}\mathcal M(t,x)$.  The ADM mass is a constant, it does not depend on time, that we can use in our simulations to check the numerical accuracy.  In addition, we can define the following quantity:
\begin{equation}
x^{}_{\mathrm{cm}} = \frac{1}{M^{}_{\rm ADM}} \int_0^{x}\!\!\! dy\, y \, \mathcal{E}(t,y) \,,
\end{equation}
which plays the role of a radial {\em center of mass}, in the sense that we can use it to track where the energy is concentrated, which is specially useful when evolving localized scalar field configurations, like for instance those corresponding to the initial conditions given in Eq.~\eqref{cauchy-initial-data_Bizon}.  There are other possible definitions of a radial center of mass, for instance we can use the radial position of the minimum of the metric function $A$, i.e. $x_{\mathrm{min}}$ such that $A(x_{\mathrm{min}}) = \mathrm{min}( A(x)) \equiv A_{\mathrm{min}}$.  We compare these two definitions of radial center of mass, $x_{\mathrm{cm}}$ and $x_{\mathrm{min}}$, in Fig.~\ref{plot_cdm}, where their evolution is compared with the evolution of $A_{\mathrm{min}}$ for two different sets of initial data, the one given in Eq.~\eqref{cauchy-initial-data_Bizon}, which collapses after five bounces off the AdS boundary, and the one given in Eq.~\eqref{cauchy-initial-data-boundary}.  As we can see in Fig.~\ref{plot_cdm}, for the initial profiles in Eq.~\eqref{cauchy-initial-data_Bizon} (left panel) the differences between $x_{\mathrm{min}}$ and $x_{\mathrm{cm}}$ are quite small although $x_{\mathrm{min}}$ presents some small abrupt features.  These features are more prominent for the more complex initial profile of Eq.~\eqref{cauchy-initial-data-boundary} (right panel), where the evolution of both $x_{\mathrm{cm}}$ and $x_{\mathrm{min}}$ is more complex, but $x_{cm}$ appears to be a much smoother indicator to track the evolution of the scalar field profile. 

Finally, regarding more technical details of the numerical implementation, we remark that all the operations involving the spatial radial direction, including the integrals, are performed within the framework of the PSC method as briefly described in Appendix~\ref{app_psc}.  Another important ingredient of the numerical implementation is how to deal with the multiple domains. In our scheme the boundaries of each domain have duplicated information because the boundary points of one domain are identified, with the exception of the global boundaries ($x=0$ and $x=\pi/2$), with the boundary points of the contiguous domains.  Although most operations are done locally at each domain we need to communicate the different domains through these boundary points.  This is the main reason why we have introduced the characteristic variables $U$ and $V$. These variables have always a well defined direction of propagation (see Appendix~\ref{app_char_var}), which is crucial in order to establish the communication between domains. The characteristic variable $U$ always propagates with positive speed (in direction to the AdS boundary) and the characteristic variable $V$ always travels with negative speed (towards the origin).  Then, the way to communicate two given contiguous domains during the Cauchy evolution is to take the value of the variable $U$ from the right boundary of the domain to the left and to copy it into the $U$-value of the left boundary of the domain to the right (see Fig.~\ref{plot_diagram_U_V}) and the equivalent procedure for $V$.  That is, we take the value of the variable $V$ from the left boundary of the domain to the right and copy it into the $V$-value of the right boundary of the domain to the left.  This way of communicating the characteristic variables ensures that we will not find discontinuities in our variables across the boundaries during the numerical evolution.  In other words, we perform the communication between domains according to the directions of propagation of the information.

\begin{figure}[t]
\begin{center}
\centerline{\resizebox{.50\textwidth}{!}{ \includegraphics{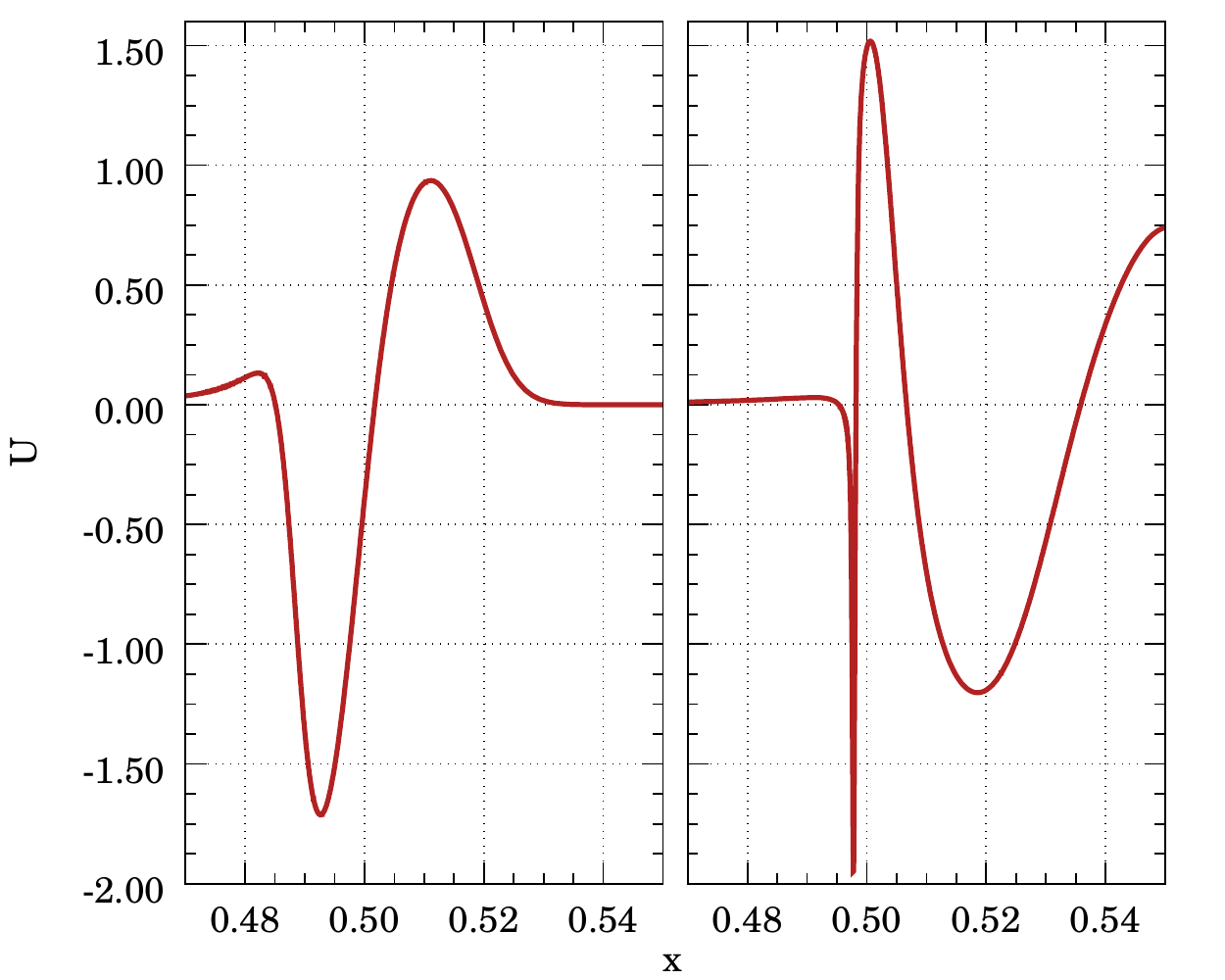}}}
\caption{Cauchy Refinement. Comparison between the $U$ profile in two different simulations at a similar time. Both of them are supposed to collapse after one bounce but the shape and the resolution needed to resolve them are very different.
\label{plot_cauchy_refinement}}
\end{center}
\end{figure}


\subsection{Adaptive Mesh Refinement for the Cauchy Evolution}\label{refinement-cauchy}

\begin{figure}[t]
\begin{center}
\centerline{\resizebox{.50\textwidth}{!}{ \includegraphics{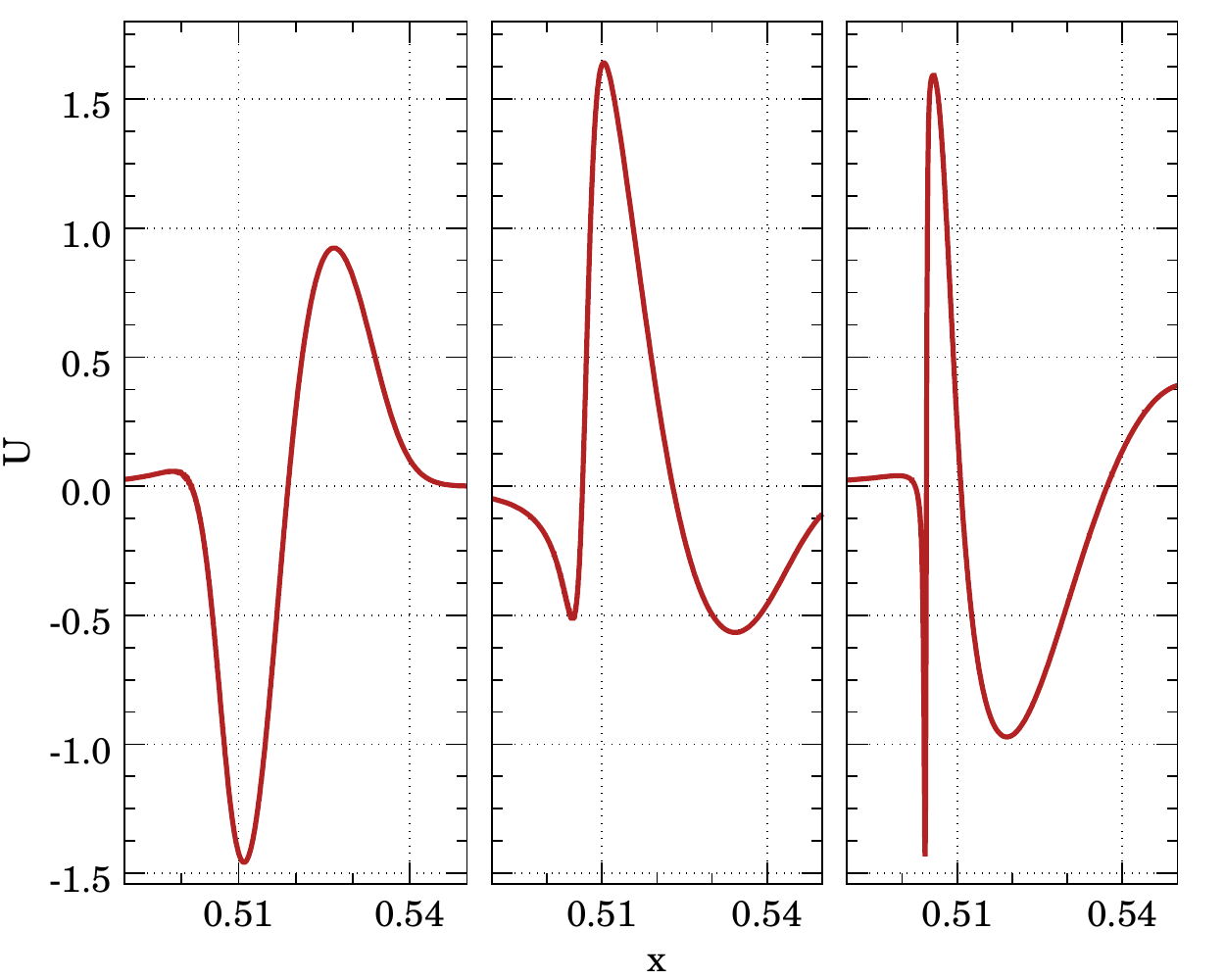}}}
\caption{Cauchy Refinement. Comparison between $U$ profiles from the same simulation at different times. This configuration collapses after three bounces. The snapshots are taken at the same position during the trip to the AdS boundary.  
\label{plot_cauchy_travel}}
\end{center}
\end{figure}

The typical scalar field configurations that we consider in this work, which are quite localized in the radial direction, follow the same evolutionary pattern, already described in Sec.~\ref{equations_cauchy}.  The scalar field attempts to collapse near the origin but if the initial amplitude is below some threshold, the scalar field bounces off the origin and disperses towards the AdS boundary.  Then, it bounces off the AdS boundary and travels again towards the origin.  Then, this sequence is repeated until the scalar field distribution is compact enough to collapse and form an AH.  This means that we need to simulate a compact scalar field distribution back and forth and some of the scalar field variables exhibit growing gradients as the evolution proceeds.  Then, in order to track the features of the scalar field during the evolution in an efficient way we resort to AMR techniques based on our multidomain PSC approach.  The aim is to design a method in which the resolution follows the field in its travel with the minimal loss of precision and without slowing down much the evolution.  In this sense, it is important to mention that although we know the evolutionary pattern the details can vary significantly as we change the initial conditions. To illustrate this, in Fig.~\ref{plot_cauchy_refinement} we show profile of the scalar field variable $U$ at a similar time for two different simulations where collapse happens after one bounce. We see that the shapes are quite different and require different grids in order to resolve them. In Fig.~\ref{plot_cauchy_travel}, we show the profile of $U$ at three different times of a simulation where collapse takes place after three bounces. The snapshots of these figures are taken when the field is traveling towards the boundary so most of the energy is concentrated in the $U$ mode (the one propagating to the right as shown in Fig.~\ref{plot_diagram_U_V}).  They illustrate the need for AMR in our simulations.  We have developed two AMR methods for our simulations.

\subsubsection{First Approach: Gradient Density Estimator}

The first AMR method for our spectral multidomain grid is based on a functional that we call {\it gradient density} and that is defined at each domain as:
\begin{equation}
\rho^{}_{\cal D} = \frac{1}{N} \int_{\cal D} dx \,| V_{,x} |\; \geq \;0\,,
\end{equation}
where ${\cal D}$ is the domain and as we can see this indicator is based on the characteristic variable $V$. The idea is to distribute the domain nodes to minimize the gradient density functional. In our numerical experiments we find a 
threshold for $\rho^{}_{\cal D}$ above which the evolution is no longer valid because of the appearance of high-frequency numerical noise.  Then, during the simulations we modify the domain structure to keep $\rho^{}_{\cal D}$ below the threshold, adding more domains if needed. 

This method works reasonably well for capturing the gradients generated during the collapse but it has several caveats. In particular, it can generate numerical noise far from the region where the scalar field is localized if we do not allow for a minimum resolution there.  It can also give problems when the scalar field presents very sharp features that have to be propagated to the AdS boundary and back, which are precisely the most relevant cases for the study of the mass gap in~\cite{SantosOlivan:2015fmy}.


\subsubsection{Second Approach: Domains over a Curve in Configuration Space}\label{section_ref_along_curve}

In order to deal with the most extreme cases, where the gradients of variables like $U$ and $V$ are very large but still these functions are smooth, we have developed an alternative method that appears to be more robust.  The starting point is to consider a combination of our Cauchy variables that reflects in a very clear way the regions where more refinement is needed, that is, where we find the largest variations in our variables, let us call it $\gamma(A,\delta,\psi,U,V)$. The curve defined in the plane $(x,\gamma(x))$ has a length that, when large gradients in our variables appear, will have  a large contribution from the relatively small interval in the $x$ direction where gradients occur.  Then, let us consider the length of this curve from the origin to a certain radial location $x$:
\begin{equation}
L(x) = \int_0^x  d\tilde{x}\,\sqrt{1 +  {\gamma'(\tilde x)}^2} \,,
\label{l_x}
\end{equation}
where $\gamma' = (\partial\gamma/\partial A)A'+ \ldots$.  The idea of this refinement method is to distribute the nodes of our domains so that they cover the same length of the curve $(x,\gamma(x))$, in contrast with the traditional choice of taking them equally distributed over the $x$ direction.  That is, we select the nodes of our domains, $\{\bar x_i\}$, as: $L (\bar x_i) = i \,L(\pi/2) / D$ ($i=0,\ldots,D$), where $D$ is the total number of domains.

In our simulations we have seen that this method does not require to establish any threshold, instead we just have to change the multidomain structure every a certain number of time steps to adapt to the changes in the variables in a quite smooth way. Every time we change the domain structure we have to interpolate the variables into the new grid. The interpolation between the old and new grids is performed via the pseudospectral representation (see Appendix~\ref{app_psc} for details), and in this way the numerical error introduced is relatively small. 

The specific choice of the function $\gamma$ is the key ingredient of this method and is quite flexible in the sense that we can tune this choice to the type of initial scalar field profiles or even to the particular state of the numerical evolution. For not very demanding simulations in terms of gradients we can choose $\gamma$ to be just $A$ and this provides a very good performance. For the more demanding simulations, a better choice is to take the scalar field characteristic variables $U$, when the field is traveling to the AdS boundary, and $V$, when it is traveling towards the origin. This is motivated by the character of this variables (see Sec.~\ref{numerics_cauchy}), $U$ is the eigenfunction that captures the movement with positive velocity and $V$ the one that captures the movement with negative velocity. In practice we have seen that these simple choices work quite well and allow us to resolve the large changes in these variables that appear during the collapse in the most extreme cases.


\subsection{Numerical Implementation of the Characteristic Evolution}
\label{numerical-implementation-characteristic-evolution}

The characteristic evolution described in Sec.~\ref{analytics_characteristic} is completely different from the Cauchy one. We need to set a grid on the initial null slide in the radial coordinate $r$.  When we evolve to the next null slide the $r$-values of each grid point change according to the ingoing null geodesics [see Eq.~\eqref{inullgeo}]. This has two main effects: First, our last grid point (largest $r$) evolves making our physical computational domain to shrink every step as we show in Fig.~\ref{plot_charac_evol}. Second, the points near the origin are swallowed because, according to the equation for ingoing null geodesics, these points should evolve to negative values of $r$, which do not have a well-defined physical meaning.  This means that we need to control the size of our grid and be careful with the computations near the origin, but other than that it is not problematic.  Actually, the reduction of the grid as we proceed with the characteristic evolution helps to focus our numerical resolution around the region where the collapse takes place and we do not need mesh refinement methods here.  In the cases where the collapse does not occur we will see that it gets scattered towards infinity as it would do in the asymptotically-flat case~\cite{Goldwirth:1987nu,Garfinkle:1994jb}. However, since we are considering AAdS spacetimes, the scalar field has to reach the AdS boundary in a finite time, but the region around the AdS boundary is not covered by our characteristic grid. This means that we have made the transition from the Cauchy to the characteristic evolution too early and therefore, we just need to continue the Cauchy evolution until we can construct an initial null slice that can cover the collapse.

To set up our initial characteristic grid it is very important to establish its size, or equivalently the $r$-coordinate of the last grid point, $r_{\rm max}$.  Once this is done, we can freely distribute the other points.  However, a uniform distribution of the grid points in the radial coordinate $r$ is not a good idea because of the CFL condition. For the characteristic evolution the CFL condition implies:
\begin{equation}
 \Delta u < \frac{1}{2} \mathrm{min} \left( \frac{r_i - r_{i-1}}{\bar g_i} \right)\,,
\end{equation}
where $r_{i}$ and $r_{i-1}$ are two contiguous grid points and $\bar g_{i}=\bar g(r_{i})$.  If we have a uniform grid,  $r_{i}-r_{i-1} = \Delta r$ is the same for any $i$. But when we go to large values of $r$ we have that the metric function $\bar g$ behaves as in pure AdS spacetime, that is: $\bar g \sim r^2$.  This means that the CFL condition in this case is controlled by the outer grid points, the ones with largest $r$, where $\Delta u$ should be too small. But on physical grounds it should be the opposite, the CFL condition should be dominated by the points where we need more resolution, around the region where the AH will form.  What we do is to construct a grid where the grid point separation is constant with respect to the radial coordinate $x$ instead of $r$, and then the outer points are well separated in $r$.  From the Cauchy evolution we extract the values of the scalar field variable $h$ at the different grid points and, from the values of $h$ we find the other variables, $\bar h$, $g$ and $\bar g$, by integration [see Eqs.~\eqref{slicedq1bis}-\eqref{slicedq3}, or Eq.~\eqref{slicedq3bis}].  The first grid point for integration is the origin, where we need to prescribe the boundary conditions: 
\begin{equation}
\begin{aligned}
\bar h (r=0) &= h (r=0) \\
\bar g (r=0) &= g (r=0) = 1\,. 
\end{aligned}
\end{equation}
The integration proceeds to the next grid points by using Simpson's rule:
\begin{eqnarray}
I_i & \equiv &\int_0^{r_i} dr\,f(r)  \\ 
   &= & I_{i-1} + \frac{ r_i - r_{i-1}  }{6} \left[ f\left( r_{i-1} \right) + 4 f\left( r_M \right)  + f \left( r_i\right) \right]\,, \nonumber
\label{simpson_rule}
\end{eqnarray}
where $r_M \equiv (r_i+r_{i-1})/2$ is the $r$-coordinate of the midpoint between $r_{i}$ and $r_{i-1}$, where the value of the integrand is evaluated using spline interpolation~\cite{Gough:2009:GSL:1538674}. 

Each grid point evolves according to the ODE system of Eqs.~\eqref{charevod} and~\eqref{inullgeo}.  To that end, we use   a standard RK4 algorithm (see Refs.~\cite{Butcher:2003jcb,Press:1992nr}).


\section{Code Validation \label{code-validation}}

In this section we show the performance of the different pieces of the numerical code that we have developed to implement the Cauchy-characteristic scheme described in the previous section to study gravitational collapse in spherically-symmetric AAdS spacetimes.


\begin{figure}[t]
\centerline{\resizebox{.5\textwidth}{!}{ \includegraphics{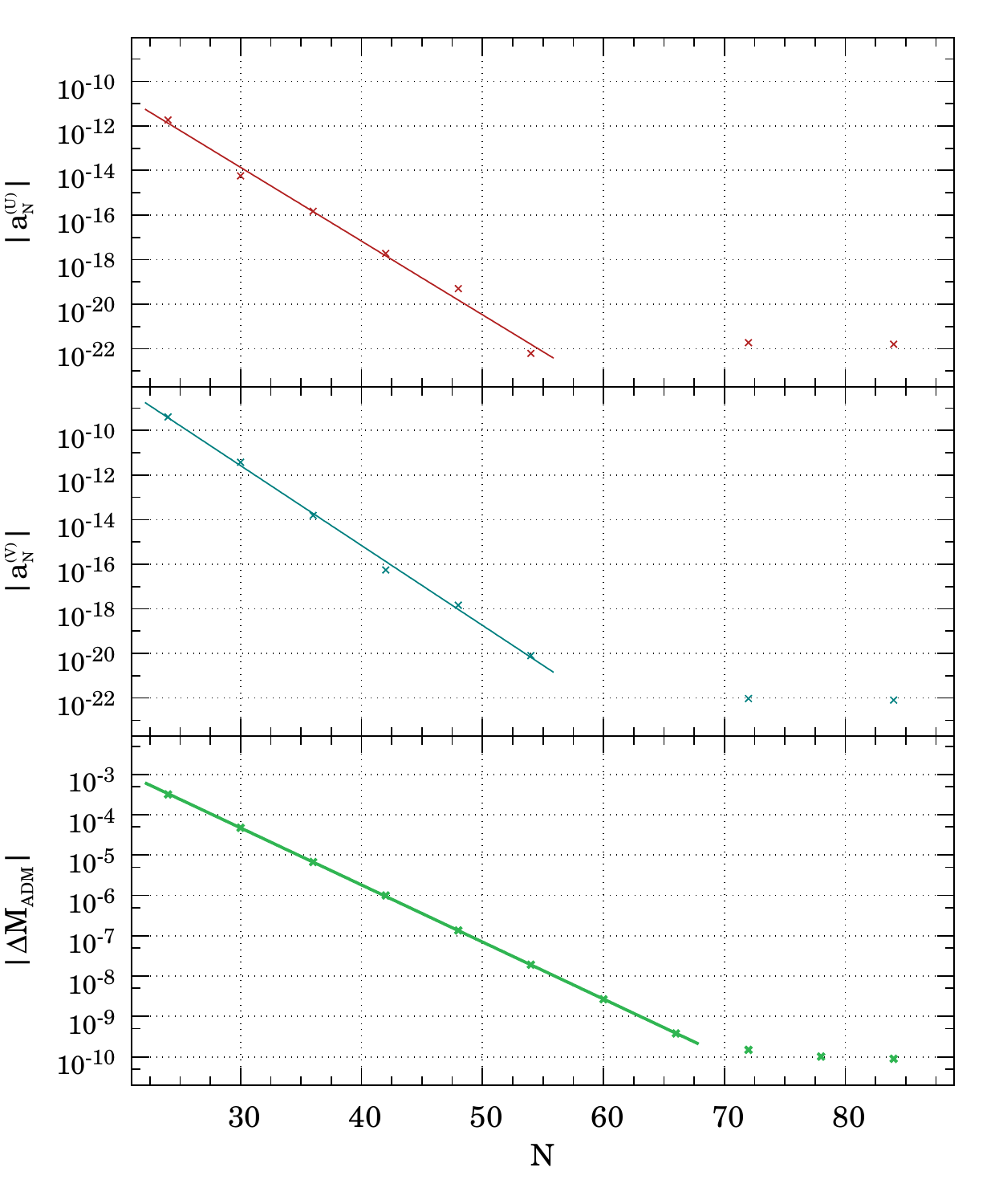}}}
\caption{The upper and middle plots show the truncation error for the variables $U$ and $V$, $|a^{(U)}_{N}|$ and $|a^{(V)}_{N}|$ respectively, as a function of the number of collocation points per domain, $N$. From all the truncation errors, one at each domain, we take the one where the variables reach their maximum values. The linear fitting with the logarithmic scale in the vertical axis shows the expected exponential convergence (see Sec.~\ref{convergence-cauchy}). The plot in the bottom shows that deviations in the ADM mass during the evolution also decrease exponentially with the number of collocation points per domain until saturation.}
\label{plot_conv_spectral}
\end{figure}

\subsection{Convergence Analysis for the Cauchy Evolution}\label{convergence-cauchy}
The Cauchy evolution uses a PSC discretization method for the radial direction with multiple domains.  At each domain we use a Chebyshev-Lobatto grid (with a linear mapping to the physical radial space; see Appendix~\ref{app_psc} for details).  The PSC method provides two representations for each variable, the spectral representation typical of general spectral methods, and the {\em physical} representation where the values of our variables at the collocation points are the unknowns to be found numerically. The truncation error, the difference between the true values of our variables and their numerical approximation, is given by the terms in spectral series that we neglect by truncating it.  We can estimate the truncation error by the absolute value of the last spectral coefficient, $|a_{N}|$ (see, e.g.~\cite{Boyd}).  For smooth functions, the convergence rate of the Chebyshev series is exponential~\cite{Boyd} (spectral convergence), i.e. the truncation error drops exponentially with the number of collocation points.  We check convergence for our Cauchy-evolution code by performing a series of runs with the same number of domains ($D=50$), uniformly distributed in the radial coordinate $x$, and with no refinement. We set the same initial conditions for all of them, from the family of configurations in Eq.~\eqref{cauchy-initial-data_Bizon}, and evolve it for a fix interval of time ($t_{f}\approx 2$, i.e. after a bounce off the AdS boundary).  Then, we look at the last spectral coefficients for the characteristic variables $U$ and $V$. Here we only show results from the domain where these variables present more features, which is in principle the most challenging one from the numerical point of view, and we have checked that we obtain equivalent results for the other domains.  In Fig.~\ref{plot_conv_spectral}, at the upper and middle panels, we show the spectral convergence for these two variables in a logarithmic plot of the absolute value of the last spectral coefficient versus the number of collocation points.  As we can see,  the linear scaling in the logarithmic plot stops at some point, followed by an almost flat profile, indicating that we have reached the round-off error of the computer and hence we cannot expect to improve the truncation error. In the bottom panel of Fig.~\ref{plot_conv_spectral}, we show the variations in the ADM mass, $M_{\rm ADM}$, with respect to its initial value, $M_{\rm ADM}(t_0)$, due to numerical inaccuracies during the Cauchy evolution (in an ideal situation this quantity should vanish for all times).  Actually, what we show in this figure is the normalized quantity:
\begin{equation}
\Delta{M_{\rm ADM}}(t) = \frac{| M_{\rm ADM}(t) - M_{\rm ADM}(t_0) |}{M_{\rm ADM}(t_0)}\,.
\end{equation}
In Fig.~\ref{plot_conv_spectral} we also see an exponential convergence of the deviations from the ADM mass ($\Delta{M_{\rm ADM}}(t_{f})$, with $t_{0}=0$) that saturate at a value around $10^{-10}$ for our particular test runs.

\begin{figure}[t]
\centerline{\resizebox{.5\textwidth}{!}{ \includegraphics{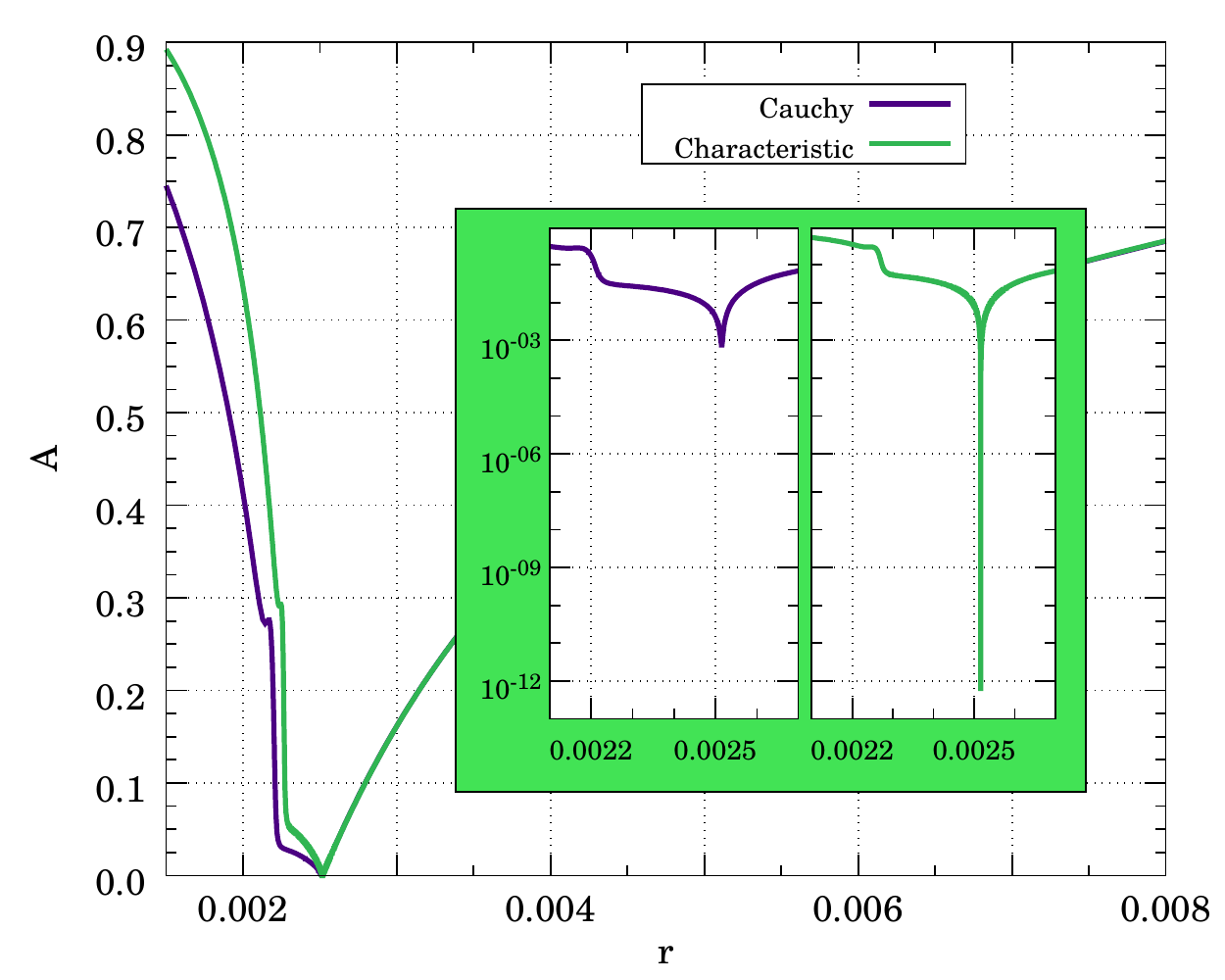}}}
\caption{Comparison of the Cauchy and characteristic evolution methods. We show a snapshot of the function $A$ just before collapse for both cases. The differences are due to the fact that in the first case $A$ is plotted from a $t=const.$ slice while in the second case comes from a $u=const.$ slice.  The plots coincide around $r = r_{\mathrm{AH}}$. The zoomed-in plot shows, by using a logarithmic scale, how close to AH formation ($A\rightarrow 0$) we can get with each evolution scheme.}
\label{plot_comparison}
\end{figure}


\subsection{Convergence Analysis for the Characteristic Evolution}\label{convergence-characteristic}

In the characteristic scheme, we have a non-uniform discretization in the radial coordinate $r$ in the initial grid, and it turns out that the evolution of the $r$-coordinate of the grid points [according to the ingoing null geodesic Eq.~\eqref{inullgeo}] makes our grid even more unequally spaced. Despite of this, the resolution increases with the number of grid points and we can study how the results converge as we increase this number. To that end, we run simulations with different initial number of grid points (the number of grid points changes along the evolution because we lose points through the origin) but with the same initial scalar field profile [see Eq.~\eqref{initial-data-characteristic}].  These initial conditions form an AH and the point of the evolution that we take to analyze the convergence is just before the formation of the AH, when $A = 10^{-8}$ [$A$ is estimated via Eq.~\eqref{Agbarg}].  That is, we monitor how the location of AH formation changes with the number of grid points, $N$.  We use the following indicator:
\begin{equation}
p = \log_2 \left( \frac{ | r_{\mathrm{AH}}^{N/4}  - r_{\mathrm{AH}}^{N/2} |  }{| r_{\mathrm{AH}}^{N/2}  - r_{\mathrm{AH}}^{N} |}\right)\,. 
\end{equation}
For $N = 120000$ we obtain $p \approx 3.0034$. This value means that the convergence of our code is third-order, in agreement with the convergence rate of the Simpson integration rule that we use.


\subsection{Comparison between the Cauchy and characteristic evolutions}

The main reason for implementing a hybrid Cauchy-characteristic evolution scheme is to bring together the best of these two methods of evolution in order to tackle interesting questions about gravitational collapse in AAdS spacetimes, taking into account that the Cauchy evolution based on the PSC method allows us to follow the possible different bounces of the matter fields (a scalar field in our case) off the AdS boundary with high precision, whereas the characteristic evolution allows us to get very close to the point of formation of an AH.  Then, although the two evolution schemes are used in different stages of the evolution, it is interesting to see how they compare when they are applied to the final moments of the collapse, when an AH forms.  This comparison is also a justification for the introduction of our hybrid scheme, which on top of the two evolution methods requires a non-trivial transition between them.  Then, we have evolved the same scalar field configurations with both evolution schemes to as close as possible to the point of AH formation, which can be monitored with the metric function $A$, which in the characteristic scheme can be computed using Eq.~\eqref{Agbarg}.  We show the results of this comparison in Fig.~\ref{plot_comparison}, where we include a zoom to the relevant region for AH formation. In the left zoomed plot, we show the metric function $A$ for a simulation with the Cauchy evolution scheme until the numerical code is not stable anymore without adding more domains, and such that if we keep adding resolution the evolution would essentially freeze because of the tiny time step allowed by the CFL condition. For the right zoomed plot, we initiated the evolution also with the Cauchy evolution scheme (in order to guarantee that we are comparing the same physical configuration) and then changed to the characteristic scheme until the point where the numerical noise becomes significant or the evolution effectively stops due a too small $\Delta u$ step.  As we can see, with the characteristic evolution we can get the function $A$ many orders of magnitude closer to the collapse than with the Cauchy scheme.  This clearly illustrates the power of our hybrid scheme to study the collapse near AH formation.

\begin{figure}[t]
\resizebox{.5\textwidth}{!}{ \includegraphics{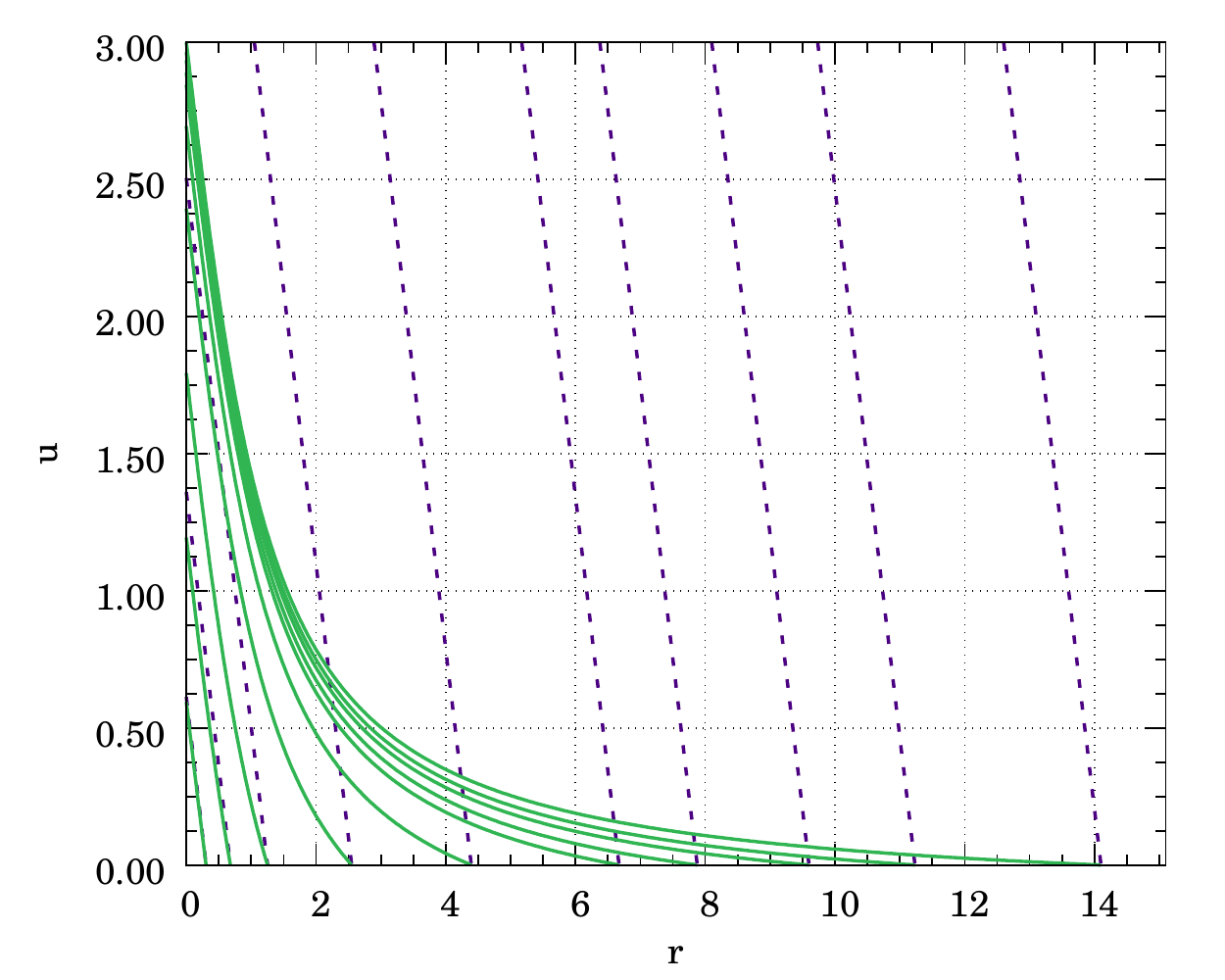}}  
\caption{Comparison of ingoing null geodesics in Mink (dashed purple lines) and AdS (green continuous lines) spacetimes.  We see the strong effect that the cosmological constant term has in the geodesics. In AdS spacetime they reach the region near the origin much faster than in Mink, as measured by the time $u$.}
\label{plot_geodesics_flat_ads}
\end{figure}


\subsection{Ingoing Null Geodesics}

In order to understand better the magnitude of the numerical challenge posed by the study of gravitational collapse in AAdS spacetimes it is interesting to analyze the ingoing null geodesics in the characteristic evolution of AAdS spacetimes and compare them with the ingoing null geodesic in asymptotically-flat spacetimes. To begin with, let us look at the difference between the ingoing null geodesics in AdS spacetime, Eq.~\eqref{charmet}, and in Mink spacetime, in the equivalent coordinate system where the metric has the same form as in Eq.~\eqref{charmet}.  The equation for the ingoing null geodesics has also the same form in both cases, i.e. Eq.~\eqref{inullgeo}, but the form of the metric function $\bar g$ is different. In AdS spacetime we have $\bar g^{}_{\mathrm{AdS}}(r) = 1 + r^2/\ell^{2}$ whereas is Mink we have $\bar g^{}_{\mathrm{Mink}}(r) = 1$. Therefore, by solving the ingoing null geodesic equation, Eq.~\eqref{inullgeo}, we get the following expressions for the ingoing null geodesics:
\begin{equation}
\begin{aligned}
 u^{}_{\mathrm{Mink}} (r) &=   2 \left( r_{0} - r\right)\,,\\
 u^{}_{\mathrm{AdS}} (r) &=  2 \left( \arctan(r_{0}) - \arctan(r)\right)\,.
\end{aligned}
\end{equation}
These geodesics have been plotted in Fig.~\ref{plot_geodesics_flat_ads}.  This illustrates what can happen with our characteristic grid in AAdS evolutions in comparison with the asymptotically-flat case.  As shown in Fig.~\ref{plot_geodesics_flat_ads}, the grid points of an initial null slide ($u=0$) move towards the origin much faster in AdS spacetime than in Mink spacetime.  The conclusion of this for our simulations is that we must be very careful in choosing the initial null slide, in particular its size, because the grid points in AdS travel very fast towards the
region near the origin, which means that our grid shrinks very fast and we may miss the interesting phenomena, in particular the formation of an AH.

\begin{figure}[t]
\resizebox{.5\textwidth}{!}{ \includegraphics{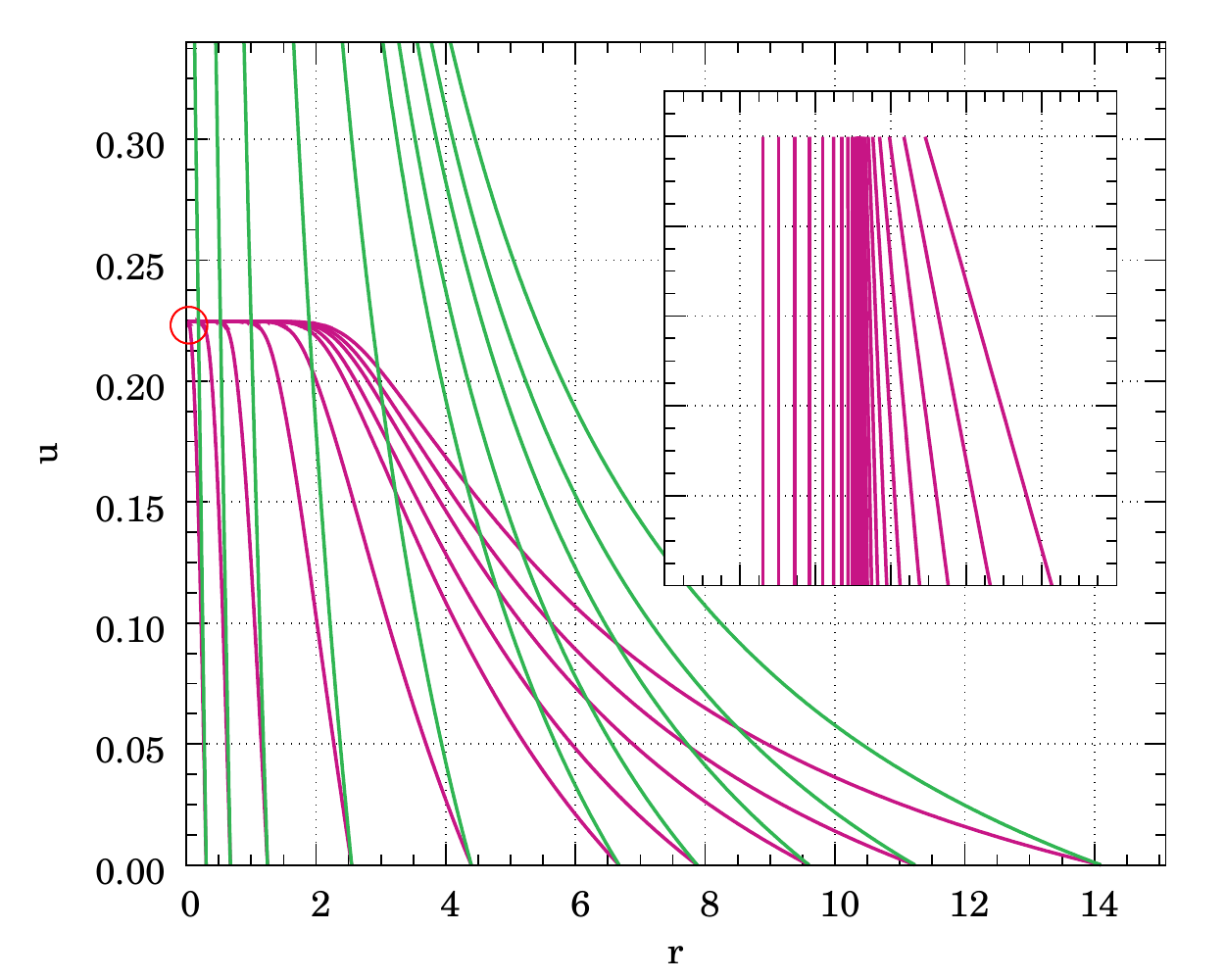}} 
\caption{Comparison between ingoing null geodesics in pure AdS spacetime (green color) and in an AAdS spacetime describing gravitational collapse of a scalar field (fuchsia color). In the second, the geodesics focus around the forming AH. The small plot is a zoom-in of the region inside the red circle where it can be seen how the geodesics behave around $r_{AH}$. This size of the zoom-in plot is $\Delta r \simeq 10^{-4}$ around $r \simeq 0.03952$ and $\Delta u \simeq 10^{-13}$ around $u \simeq 0.224673828858275$.}
\label{plot_geodesics_ads_bh}
\end{figure}

On the other hand, in Fig.~\ref{plot_geodesics_ads_bh} we show the comparison between ingoing null geodesics in AdS spacetime and the ones of an AAdS spacetimes where the scalar field collapses forming an AH. We compute these geodesics numerically as the solution of Eq.~\eqref{inullgeo}. As soon as the geodesics approach to the spacetime point in the $(u,r)$-plane where the AH forms, all the ingoing null geodesics with $r>r_{AH}$ focus at that point as it can be seen in the zoom-in area of this figure, while those with $r<r_{AH}$ follow a different pattern.

\section{Results from the Numerical Evolution}\label{results}

In this section we present results of our evolutions of the EKG system, Eqs.~\eqref{efes} and~\eqref{kg_eq}, in spherically-symmetric AAdS spacetimes.  The families of initial configurations that we use for our evolutions are shown below: Eqs.~\eqref{cauchy-initial-data_Bizon} and~\eqref{cauchy-initial-data-boundary} are initial data for Cauchy-only and Cauchy-characteristic evolutions, whereas Eq.~\eqref{initial-data-characteristic} shows initial data used for characteristic-only evolutions.   

\begin{figure}[t]
\centerline{\resizebox{.5\textwidth}{!}{ \includegraphics{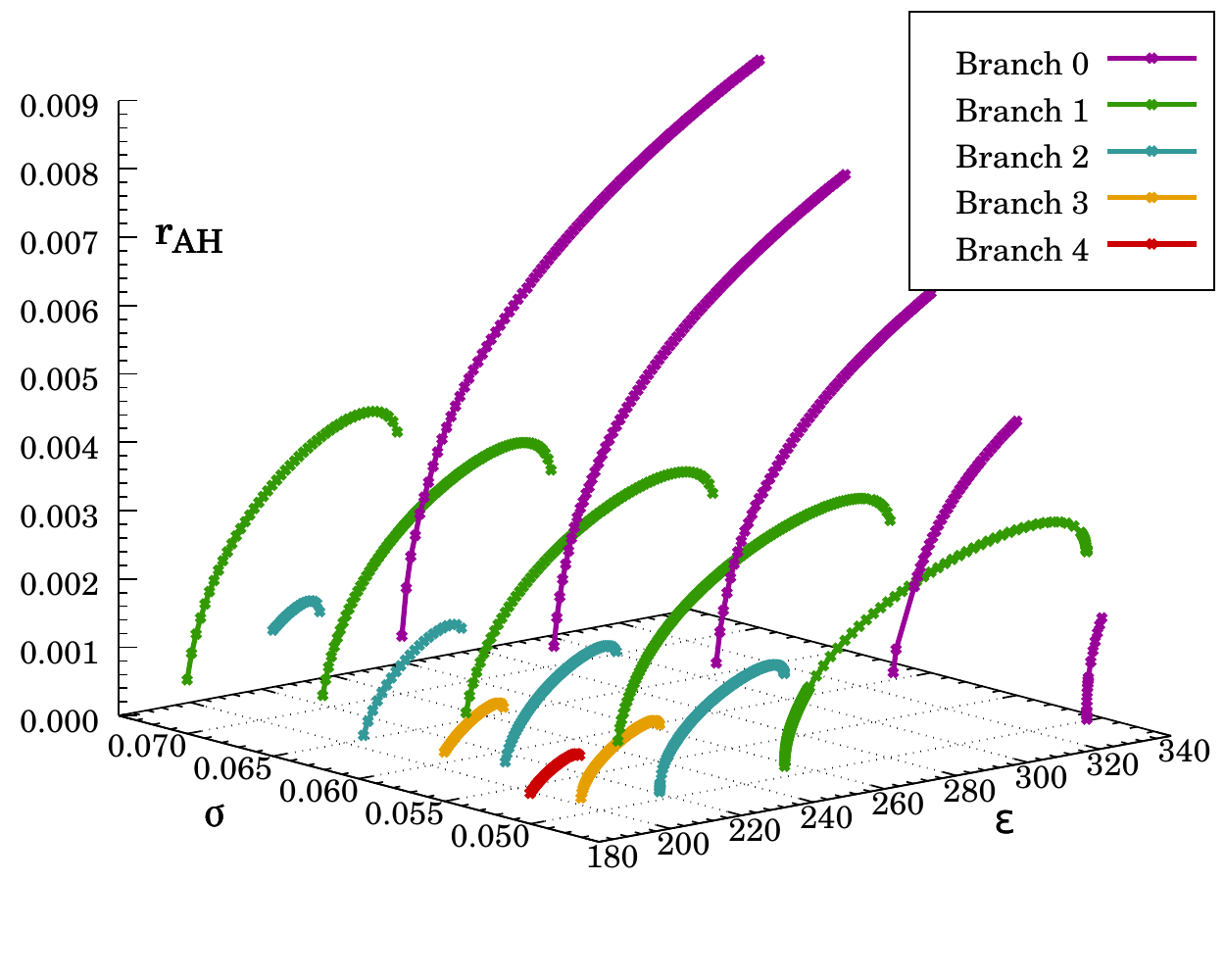}}}
\caption{AH location in a 2-parameter phase space. We show the radial location of the AH formed for initial configurations from the family in Eq.~\eqref{cauchy-initial-data_Bizon} for different values of the amplitude, $\varepsilon$, and the width, $\sigma$.}
\label{plot_3d}
\end{figure}

The landscape of the gravitational collapse that emerged after the pioneer work of Ref.~\cite{Bizon:2011gg} can be summarized by saying that initially-compact scalar field configurations will sooner or later form an AH, depending on how many bounces off the AdS boundaries are required for the AdS {\em turbulent instability} to convert long-wavelength modes into short-wavelength ones so that the scalar field profile gets compressed enough to form a BH.  This is illustrated in Fig.~\ref{plot_3d} where show the AH radius, $r_{AH}$, obtained by evolving a number of initial configurations from the family of Eq.~\eqref{cauchy-initial-data_Bizon} with our Cauchy-characteristic evolution scheme. This three-dimensional plot has been obtained by varying both the amplitude, $\varepsilon$, and the width, $\sigma$, of the initial configurations.  It shows the different branches that appear and that represent configurations that have bounced off the AdS boundary a fixed number of times (indicated by the color and branch number in Fig.~\ref{plot_3d})  before collapsing and forming an AH, in contrast with the asymptotically-flat case where we have a single branch.  The branches are clearly seen in the direction of the amplitude $\varepsilon$, where we have a high number of points, but it can be seen that it also happens in the direction of the width $\sigma$.  The same should happen if we look at any direction in the plane $(\varepsilon,\sigma)$.

In what follows we describe new results regarding the critical collapse, that is, analyzing the configurations in Fig.~\ref{plot_3d} near the plane $r_{AH} = 0$, and we also describe results about the mass gap between branches and the power-law scaling found in our recent study~\cite{SantosOlivan:2015fmy} for the AH mass of the near subcritical configurations.  These results consolidate further the conclusion reached in~\cite{SantosOlivan:2015fmy}.

\subsection{Critical Phenomena in AAdS Gravitational Collapse \label{subs_results_origin_cent_IC}}

In~\cite{Bizon:2011gg} it was concluded that at the critical points separating the branches, the supercritical configurations form an AH with mass going to zero with the same scaling as in the case of asymptotically-flat spacetimes~\cite{Choptuik:1992jv}, that is: $r_{\mathrm{AH}} \sim (p - p_n)^\gamma$ where $p$ is an arbitrary parameter of the family of initial configurations, $p_n$ is the n-th critical value at which the AH mass goes to zero, and $\gamma$ is an universal exponent that in $d=3$ has the expected value: $\gamma \simeq 0.374$.  This conclusion was confirmed in~\cite{SantosOlivan:2015fmy} for the branch $0$ in Fig.~\ref{plot_3d} by fitting a large number of simulations near the critical point.  In the present work we extend this result for the next five branches (up to branch $5$) using the same family of initial data (similar to the one used in~\cite{Bizon:2011gg}):
\begin{equation}
\begin{aligned}
U(t^{}_{o},x) &=  \varepsilon \exp{\left(-\frac{4\tan^{2}x}{\pi^{2}\sigma^{2}} \right)}\,, \\
V(t^{}_{o},x) &=  - U(t^{}_{o},x)\,.
\label{cauchy-initial-data_Bizon}
\end{aligned}
\end{equation}
that represents a profile centered around the origin at the initial time, and characterized by the amplitude $\varepsilon$ and width $\sigma$. This particular choice, and anyone that fulfils the condition $V(t^{}_{o},x) =  - U(t^{}_{o},x)$, directly implies: $(\partial^{}_{x}\phi)(t^{}_{o},x) = 0\,$.

To obtain the scaling of the AH mass near the critical points of the different branches we have used the method introduced in~\cite{Garfinkle:1998va}, which consists in following the evolution of subcritical configurations very near the critical point and tracking the behaviour of the curvature scalar $R$ at the origin ($x=0$).  The subcritical character of these evolutions allows us to make the computation accurately using only the Cauchy evolution.
The curvature scalar $R$ at $x=0$, as well as other curvature scalars, starts to grow when the scalar field attempts to form an AH, reaching a maximum value (in absolute value), and then disperses towards the AdS boundary. This maximum will be higher the closer we approach the critical point $p_n$, being infinite at that point. Actually, it follows an scaling law of the kind~\cite{Garfinkle:1998va}:
\begin{equation}
\left. R_{\mathrm{max}} \right |_{x=0}  \sim (p_n - p)^{-2 \gamma}\,.
\label{rmax-scaling}
\end{equation}
The scalar field near the critical point exhibits discrete self-similarity (type II critical behaviour) with an echoing period $\Delta$, which for the same reasons than $\gamma$ should be the same as in the asymptotically-flat case: $\Delta\approx 3.44$. In ~\cite{Gundlach:1996eg,Hod:1996az} it was shown that on top of the scaling of Eq.~\eqref{rmax-scaling}, a finer structure can be seen as oscillations of the following form:
\begin{equation}
\begin{aligned}
\ln \left. R_{\mathrm{max}} \right|^{}_{x=0} &= \\
 = (-2 \gamma)& \, \ln(p_n-p) + b_0 + F(\ln(p_n - p))\,,
\end{aligned}
\end{equation}
where $F$ is a periodic function with a period depending on the universal values of the exponent $\gamma$ and the echoing period $\Delta$ and equal to $\Delta/2\gamma$.  In terms of the Cauchy-evolution variables, the scalar of curvature at the origin can be computed as:
\begin{equation}
\left. R \right |_{x=0} = - \frac{12}{\ell^2} - \frac{1}{2 \ell^2} \left (V - U\right)^2\,. 
\end{equation}
%

\renewcommand{\arraystretch}{1.5}
\renewcommand{\tabcolsep}{8pt}
\begin{table*}
\begin{ruledtabular}
\caption{Critical Exponents for fixed width ($\sigma = 0.05$). Fitting values of the critical exponents obtained in Sec.~\ref{subs_results_origin_cent_IC} corresponding to the critical parameters for the six first branches as it is shown in Fig.~\ref{plot_scaling_b05}. }
\begin{tabular}{l l l l}
Branch &  Critical Value~($\varepsilon_n$)   &  Critical Exponent~($\gamma$)     & Echoing Period~($\Delta$)  \\
\hline
$n=0$ & $ 335.572231    \pm 0.000005$	&	$0.374  \pm 0.006$	&	$3.33 \pm 0.15$	\\
$n=1$ & $ 251.09427729  \pm 0.00000001$	&	$0.3746 \pm 0.0008$	&	$3.45 \pm 0.02$	\\
$n=2$ & $ 216.208077165 \pm 0.000000001$	&	$0.3743 \pm 0.0004$	&	$3.45 \pm 0.02$	\\
$n=3$ & $ 193.9755275   \pm 0.0000001$	&	$0.377  \pm 0.007$	&	$3.43 \pm 0.04$	\\
$n=4$ & $ 178.070915    \pm 0.000001$	&	$0.376  \pm 0.007$	&	$3.42 \pm 0.06$	\\
$n=5$ & $ 165.946674    \pm 0.000004$	&	$0.377  \pm 0.010$	&	$3.46 \pm 0.10$	
\end{tabular}
\label{table_fittings}
\end{ruledtabular}
\end{table*}

\begin{figure}[t]
\centerline{\resizebox{.5\textwidth}{!}{ \includegraphics{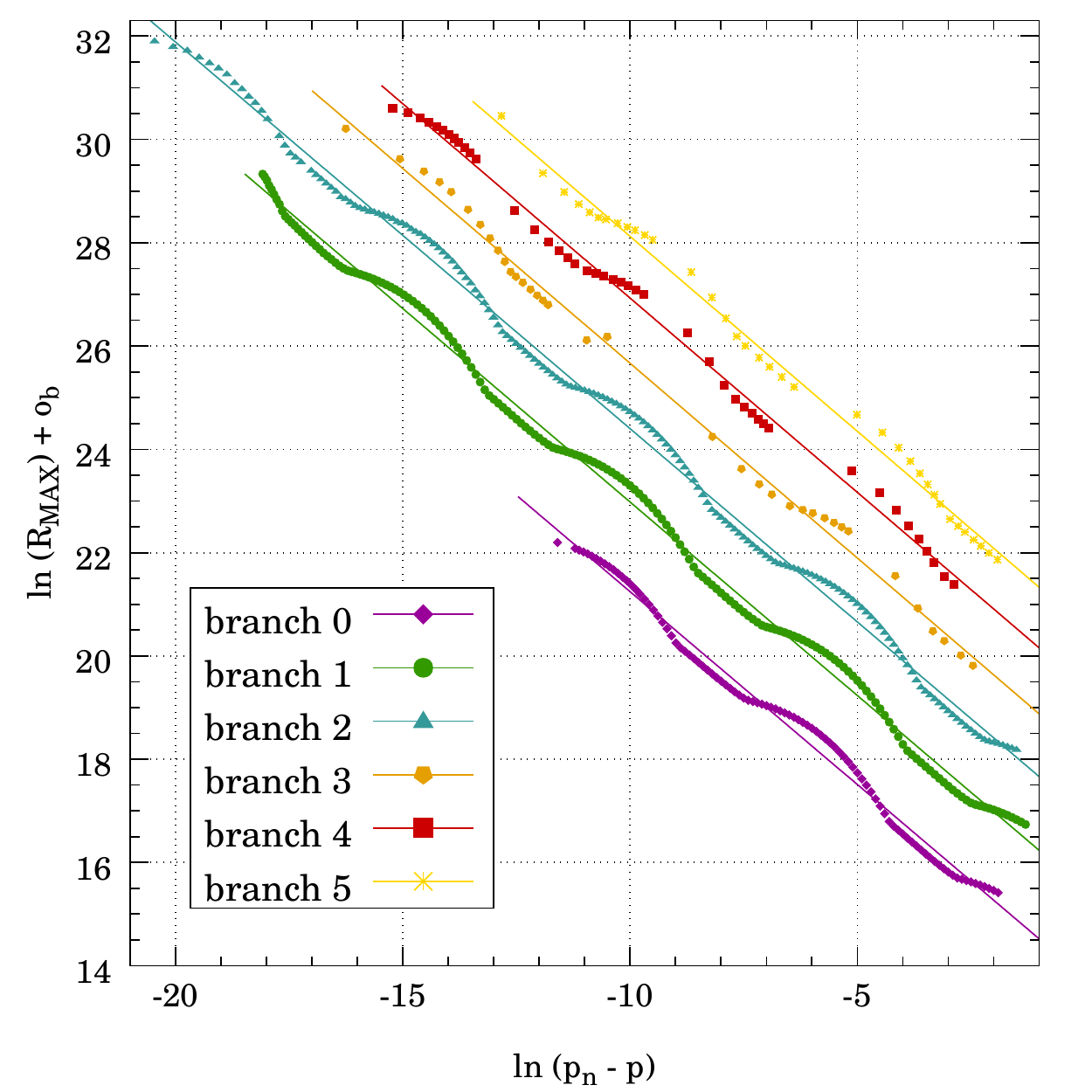}}}
\caption{Critical Exponents for fixed width ($\sigma = 0.05$). Scaling of the scalar of curvature for subcritical configuration near the critical point for the different branches from the branch of direct collapse ($b=0$, bottom) to the branch with five bounces ($b=5$, top). An offset $o_b$ have been added to the $y$ axis to make the plot more clear. The values of the offset are:  $o_b = -2, -1, 0, +1, +2, +3$, starting from $b=0$. The results from the fittings are given in Table~\ref{table_fittings}.}
\label{plot_scaling_b05}
\end{figure}

We have carried out a series of Cauchy evolutions of initial configurations of the family in Eq.~\eqref{cauchy-initial-data_Bizon} with fixed width $\sigma = 0.05$ and amplitudes chosen to be subcritical for the first six critical points (see Fig.~\ref{plot_3d}).  The results of these simulations are plotted in Fig.~\ref{plot_scaling_b05} with the corresponding fittings.  The values of the critical amplitudes, $\varepsilon_{n}$, the critical exponents, $\gamma$, and the echoing periods, $\Delta$, are presented in Table~\ref{table_fittings}.  We can see that the values obtained for $\gamma$ and $\Delta$ are consistent with the known values for the collapse of massless scalar fields in asymptotically-flat spacetimes~\cite{Choptuik:1992jv,Gundlach:1996eg}. This was already shown for the first branch in~\cite{Husain:2002nk} and also checked in~\cite{SantosOlivan:2015fmy}.

\begin{table*}
\begin{ruledtabular}
\caption{Critical Exponents for branch zero in the characteristic method. Fitting values of the critical exponents obtained in Sec.~\ref{subs_results_origin_cent_IC} corresponding to the critical parameters for the zero branch using the characteristic method as it is shown in Fig.~\ref{plot_charac_ce0}}
\begin{tabular}{l l l l}
Width $(\sigma)$ &  Critical Value~($\varepsilon_0$)   &  Critical Exponent~($\gamma$)     & Echoing Period~($\Delta$)  \\
\hline
$0.01$ & $ 7.828039    \pm 0.000002$    	&	$0.376  \pm 0.006$	&	$3.2 \pm 0.4$	\\
$0.05$ & $ 25.907772996  \pm 0.000000003$	&	$0.3748 \pm 0.0004$	&	$3.33 \pm 0.10$	\\
$0.10$ & $ 23.8595911 \pm 0.0000001$	    &   $0.375 \pm 0.005$	&	$3.45 \pm 0.10$	
\end{tabular}
\label{table_fittings2}
\end{ruledtabular}
\end{table*}

\begin{figure}[t]
\centerline{\resizebox{.5\textwidth}{!}{ \includegraphics{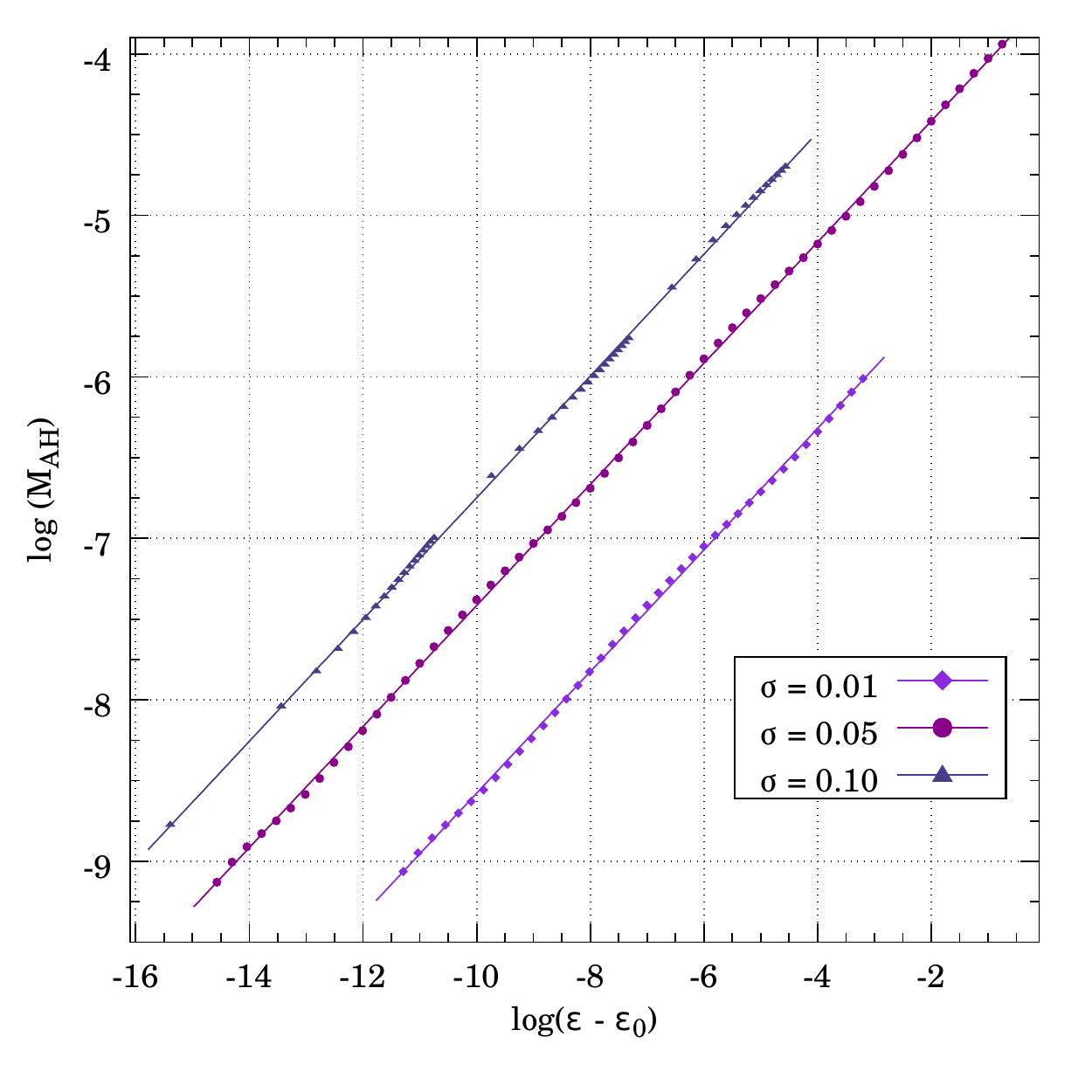}}}
\caption{Critical Exponents for branch zero from the characteristic method. We show the results for the AH mass versus initial amplitude for three different families of initial configurations with fixed width [see Eq.~\eqref{initial-data-characteristic}] of the branch $0$ (direct collapse). The values of the critical amplitudes, critical exponent and echoing period are given in Table~\ref{table_fittings2}.}
\label{plot_charac_ce0}
\end{figure}

We have already mentioned that the characteristic method of Sec.~\ref{analytics_characteristic} cannot be used to follow the full evolution of the scalar field because the characteristic grid shrinks with time and hence we cannot track bounces of the scalar field off the AdS boundary.  However, we can in principle use the characteristic evolution for the particular cases in which the scalar field collapses directly, or in other words, we can in principle study the branch 0 with the characteristic evolution.  Actually, this was already done in~\cite{Husain:2002nk} using a double-null characteristic scheme.  Here we repeat this analysis as a way of confirming this result and, at the same time, as a way of testing further our characteristic evolution. To that end, we have to prescribe initial data on a null slice $u=const.$ for the scalar field variable $h$. We choose the following family of initial conditions:
\begin{equation}
\phi(r) = \bar h(r) = \varepsilon\, \frac{r^2}{\ell^{2}} \exp \left(  - \frac{(r-r_0)^2}{\ell^{2}\,\sigma^2} \right)\,. 
\label{initial-data-characteristic}
\end{equation}
We have performed a series of characteristic evolutions varying the amplitude $\varepsilon$ for three fixed values of the width $\sigma = 0.01, 0.05$ and $0.10$. We compute the critical exponent $\gamma$ and echoing period $\Delta$ from the AH mass, in AAdS spacetimes $M_{\mathrm{AH}} = r_{\mathrm{AH}}\, \left( 1 + r^2_{\mathrm{AH}}/\ell^2 \right) /2$ (the values of this mass that we quote in this paper are in units of $\ell$); that is, we fit our characteristic-only simulations to the formula:
\begin{equation}
\begin{aligned}
\ln M_{\mathrm{AH}} = \gamma \, \ln(p-p_n) + b_0 + F(\ln(p - p_n))\,,
\end{aligned}
\label{characteristic-fitting}
\end{equation}
where $F$ is again a periodic function with period $\Delta/2\gamma$~\cite{Gundlach:1996eg,Garfinkle:1998va}.  The results obtained from these simulations are plotted in Fig.~\ref{plot_charac_ce0} with the fittings to Eq.~\eqref{characteristic-fitting}. The critical values of the amplitude, $\varepsilon_{0}$, the critical exponent $\gamma$, and the echoing period $\Delta$ are given in Table~\ref{table_fittings}.  Again, the results are consistent with the predictions for the asymptotically-flat case.


\subsection{Power-law Behavior near the Mass Gaps}

The second application of our Cauchy-characteristic evolution scheme is to study the mass gap between the branches of collapsed scalar field configurations [see Fig.~\ref{plot_3d}].  In a previous work~\cite{SantosOlivan:2015fmy}, using already this evolution scheme, we found that the subcritical solutions that are very close to the critical points form BHs with an AH whose mass obeys an scaling law of the form:
\begin{equation}
M_{\mathrm{AH}}-M^{n+1}_{g} \propto (p_{n}-p)^{\xi}\,,
\label{massgap_scaling}
\end{equation}
where $p_n$ denotes the critical value of the initial-data parameter $p$ for the $n$-th branch; $M^{n+1}_g$ is the mass of the $(n+1)$-th gap, between the branches $n$ and $n+1$, corresponding to the minimum mass of the AH formed by subcritical configurations; and $\xi$ is the power-law exponent. In~\cite{SantosOlivan:2015fmy} we found that the exponent $\xi$ has a value $\xi \simeq 0.70$ and it was conjectured that this value is universal, the same for all families of initial configurations and for all branches/critical points.  The numerical support for this conjecture given in~\cite{SantosOlivan:2015fmy} came from the evolution of two different one-parameter families of initial configurations and for the first two mass gaps, the one between branches 0 and 1, and the one between branches 1 and 2.  The two one-parameter families of initial configurations both came from the same larger family of initial conditions given in Eq.~\eqref{cauchy-initial-data_Bizon}, one by fixing the width $\sigma$ and the other one by fixing the amplitude $\varepsilon$.

In this work we give new evidence for this conjecture.  First by considering a completely new different family of initial configurations, in the sense that it is functionally different, and second, by extending the study to the third mass gap, between the branches 2 and 3. 

\begin{figure}[t]
\centerline{\resizebox{.5\textwidth}{!}{ \includegraphics{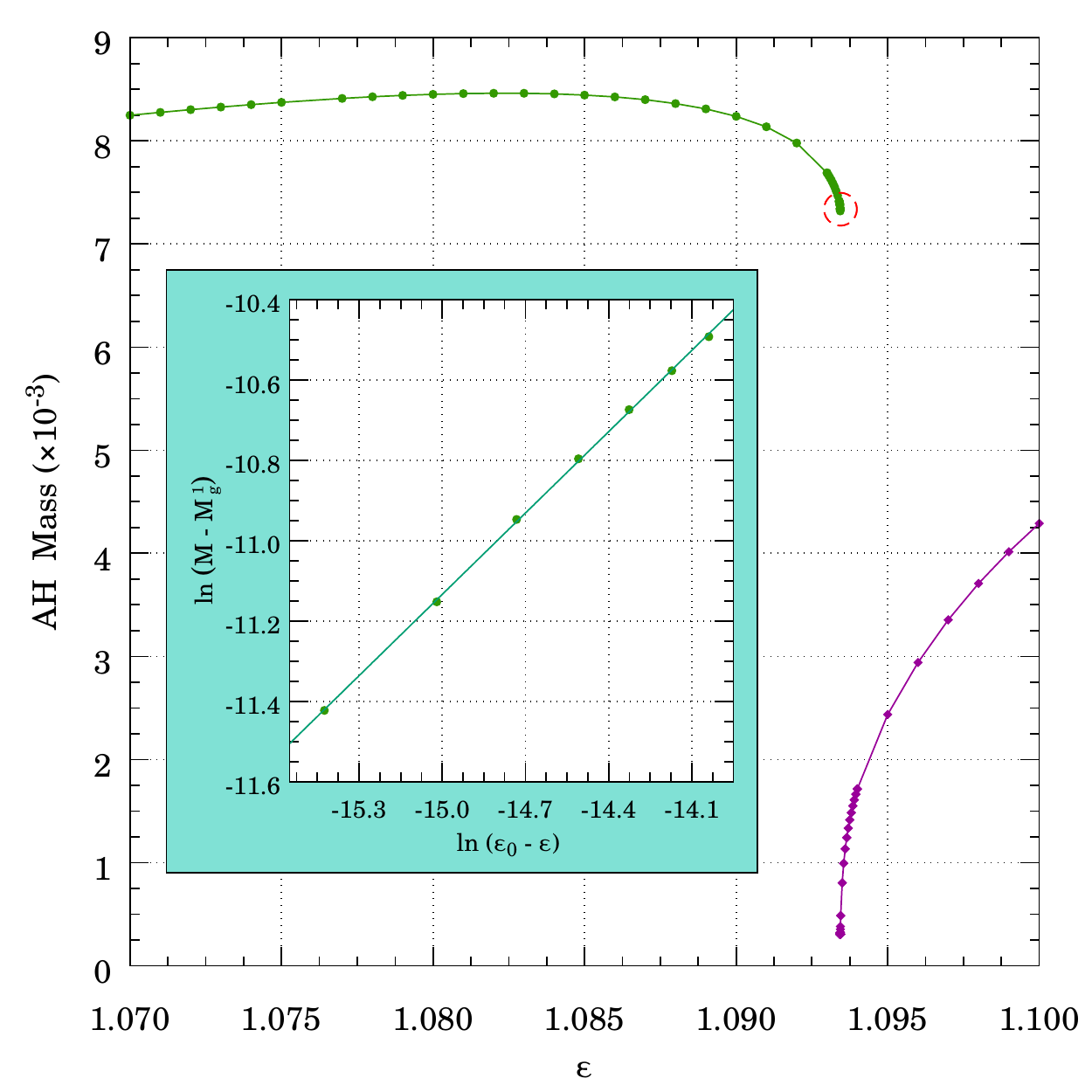}}}
\caption{Mass gap 1, between the branches $0$ and $1$. The behavior near the subcritical solutions marked by the red circle follows the power law in Eq.~\eqref{massgap_scaling}. The zoom-in plot shows the fitting of the data.  The critical point is located at an amplitude $\varepsilon_0 = 1.093435 \pm 0.000001$.  The values for the mass gap, $M^{1}_g$, and power law exponent, $\xi$, are given in Eqs.~\eqref{massgap-boundary-data} and~\eqref{xi-value-different-id} respectively.}
\label{plot_massgap}
\end{figure}

The new family of initial conditions that we consider is inspired by a proposal in~\cite{Abajo-Arrastia:2014fma,daSilva:2014zva} of setting the initial profile far from the origin and centered close to the AdS boundary. The form for the initial values of our Cauchy variables is given by
\begin{equation}
\begin{aligned}
U(t^{}_{o},x) &= \varepsilon \cosh^{-1} \left(  \frac{\tan(x) - \tan(x_0)}{\sigma} \right)  \,,\\
V(t^{}_{o},x) &= -U(t^{}_{o},x)\,.
\label{cauchy-initial-data-boundary}
\end{aligned}
\end{equation}
This family has also two parameters, an amplitude $\varepsilon$ and a width $\sigma$. These configurations are evolved using the Cauchy-characteristic evolution scheme.  It turns out that the simulations for the configurations within the parameter region of interest, those that lead to subcritical scalar field collapse, are numerically more challenging than in the case of the initial conditions from the family in Eq.~\eqref{cauchy-initial-data_Bizon}.  The reason for this is that the energy distribution in the new family of configurations is not as compact as in the old one.  This has already been illustrated in the evolutions tracking the {\em center of mass} of the scalar field profile shown in Fig.~\ref{plot_cdm}.   For our simulations we set $x_0 = 1.2$ and $\sigma = 0.2$. With this choice the first critical point is found at $\varepsilon = \varepsilon_0 = 1.093435 \pm 0.000001$. In Fig.~\ref{plot_massgap} we show, in the plane $\varepsilon - M_{\rm AH}$, the region near this first critical point.   The red circle in the figure indicates the area from where we have taken the data to fit the power law of Eq.~\eqref{massgap_scaling}, which is shown in the zoom-in plot of the same figure. In this case we find that the mass gap is:
\begin{equation}
M^{1}_g = (7.2954 \pm 0.0008)\times 10^{-3}\,,
\label{massgap-boundary-data}
\end{equation}
and the power-law exponent has a value consistent with the conjectured universal character: 
\begin{equation}
\xi = 0.68 \pm 0.07\,.
\label{xi-value-different-id}
\end{equation}
%

\begin{figure}[t]
\centerline{\resizebox{.5\textwidth}{!}{ \includegraphics{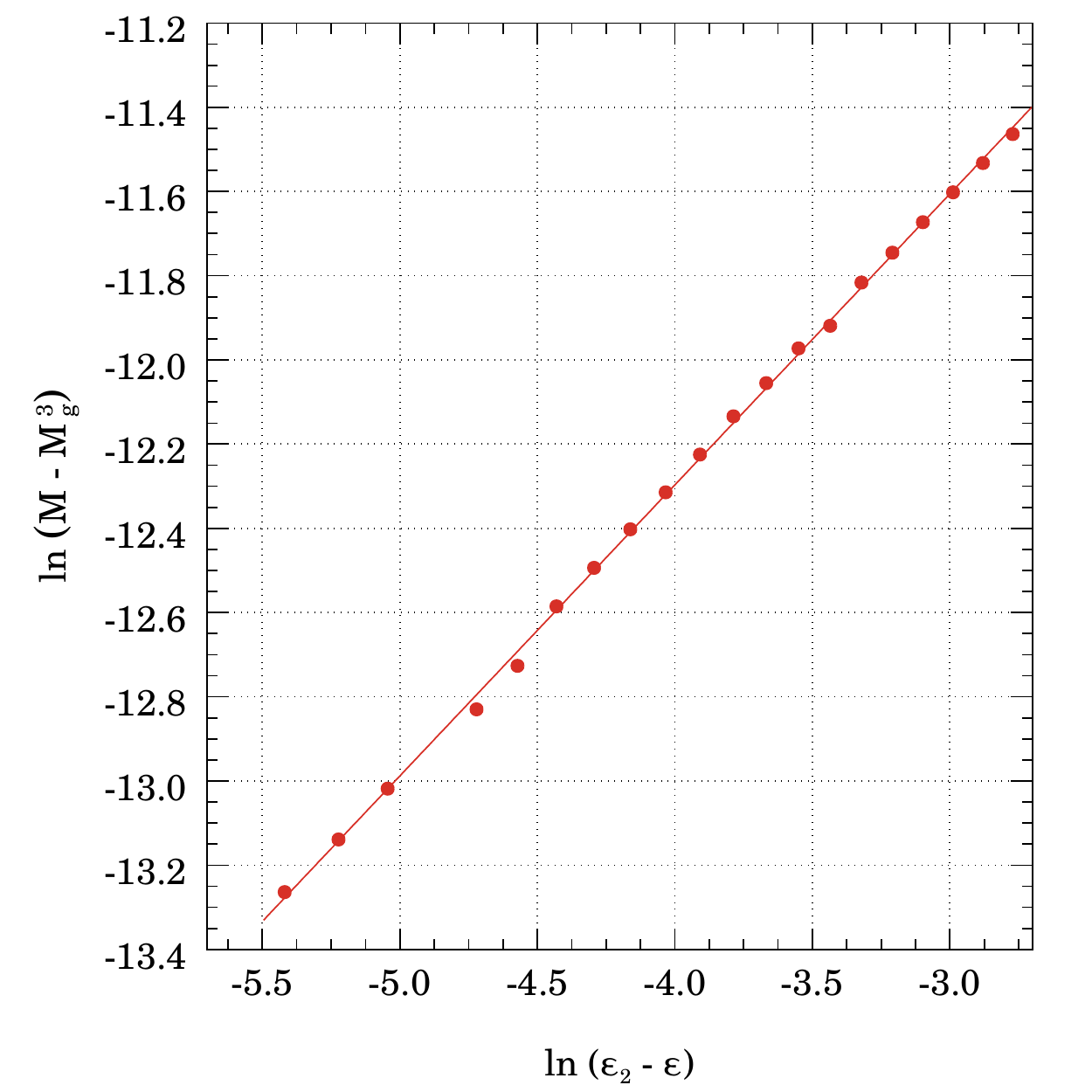}}}
\caption{Subcritical solutions associated with the third mass gap (between branches 2 and 3).  The fit of the data to the power law in Eq.~\eqref{massgap_scaling} is shown.  The critical point is found to be located at an amplitude $\varepsilon = \varepsilon_2 = 216.203 \pm 0.009$ and values for the mass gap, $M^{3}_g$, and power law exponent, $\xi$, are given in Eqs.~\eqref{mass-gap-number-3} and~\eqref{xi-mass-gap-number-3} respectively.}
\label{plot_massgap_g3}
\end{figure}

On the other hand, we have also studied the power law of Eq.~\eqref{massgap_scaling} for additional mass gaps, beyond the gaps between branches 0-1 and 1-2, already studied in~\cite{SantosOlivan:2015fmy}.  This is a particularly challenging goal since it involves a number of simulations where we have to track the scalar field through two bounces off the AdS boundary.  Then, the problem is challenging from the point of view of accuracy because the subcritical configurations are quite challenging in terms of resolution requirements since we have to evolve the sharp features originated during the quasi-collapse stage to the AdS boundary and back.  It is also challenging from the point of view of computational cost since each of these simulations takes significantly much more time than the previous ones. We need to perform many of them in order to locate the critical point and to have enough subcritical configurations close to it in order to extract the values of the mass gap and power-law exponent with a good precision.  Again, considering the conjecture established in~\cite{SantosOlivan:2015fmy} we expect to find the same power law [Eq.~\eqref{massgap_scaling}] around all the mass gaps. We have analyzed the situation around the third mass gap, between the branches 2 and 3, using the initial configurations  in Eq.~\eqref{cauchy-initial-data_Bizon} fixing the value of the width to $\sigma = 0.05$. The critical point associated with the second branch is located at a value of $\varepsilon = \varepsilon_2 = 216.203 \pm 0.009$.  Then, from the evolution of subcritical configurations associated with this critical point we obtained the $\varepsilon - M_{\rm AH}$ plot shown in Fig.~\ref{plot_massgap_g3}.  The fit to this data gives us the value of the mass gap:
\begin{equation}
M^{3}_g = (5.44 \pm 0.02) \times 10^{-4}\,,
\label{mass-gap-number-3}
\end{equation}
and the power-law exponent of Eq.~\eqref{massgap_scaling} is found to be:
\begin{equation}
\xi = 0.69 \pm 0.04\,.
\label{xi-mass-gap-number-3}
\end{equation}
This value is also consistent with the values found in~\cite{SantosOlivan:2015fmy}, for the first two mass gaps, and with the value found here [see Eq.~\ref{xi-value-different-id}], for the first mass gap, using a functionally different family of initial data, namely the one given by Eq.~\eqref{cauchy-initial-data-boundary}.  Therefore, these results give more numerical support to the conjecture about the universality of the power law of Eq.~\eqref{massgap_scaling}, with an exponent of $\xi \simeq 0.70$.

\section{Conclusions}\label{conclusions}

The collapse of a massless scalar field in a spherically-symmetric AAdS spacetime shows a much richer phenomenology than the analogous problem in asymptotically-flat spacetimes as it was realized for the first time in~\cite{Bizon:2011gg} and is illustrated by our Fig.~\ref{plot_3d}.  Both the long-term evolution and the dynamics of gravitational collapse present distinctive features that are not yet fully understood.  In this work we have focussed on the dynamics near collapse, when an AH is forming.  To study this question we need to resort to numerical methods, taking into account that we are dealing with a problem that represents an important challenge for the design and performance of a numerical code that solves the PDEs describing the system, despite the fact that we are dealing with a 1+1 problem (spherical symmetry).  The main reason for this lies behind the causal structure of global AdS, where light-like signals can reach the AdS boundary in a finite time.  As a consequence, an scalar field configurations that fails to form an AH in a first attempt will travel to the AdS in a finite time, bounce and travel back towards the origin, where it will have a second chance to form an AH.  This process will be repeated until, after a number of bounces, the scalar field will collapse forming an AH.  This is not the whole story as we have evidence of stable scalar field configurations, but the whole picture is not yet completely understood.  The near subcritical configurations are very challenging since they are very close to collapse, which induce large gradients in the field variables that we have to propagate to the AdS boundary and back.  In this sense, AAdS spacetimes constitute an excellent arena for the development of new numerical relativity methods and tools.

In this paper we have presented a new numerical scheme to study these situations which, in essence, is a hybrid Cauchy-characteristic evolution scheme.  The Cauchy evolution uses a multidomain PSC method for the spatial discretization and the characteristic evolution follows the ingoing null geodesics, which allows us to get much closer to the point of AH formation than with the Cauchy evolution.  An additional crucial part of this method is the transition between the two schemes.  We have described in detail all the analytic and numerical ingredients of this Cauchy-characteristic evolution scheme.  In doing so we have also analyzed the differences between evolution in AAdS and asymptotically-flat spacetimes, pointing out how the effect of the cosmological constant makes our simulations more challenging.  We have also shown the convergence properties of the different parts of the scheme and how we implement AMR techniques for the Cauchy-evolution sector.  Given that the scalar field configurations that we have considered are localized in the radial direction, in the sense that the energy density is concentrated within a single radial interval, we have studied how two definitions of center of mass can track the evolution of the field and how by using them we can also have a sense of how compact a certain scalar field configuration is.  

With this numerical scheme we studied in~\cite{SantosOlivan:2015fmy} the subcritical scalar field configurations near the different branches that appear depending on the number of times that field bounced off the AdS boundary.  We found evidence that this configurations follow the power law: $M_{\mathrm{AH}}-M^{n+1}_{g} \propto (p_{n}-p)^{\xi}$, with the mass gap between branches (separated by the location of the critical point, $p_{n}$) given by $M^{n+1}_{g}$ and the exponent $\xi$ was conjectured to be universal, independent of the initial data and also the same for all the mass gaps/branches.  The numerical support provided in~\cite{SantosOlivan:2015fmy} comes from the first two mass gaps using the initial conditions of Eq.~\eqref{cauchy-initial-data_Bizon} varying both the amplitude and the width.  In this paper we have further evidence for this conjecture, first by using a completely different class of initial conditions, the one in Eq.~\eqref{cauchy-initial-data-boundary}, and second, by looking at the third mass gap.  All these results support our conjecture for the power law at the mass gaps and the universal character of the exponent which, in all cases, has been found to be consistent with a value of $\xi \approx 0.7$.

In addition, we have also obtained the critical exponents associated with the multiple critical points that appear in the case of AAdS spacetimes.  By tracking supercritical configurations using only the Cauchy evolution we have been able to find the critical points associated with the branches 0 to 5.  We have confirmed the expected result~\cite{Bizon:2011gg} that at AH formation the presence of the negative cosmological constant is irrelevant.  Indeed, the critical exponents and echoing periods that we have found are consistent with the values of the asymptotically-flat case.

In summary, we have introduced a hybrid Cauchy-characteristic method that is particularly suited to study the dynamics in spherically-symmetric spacetimes near the point of formation of an AH.  This has allowed us to find new features and understand better gravitational collapse in AAdS spacetimes. This evolution scheme is quite general and it can be applied to other scenarios of physical interest, including different spacetime dimension, other matter fields, or even for a different spacetime causal structure.


\begin{acknowledgments}
The authors thank Sascha Husa, Esperanza L\'opez, and Ulrich Sperhake for discussions on different aspects of this work. The authors acknowledge the high-performance computing resources provided by the Consorci de Serveis Universitaris de Catalunya (CSUC) and the Galicia Supercomputing Center (CESGA) under projects ICTS-CESGA-249 and ICTS-CESGA-266. They also acknowledge support from contracts ESP2013-47637-P and ESP2015-67234-P (Spanish Ministry of Economy and Competitivity of Spain, MINECO). DS acknowledges support from a FPI doctoral contract BES-2012-057909 from MINECO.
\end{acknowledgments}


\appendix


\section{Hyperbolic structure of the scalar field equations}\label{app_char_var}
The PSC method used for the numerical implementation of the Cauchy evolution requires a first-order hyperbolic formulation of the scalar field equation, the Klein-Gordon equation~\eqref{kg_eq}.  This equation, for the metric in Eq.~\eqref{aadstx}, becomes the following second-order PDE:
\begin{eqnarray}
&& \ddot{\phi} - A^2e^{-2\delta} \phi'' = \frac{\dot{A}}{A}\dot{\phi} - \dot{\delta}\dot{\phi} 
+ A e^{-2\delta} \phi'A' - A^2e^{-2\delta} \delta' \phi' \nonumber \\
&& +\;(d-1) \frac{ A^2 e^{-2\delta }}{\cos x \sin x}\phi' \,.  \label{ddot_phi}
\end{eqnarray}
To reduce the order of the equation we introduce the variables:
\begin{equation}
\Pi = \dot{\phi} \,,\qquad  \Phi =  \phi'\,. \label{def-Pi-Phi} 
\end{equation}
The equations for $\bm{W} = (\phi,\Pi,\Phi)$ constitute a first-order system of PDEs that can be derived from these definitions and from Eq.~\eqref{ddot_phi}:
\begin{eqnarray}
\dot{\phi} & = & \Pi\,, \label{pp1} \\
\dot{\Pi} - A^2e^{-2\delta} \Phi' & = & \frac{\dot{A}}{A}\,\Pi -\dot{\delta}\,\Pi + A e^{-2\delta}A' \Phi \nonumber\\
 &-& A^2 e^{-2\delta} \delta' \Phi  + (d-1) \frac{ A^2 e^{-2\delta }}{\cos x \sin x}\Phi \,,\qquad  \label{pp2} \\
\dot \Phi -  \Pi'  & = & 0\,. \label{pp3}
\end{eqnarray}
As expected, this is a set of first-order PDEs that admits the following compact form: 
\begin{equation}
\partial^{}_t \bm{W} + {\cal A}[x,\bm{W}]\cdot \partial^{}_x \bm{W} =  \bm{S}[x,\bm{W}] \,,\label{ads_charac_pde}
\end{equation}
where ${\cal A}$ is a matrix and $\bm{S}$ a vector that depend on the radial coordinate $x$ and our variables $\bm{W}$, but they do not depend explicitly on the time $t$.  Here, the metric functions $\delta$ and $A$ have to be taken as functionals of our variables $\bm{W}$ since they are the result of integrating Eqs.~\eqref{A_prime} and~\eqref{delta_prime}.  From Eqs.~\eqref{pp1}-\eqref{pp3}, the components of the matrix ${\cal A}$ are:
\begin{equation}
{\cal A} = \begin{pmatrix}
   0 & 0 & 0 \\
   0 & 0 & -{A^2 e^{-2\delta}}  \\
   0 & -1 & 0 
\end{pmatrix}\,, 
\end{equation}
and the components of the vector $\bm{S}$ are:
\begin{equation}
\bm{S} = \begin{pmatrix}
   \Pi \\
   \frac{\dot{A}}{A}\,\Pi -\dot{\delta}\,\Pi + A e^{-2\delta}A' \Phi \\
 - A^2 e^{-2\delta} \delta' \Phi  + (d-1) \frac{ A^2 e^{-2\delta }}{\cos x \sin x}\Phi   \\
   0 
\end{pmatrix}\,, 
\end{equation}
The characteristic structure of this hyperbolic system of first-order PDEs is determined exclusively by the matrix ${\cal A}$ (see, e.g.~\cite{Courant:1989aa} for details), in such a way that the eigenvectors of ${\cal A}$ correspond to the different characteristic fields of the system and the eigenvalues to the characteristic speeds associated with the eigenvectors.  Strongly hyperbolic systems are those that have a complete set of eigenvalues and eigenvectors (see, e.g.~\cite{Courant:1989aa,John:1991fj,Gustafsson:1995tb} for a description of PDEs with hyperbolic structure), which is a key property for showing existence and uniqueness of solutions, and also a condition for the stability of algorithms to evolve the system.  To find out what happens in our case we have to solve the eigenvalue problem:
\begin{equation}
{\cal A} \, \bm{E} = \sigma \bm{E} \,, 
\label{eigenvalue-problem}
\end{equation}
where $\bm{E}$ is any vector in the space $\{(\phi,\Pi,\Phi)\}$. By analyzing Eq.~\eqref{eigenvalue-problem} we find that we have a complete set of real eigenvalues and eigenvectors, so our system is strongly hyperbolic, as expected for a system of PDEs that is equivalent to the Klein-Gordon equation.  The resulting set of eigenvalues and eigenvectors for Eq.~\eqref{eigenvalue-problem} is:
\begin{eqnarray}
\sigma^{}_{1} = 0  & \quad\longrightarrow\quad & \bm{E}^{}_{1} = (1,0,0) \,,  \label{eigen1} \\
\sigma^{}_{2} = + A e^{-\delta} & \quad\longrightarrow\quad & \bm{E}^{}_{2} = (0,-Ae^{-\delta},1)\,, \label{eigen2} \\
\sigma^{}_{3} = - A e^{-\delta} & \quad\longrightarrow\quad & \bm{E}^{}_{3} = (0,+Ae^{-\delta},1)\,. \label{eigen3}
\end{eqnarray}
The meaning of the eigenvalues is that they are the characteristic speeds of the characteristic variables.  Then, we have a quantity, $\phi$, that does not propagate (or in other words, it propagates with zero speed) and two that propagate with speed $\pm v \equiv \pm A\,e^{-\delta}$ [this is the same speed defined in Eq.~\eqref{outgoing-null-geodesics}].  We can diagonalize the matrix ${\cal A}$ by the matrix transformation: ${\cal A} = {\cal K}\cdot {\cal D}^{}_{\cal A}\cdot{\cal K}^{-1}$, where ${\cal D}^{}_{\cal A} = {\rm diag}(0,v,-v)$ and ${\cal K} = (\bm{E}^{}_{1},\bm{E}^{}_{2},\bm{E}^{}_{3})$.  At this point we can define a new set of variables $\bm{Y}$ as follows: $\bm{Y} = {\cal K}^{-1}\cdot\bm{W}$.  We can see that the principal part of our set of equations becomes completely decoupled for the variables $\bm{Y}$.  These are the characteristic variables.  The first one is $\phi$, with zero associated propagation speed (eigenvalue), and the other two are:
\begin{eqnarray}
Y^{}_{+} & = & \Phi - \frac{\Pi}{v} \,, \\               
Y^{}_{-} & = & \Phi + \frac{\Pi}{v} \,.                
\end{eqnarray}
The characteristic variable $Y^{}_{+}$ propagates with speed $v$ and $Y^{}_{-}$ with speed $-v$ respectively.  Since $v>0$, $Y^{}_{+}$ is a field propagating to the right with speed $v$ and $Y^{}_{-}$ is a field propagating to the left with the same speed.  From these characteristic variables we have introduced the variables in Eqs.~\eqref{Udef}, \eqref{Vdef}, and~\eqref{psi-def}, which are also characteristic variables as it can be seen from the evolution equations~\eqref{evol_psi}-\eqref{evol_V}. Actually, we can see that
\begin{equation}
U = \frac{Y^{}_{+}}{\cos^{d-2}x}\,,\qquad
V = \frac{Y^{}_{-}}{\cos^{d-2}x}\,. 
\label{characteristic-variables-u-v}
\end{equation}
The use of this variables is very important in this work for several reasons, but mainly in order to use the PSC method with a multidomain grid.  The communication between subdomains becomes very clear in terms of the characteristic variables.


\section{Basic ingredients of the PSC Method}\label{app_psc}
Broadly speaking, spectral methods can approximate solutions of PDEs by finite expansions of the variables using a given basis of functions.  The coefficients of such expansion are determined by imposing an appropriate criterium that forces this expansion to approach the exact solution as we increase the number of terms in the expansion. In the case of the PSC method, the criterium consists in imposing the expansion to be exact at a set of {\em collocation} points (see, e.g.~\cite{Boyd,Canutoetal:2006sm1,Fornberg:1996psc}). Here we use the Chebyshev polynomials, $\{T^{}_{n}(X)\}$ ($X \in [-1,1]$, the {\em spectral} domain), as the basis functions, which can be expressed in the following form:
\begin{eqnarray}
T^{}_n(X) =\cos\left(n\cos^{-1}(X) \right)\,. \label{chebyshev-polynomials}
\end{eqnarray}
They are orthogonal in the {\em continuum} in the following sense:
\begin{equation}
\left(T^{}_n,T^{}_m\right) = \int^{1}_{-1} \frac{dX}{\sqrt{1-X^2}} T^{}_n(X)\,T^{}_m(X) = \frac{\pi c^{}_{n}}{2}\delta^{}_{nm}\,,
\end{equation}
where the coefficients $c^{}_{n}$ are 
\begin{eqnarray}
{c}^{}_n=\left\{\begin{array}{ll} 2  & \text{for}~n=0\,, \\[2mm]
                                  1  & \text{otherwise\,.}
\end{array}\right. \label{cns}
\end{eqnarray}

The set of collocation points that we use are those of a {\em Lobatto-Chebyshev} grid.  The spectral coordinates of these points are the zeros of the polynomial $(1-X^{2})T'_{N}(X) = 0\,$, where the prime here indicates differentiation with respect to $X$. The zeros can be written as 
\begin{eqnarray}
 X^{}_i = -\cos\left(\frac{\pi\,i}{N}\right)\qquad (i=0,1,\ldots,N)\,,  \label{chebyshevlobattogrid}
\end{eqnarray}
which means that the boundary points $X=\pm 1$ are included in the grid, in contrast with other collocation grids, like for instance the {\em Gauss-Chebyshev} grid (see, e.g.~\cite{Boyd}). Taking into account the properties of the Gauss-Lobatto-Chebyshev quadratures (see, e.g.~\cite{Boyd}), the Chebyshev polynomials have another orthogonality relation, in the {\em discrete}, in the following sense ($n,m = 0,\ldots,N$):
\begin{eqnarray}
\left[ T^{}_n, T^{}_m\right] = \sum_{i=0}^{N} w^{}_i\, T^{}_n(X^{}_i)\, T^{}_m(X^{}_{i}) = 
\nu_n^2\,\delta^{}_{nm}\,,  \label{scalarp}
\end{eqnarray}
where $w^{}_i$ are the weights associated with the Chebyshev-Lobatto grid, $w^{}_i = \pi/(N\,\bar{c}^{}_{i})$, and where the $\bar{c}^{}_i$'s are normalization coefficients given by
\begin{eqnarray}
\bar{c}^{}_i=\left\{\begin{array}{ll} 2  & \text{for}~i=0,N\,, \\[2mm]
                                      1  & \text{otherwise\,.}
\end{array}\right.
\end{eqnarray}
Finally, the constants $\nu^{}_n$ in Eq.~\eqref{scalarp} are given by $\nu^{2}_n=\pi \bar{c}^{}_n/2$.

In general the computational domain, say $[x^{}_{L},x^{}_{R}]$, does not coincide with the spectral one, $[-1,1]$ and we need a one-to-one mapping between them.  The simplest choice, and the one we use, is the linear mapping:
\begin{eqnarray}
x \longrightarrow X(x) = \frac{2x- x^{}_{L}- x^{}_{R}}{ x^{}_{R} - x^{}_{L} } \,.    \label{lmap1}
\end{eqnarray} 
and the inverse one is:
\begin{eqnarray}
X  \longrightarrow x(X) = \frac{x^{}_{R}-x^{}_{L}}{2}X+\frac{x^{}_{L}+ x^{}_{R}}{2}\,. \label{lmap2}
\end{eqnarray}

For mesh refinement purposes (see Sec.~\ref{refinement-cauchy}) we use a multidomain PSC method consisting in the split the computational domain, $\Omega=[x^{}_{L},x^{}_{R}]$, into a number of disjoint subdomains ($D$): 
\begin{equation}
\Omega = \bigcup^{D}_{a=1} \Omega^{}_a\,, ~~\Omega^{}_a = \left[ x^{}_{a,L}, x^{}_{a,R}\right]\,,
\end{equation}
where $x^{}_{a,L}$ and $x^{}_{a,R}$ are the left and right boundaries of the subdomain $\Omega^{}_{a}$ ($x^{}_{1,L}=x^{}_{L}$ and $x^{}_{D,R}=x^{}_{R}$).  Since they are disjoint subdomains we have that $x^{}_{a,L}=x^{}_{a-1,R}$ ($a=1,\dots,D$).  We apply the PSC method to each subdomain, and hence our variables have different expansions in Chebyshev polynomials in each subdomain.  Then, each {\em physical} subdomain is mapped to the {\em spectral} domain $[-1,1]$ using the linear mappings of Eqs.~\eqref{lmap1} and~\eqref{lmap2}, which we call $x^{}_{a}(X)$ and $X^{}_{a}(x)$.  The different expansion for the different subdomains are then {\em matched} by using the appropriate boundary conditions (see the description in Sec.~\ref{numerics_cauchy}).

Let us now look at the spectral approximation for the variables of our problem, which we arranged in the vector $\bm{Z}$. At a given subdomain $\Omega^{}_a$, in the PSC method, we have two representations of the approximation for our variables.  First, we have the standard spectral representation of the approximation to our variables in a given subdomain $a$, $\bm{U}^{}_{a,N}(t,x)$:
\begin{equation}
\bm{U}^{}_{a,N}(t,x) = \sum_{n=0}^N \bm{a}^{}_{a,n}(t)\, T^{}_n(X^{}_{a}(x)) \,, \label{spectralrepresentation}
\end{equation}
where the $\bm{a}^{}_{a,n}$ are (time-dependent) vectors that contain the spectral coefficients of the expansion of our variables.  In the PSC method, we have also a {\em physical} expansion, which looks as follows:
\begin{equation}
\bm{U}^{}_{a,N}(t,x) = \sum_{i=0}^N \bm{U}^{}_{a,i}(t)\, {\cal C}^{}_i(X^{}_{a}(x))\,, \label{physicalrepresentation}
\end{equation}
where ${\cal C}^{}_i(X)$ are the {\em cardinal} functions~\cite{Boyd} associated with our choice of basis functions (Chebyshev polynomials) and set of collocation points (Lobatto-Chebyshev grid).  Their expression is
\begin{equation}
{\cal C}^{}_i(X) = \frac{(1-X^2){T'_N}(X)}{(1-X^{2}_i)(X-X^{}_i){T''_N}(X^{}_i)}\,.
\end{equation}
The cardinal functions have the following remarkable property:
\begin{equation}
{\cal C}^{}_i(X^{}_{j}) = \delta^{}_{ij} \qquad (i,j=0,\ldots,N)\,,
\end{equation}
so that the time-dependent (vector) coefficients, $\{\bm{U}^{}_{i}\}$, of the expansion in Eq.~\eqref{physicalrepresentation} are the values of our variables at the collocation points
\begin{equation}
\bm{U}^{}_{a,N}(t,x^{}_{a}(X^{}_{i})) = \bm{U}^{}_{a,i}(t)\,.
\end{equation}
These are the unknowns that one looks for in the PSC method. The spectral and physical representations [Eqs.~\eqref{spectralrepresentation} and~\eqref{physicalrepresentation} respectively] are related via a matrix transformation~\cite{Boyd}. The computations (float-point operations) required to change representation using the matrix transformations increase, with the number of collocation points, as $\sim N^{2}$.  Nevertheless, we can introduce a new spectral coordinate via $X =\cos\theta$ (with $\theta\in [0,\pi]$), in such a way that the Chebyshev polynomials become
\begin{equation}
T^{}_{n}(\cos\theta) = \cos(n\theta)\,. \label{changespectralcoord}
\end{equation}
Then, an spectral expansion in Chebyshev polynomials like the one in Eq.~\eqref{spectralrepresentation} can be mapped to a cosine series.  Then, we can perform the change of representation by means of a discrete Fourier transform using a Fast-Fourier Transform (FFT) algorithm.  In our numerical codes, we use the routines of the FFTW library~\cite{fftw:2005}.  Then, the number of computations required for a change of representations increases as $\sim N\ln N$ with the number of collocation points.

Changing between representations is useful in order to compute derivatives and nonlinear terms.  In the case of derivatives, they are much simpler to be computed in the spectral representation. Then, we can transform from the physical to the spectral representation, compute derivatives there, and finally transform back to the physical representation. In the case of a Chebyshev PSC method, the differentiation process can be described by the following scheme
\begin{eqnarray}
\partial^{}_{x} :\,\{\bm{U}^{}_i\} \stackrel{FFT}{\longrightarrow}
\{\bm{a}^{}_n\} \stackrel{\partial^{}_{x}}{\longrightarrow}
\{\bm{b}^{}_n\} \stackrel{FFT}{\longrightarrow}
\{(\partial^{}_{x}{\bm{U}})^{}_i\}\,,\qquad \label{pscdifferentiation}
\end{eqnarray}
where $\{\bm{b}^{}_{n}\}$ are the spectral coefficients associated with the spatial derivative $\partial^{}_{x}$, and their relation to the spectral coefficients of the variables, $\{\bm{a}^{}_{n}\}$, is given by (see, e.g.~\cite{Boyd})
\begin{eqnarray}
&& \bm{b}^{}_N = \bm{b}^{}_{N-1}=0\,, \\
&& \bm{b}^{}_{n-1} = \frac{1}{c^{}_n}\left\lbrace 2n\,\bm{a}^{}_n+\bm{b}^{}_{n+1}\right\rbrace~(n=N-1,\ldots,1)\,,\qquad
\end{eqnarray}
where the coefficients $c^{}_{n}$ are given in Eq.~\eqref{cns}.

Another important operation where the changing between representation is very useful is integration.  Let us assume we want to integrate the function $g(X)$ (we assume we have already changed to the spectral coordinate $X$) from the right, that is, $f(X) = \int^{X_{N}=1}_{X}dX'\,g(X')$, which assumes that an integration constant/boundary condition is imposed on the right boundary, $f(X^{}_{N}) = f^{}_{N}$. Then, we can follow the scheme
\begin{eqnarray}
\begingroup\textstyle\int\endgroup^{1}_{X} :\,\{\bm{U}^{}_i\} \stackrel{FFT}{\longrightarrow}
\{\bm{a}^{}_n\} \stackrel{\partial^{}_{x}}{\longrightarrow}
\{\bm{b}^{}_n\} \stackrel{FFT}{\longrightarrow}
\{(\int^{1}_{X}{\bm{U}})^{}_i\}\,,\qquad \label{pscintegration}
\end{eqnarray}
where $\{{b}^{}_{n}\}$ are the spectral coefficients associated with the integral from the right, $f(X)$, and their relation to the spectral coefficients of the function $g(X)$, $\{{a}^{}_{n}\}$, is given by 
\begin{eqnarray}
&& {b}^{}_N = \frac{a^{}_{N-1}}{2N}\,, \\
&& {b}^{}_{n} = \frac{1}{2n}\left\lbrace \bar{c}^{}_{n-1}\,{a}^{}_{n-1} - a^{}_{n+1}\right\rbrace~(n=N-1,\ldots,1)\,,\qquad \\
&& {b}^{}_{0} = f^{}_{N} - \sum^{N}_{n=1} {b}^{}_{n}\,.
\end{eqnarray}
The process to integrate from the left, $f(X) = \int^{X}_{X_{0}=-1}dX'\,g(X')$, is very similar. In this paper we use both since some variables that are obtained via integration with respect to $x$ require a boundary condition at the origin and others on the AdS boundary.  It is simple to extend these rules to our multidomain scheme.

Finally, in the PSC method, we find a discretization of our system of equations in Eqs.~\eqref{evol_psi}-\eqref{delta_prime} by imposing them at every collocation point.  In practice, this is done by introducing the expansion~\eqref{physicalrepresentation} into the Eqs.~\eqref{evol_psi}-\eqref{delta_prime}, and then we evaluate the result at every collocation point of our Chebyshev-Lobatto grid~\eqref{chebyshevlobattogrid}.  We obtain a system of ODEs for the variables $\{\bm{U}^{}_{i}(t)\}$
\begin{eqnarray}
\dot{\bm{U}}^{}_i = \mathbb{A}\cdot (\partial^{}_{x}\bm{U})^{}_{i} + \mathbb{B}\cdot\bm{U}^{}_i + \bm{S}^{}_i \,, \label{odes}
\end{eqnarray}
where the dot denotes differentiation with respect to the time coordinate $t$, and $(\partial^{}_{x}\bm{U})^{}_{i}$ has to be interpreted according to the scheme in Eq.~\eqref{pscdifferentiation}.



\begin{thebibliography}{96}%
\makeatletter
\providecommand \@ifxundefined [1]{%
 \@ifx{#1\undefined}
}%
\providecommand \@ifnum [1]{%
 \ifnum #1\expandafter \@firstoftwo
 \else \expandafter \@secondoftwo
 \fi
}%
\providecommand \@ifx [1]{%
 \ifx #1\expandafter \@firstoftwo
 \else \expandafter \@secondoftwo
 \fi
}%
\providecommand \natexlab [1]{#1}%
\providecommand \enquote  [1]{``#1''}%
\providecommand \bibnamefont  [1]{#1}%
\providecommand \bibfnamefont [1]{#1}%
\providecommand \citenamefont [1]{#1}%
\providecommand \href@noop [0]{\@secondoftwo}%
\providecommand \href [0]{\begingroup \@sanitize@url \@href}%
\providecommand \@href[1]{\@@startlink{#1}\@@href}%
\providecommand \@@href[1]{\endgroup#1\@@endlink}%
\providecommand \@sanitize@url [0]{\catcode `\\12\catcode `\$12\catcode
  `\&12\catcode `\#12\catcode `\^12\catcode `\_12\catcode `\%12\relax}%
\providecommand \@@startlink[1]{}%
\providecommand \@@endlink[0]{}%
\providecommand \url  [0]{\begingroup\@sanitize@url \@url }%
\providecommand \@url [1]{\endgroup\@href {#1}{\urlprefix }}%
\providecommand \urlprefix  [0]{URL }%
\providecommand \Eprint [0]{\href }%
\providecommand \doibase [0]{http://dx.doi.org/}%
\providecommand \selectlanguage [0]{\@gobble}%
\providecommand \bibinfo  [0]{\@secondoftwo}%
\providecommand \bibfield  [0]{\@secondoftwo}%
\providecommand \translation [1]{[#1]}%
\providecommand \BibitemOpen [0]{}%
\providecommand \bibitemStop [0]{}%
\providecommand \bibitemNoStop [0]{.\EOS\space}%
\providecommand \EOS [0]{\spacefactor3000\relax}%
\providecommand \BibitemShut  [1]{\csname bibitem#1\endcsname}%
\let\auto@bib@innerbib\@empty
\bibitem [{\citenamefont {Choptuik}(1993)}]{Choptuik:1992jv}%
  \BibitemOpen
  \bibfield  {author} {\bibinfo {author} {\bibfnamefont {M.~W.}\ \bibnamefont
  {Choptuik}},\ }\bibfield  {title} {\enquote {\bibinfo {title} {{Universality
  and scaling in gravitational collapse of a massless scalar field}},}\ }\href
  {\doibase 10.1103/PhysRevLett.70.9} {\bibfield  {journal} {\bibinfo
  {journal} {Phys. Rev. Lett.}\ }\textbf {\bibinfo {volume} {70}},\ \bibinfo
  {pages} {9--12} (\bibinfo {year} {1993})}\BibitemShut {NoStop}%
\bibitem [{\citenamefont {Christodoulou}(1986)}]{Christodoulou:1986zr}%
  \BibitemOpen
  \bibfield  {author} {\bibinfo {author} {\bibfnamefont {D.}~\bibnamefont
  {Christodoulou}},\ }\bibfield  {title} {\enquote {\bibinfo {title} {{The
  Problem of a Selfgravitating Scalar Field}},}\ }\href {\doibase
  10.1007/BF01205930} {\bibfield  {journal} {\bibinfo  {journal} {Commun. Math.
  Phys.}\ }\textbf {\bibinfo {volume} {105}},\ \bibinfo {pages} {337--361}
  (\bibinfo {year} {1986})}\BibitemShut {NoStop}%
\bibitem [{\citenamefont {Gundlach}(1995)}]{Gundlach:1995kd}%
  \BibitemOpen
  \bibfield  {author} {\bibinfo {author} {\bibfnamefont {C.}~\bibnamefont
  {Gundlach}},\ }\bibfield  {title} {\enquote {\bibinfo {title} {{The Choptuik
  space-time as an eigenvalue problem}},}\ }\href {\doibase
  10.1103/PhysRevLett.75.3214} {\bibfield  {journal} {\bibinfo  {journal}
  {Phys. Rev. Lett.}\ }\textbf {\bibinfo {volume} {75}},\ \bibinfo {pages}
  {3214--3217} (\bibinfo {year} {1995})},\ \Eprint
  {http://arxiv.org/abs/gr-qc/9507054} {arXiv:gr-qc/9507054 [gr-qc]}
  \BibitemShut {NoStop}%
\bibitem [{\citenamefont {Gundlach}(1997)}]{Gundlach:1996eg}%
  \BibitemOpen
  \bibfield  {author} {\bibinfo {author} {\bibfnamefont {C.}~\bibnamefont
  {Gundlach}},\ }\bibfield  {title} {\enquote {\bibinfo {title} {{Understanding
  critical collapse of a scalar field}},}\ }\href {\doibase
  10.1103/PhysRevD.55.695} {\bibfield  {journal} {\bibinfo  {journal} {Phys.
  Rev.}\ }\textbf {\bibinfo {volume} {D55}},\ \bibinfo {pages} {695--713}
  (\bibinfo {year} {1997})},\ \Eprint {http://arxiv.org/abs/gr-qc/9604019}
  {arXiv:gr-qc/9604019 [gr-qc]} \BibitemShut {NoStop}%
\bibitem [{\citenamefont {Gundlach}(2003)}]{Gundlach:2002sx}%
  \BibitemOpen
  \bibfield  {author} {\bibinfo {author} {\bibfnamefont {C.}~\bibnamefont
  {Gundlach}},\ }\bibfield  {title} {\enquote {\bibinfo {title} {{Critical
  phenomena in gravitational collapse}},}\ }\href {\doibase
  10.1016/S0370-1573(02)00560-4} {\bibfield  {journal} {\bibinfo  {journal}
  {Phys. Rept.}\ }\textbf {\bibinfo {volume} {376}},\ \bibinfo {pages}
  {339--405} (\bibinfo {year} {2003})},\ \Eprint
  {http://arxiv.org/abs/gr-qc/0210101} {arXiv:gr-qc/0210101 [gr-qc]}
  \BibitemShut {NoStop}%
\bibitem [{\citenamefont {Gundlach}\ and\ \citenamefont
  {Martin-Garcia}(2007)}]{Gundlach:2007gc}%
  \BibitemOpen
  \bibfield  {author} {\bibinfo {author} {\bibfnamefont {C.}~\bibnamefont
  {Gundlach}}\ and\ \bibinfo {author} {\bibfnamefont {J.~M.}\ \bibnamefont
  {Martin-Garcia}},\ }\bibfield  {title} {\enquote {\bibinfo {title} {{Critical
  phenomena in gravitational collapse}},}\ }\href {\doibase
  10.12942/lrr-2007-5} {\bibfield  {journal} {\bibinfo  {journal} {Living Rev.
  Rel.}\ }\textbf {\bibinfo {volume} {10}},\ \bibinfo {pages} {5} (\bibinfo
  {year} {2007})},\ \Eprint {http://arxiv.org/abs/0711.4620} {arXiv:0711.4620
  [gr-qc]} \BibitemShut {NoStop}%
\bibitem [{\citenamefont {{Chandrasekhar}}(1931)}]{Chandrasekhar:1931sc}%
  \BibitemOpen
  \bibfield  {author} {\bibinfo {author} {\bibfnamefont {S.}~\bibnamefont
  {{Chandrasekhar}}},\ }\bibfield  {title} {\enquote {\bibinfo {title} {{The
  Maximum Mass of Ideal White Dwarfs}},}\ }\href {\doibase 10.1086/143324}
  {\bibfield  {journal} {\bibinfo  {journal} {Astrophys. J.}\ }\textbf
  {\bibinfo {volume} {74}},\ \bibinfo {pages} {81} (\bibinfo {year}
  {1931})}\BibitemShut {NoStop}%
\bibitem [{\citenamefont {Goldwirth}\ and\ \citenamefont
  {Piran}(1987)}]{Goldwirth:1987nu}%
  \BibitemOpen
  \bibfield  {author} {\bibinfo {author} {\bibfnamefont {D.~S.}\ \bibnamefont
  {Goldwirth}}\ and\ \bibinfo {author} {\bibfnamefont {T.}~\bibnamefont
  {Piran}},\ }\bibfield  {title} {\enquote {\bibinfo {title} {{Gravitational
  Collapse of Massless Scalar Field and Cosmic Censorship}},}\ }\href {\doibase
  10.1103/PhysRevD.36.3575} {\bibfield  {journal} {\bibinfo  {journal} {Phys.
  Rev.}\ }\textbf {\bibinfo {volume} {D36}},\ \bibinfo {pages} {3575} (\bibinfo
  {year} {1987})}\BibitemShut {NoStop}%
\bibitem [{\citenamefont {Garfinkle}(1995)}]{Garfinkle:1994jb}%
  \BibitemOpen
  \bibfield  {author} {\bibinfo {author} {\bibfnamefont {D.}~\bibnamefont
  {Garfinkle}},\ }\bibfield  {title} {\enquote {\bibinfo {title} {{Choptuik
  scaling in null coordinates}},}\ }\href {\doibase 10.1103/PhysRevD.51.5558}
  {\bibfield  {journal} {\bibinfo  {journal} {Phys. Rev.}\ }\textbf {\bibinfo
  {volume} {D51}},\ \bibinfo {pages} {5558--5561} (\bibinfo {year} {1995})},\
  \Eprint {http://arxiv.org/abs/gr-qc/9412008} {arXiv:gr-qc/9412008 [gr-qc]}
  \BibitemShut {NoStop}%
\bibitem [{\citenamefont {Lehner}(2001)}]{Lehner:2001wq}%
  \BibitemOpen
  \bibfield  {author} {\bibinfo {author} {\bibfnamefont {L.}~\bibnamefont
  {Lehner}},\ }\bibfield  {title} {\enquote {\bibinfo {title} {{Numerical
  relativity: A Review}},}\ }\href {\doibase 10.1088/0264-9381/18/17/202}
  {\bibfield  {journal} {\bibinfo  {journal} {Class. Quant. Grav.}\ }\textbf
  {\bibinfo {volume} {18}},\ \bibinfo {pages} {R25--R86} (\bibinfo {year}
  {2001})},\ \Eprint {http://arxiv.org/abs/gr-qc/0106072} {arXiv:gr-qc/0106072
  [gr-qc]} \BibitemShut {NoStop}%
\bibitem [{\citenamefont {Bona}\ \emph {et~al.}(2009)\citenamefont {Bona},
  \citenamefont {Palenzuela-Luque},\ and\ \citenamefont
  {Bona-Casas}}]{Bona:2009bo}%
  \BibitemOpen
  \bibfield  {author} {\bibinfo {author} {\bibfnamefont {C.}~\bibnamefont
  {Bona}}, \bibinfo {author} {\bibfnamefont {C.}~\bibnamefont
  {Palenzuela-Luque}}, \ and\ \bibinfo {author} {\bibfnamefont
  {C.}~\bibnamefont {Bona-Casas}},\ }\href@noop {} {\emph {\bibinfo {title}
  {Elements of Numerical Relativity and Relativistic Hydrodynamics: From
  Einstein's equations to Astrophysical Simulations}}}\ (\bibinfo  {publisher}
  {Springer},\ \bibinfo {address} {Berlin},\ \bibinfo {year}
  {2009})\BibitemShut {NoStop}%
\bibitem [{\citenamefont {Baumgarte}\ and\ \citenamefont
  {Shapiro}(2010)}]{Baumgarte:2010bs}%
  \BibitemOpen
  \bibfield  {author} {\bibinfo {author} {\bibfnamefont {T.~W.}\ \bibnamefont
  {Baumgarte}}\ and\ \bibinfo {author} {\bibfnamefont {S.~L.}\ \bibnamefont
  {Shapiro}},\ }\href@noop {} {\emph {\bibinfo {title} {Numerical Relativity:
  Solving Einstein's Equations on the Computer}}}\ (\bibinfo  {publisher}
  {Cambridge University Press},\ \bibinfo {address} {Cambridge},\ \bibinfo
  {year} {2010})\BibitemShut {NoStop}%
\bibitem [{\citenamefont {Lehner}\ and\ \citenamefont
  {Pretorius}(2014)}]{Lehner:2014asa}%
  \BibitemOpen
  \bibfield  {author} {\bibinfo {author} {\bibfnamefont {L.}~\bibnamefont
  {Lehner}}\ and\ \bibinfo {author} {\bibfnamefont {F.}~\bibnamefont
  {Pretorius}},\ }\bibfield  {title} {\enquote {\bibinfo {title} {{Numerical
  Relativity and Astrophysics}},}\ }\href {\doibase
  10.1146/annurev-astro-081913-040031} {\bibfield  {journal} {\bibinfo
  {journal} {Ann. Rev. Astron. Astrophys.}\ }\textbf {\bibinfo {volume} {52}},\
  \bibinfo {pages} {661--694} (\bibinfo {year} {2014})},\ \Eprint
  {http://arxiv.org/abs/1405.4840} {arXiv:1405.4840 [astro-ph.HE]} \BibitemShut
  {NoStop}%
\bibitem [{\citenamefont {Sperhake}(2015)}]{Sperhake:2014wpa}%
  \BibitemOpen
  \bibfield  {author} {\bibinfo {author} {\bibfnamefont {U.}~\bibnamefont
  {Sperhake}},\ }\bibfield  {title} {\enquote {\bibinfo {title} {{The numerical
  relativity breakthrough for binary black holes}},}\ }\href {\doibase
  10.1088/0264-9381/32/12/124011} {\bibfield  {journal} {\bibinfo  {journal}
  {Class. Quant. Grav.}\ }\textbf {\bibinfo {volume} {32}},\ \bibinfo {pages}
  {124011} (\bibinfo {year} {2015})},\ \Eprint {http://arxiv.org/abs/1411.3997}
  {arXiv:1411.3997 [gr-qc]} \BibitemShut {NoStop}%
\bibitem [{\citenamefont {Hawking}\ and\ \citenamefont
  {Ellis}(1973)}]{Hawking:1973uf}%
  \BibitemOpen
  \bibfield  {author} {\bibinfo {author} {\bibfnamefont {S.~W.}\ \bibnamefont
  {Hawking}}\ and\ \bibinfo {author} {\bibfnamefont {G.~F.~R.}\ \bibnamefont
  {Ellis}},\ }\href@noop {} {\emph {\bibinfo {title} {{The Large scale
  structure of space-time}}}}\ (\bibinfo  {publisher} {Cambridge University
  Press},\ \bibinfo {address} {Cambridge},\ \bibinfo {year} {1973})\BibitemShut
  {NoStop}%
\bibitem [{\citenamefont {Maldacena}(1999)}]{Maldacena:1997re}%
  \BibitemOpen
  \bibfield  {author} {\bibinfo {author} {\bibfnamefont {J.~M.}\ \bibnamefont
  {Maldacena}},\ }\bibfield  {title} {\enquote {\bibinfo {title} {{The Large N
  limit of superconformal field theories and supergravity}},}\ }\href {\doibase
  10.1023/A:1026654312961} {\bibfield  {journal} {\bibinfo  {journal} {Int. J.
  Theor. Phys.}\ }\textbf {\bibinfo {volume} {38}},\ \bibinfo {pages}
  {1113--1133} (\bibinfo {year} {1999})},\ \Eprint
  {http://arxiv.org/abs/hep-th/9711200} {arXiv:hep-th/9711200 [hep-th]}
  \BibitemShut {NoStop}%
\bibitem [{\citenamefont {Witten}(1998)}]{Witten:1998qj}%
  \BibitemOpen
  \bibfield  {author} {\bibinfo {author} {\bibfnamefont {E.}~\bibnamefont
  {Witten}},\ }\bibfield  {title} {\enquote {\bibinfo {title} {{Anti-de Sitter
  space and holography}},}\ }\href@noop {} {\bibfield  {journal} {\bibinfo
  {journal} {Adv. Theor. Math. Phys.}\ }\textbf {\bibinfo {volume} {2}},\
  \bibinfo {pages} {253--291} (\bibinfo {year} {1998})},\ \Eprint
  {http://arxiv.org/abs/hep-th/9802150} {arXiv:hep-th/9802150 [hep-th]}
  \BibitemShut {NoStop}%
\bibitem [{\citenamefont {Chesler}\ and\ \citenamefont
  {Yaffe}(2014)}]{Chesler:2013lia}%
  \BibitemOpen
  \bibfield  {author} {\bibinfo {author} {\bibfnamefont {P.~M.}\ \bibnamefont
  {Chesler}}\ and\ \bibinfo {author} {\bibfnamefont {L.~G.}\ \bibnamefont
  {Yaffe}},\ }\bibfield  {title} {\enquote {\bibinfo {title} {{Numerical
  solution of gravitational dynamics in asymptotically Anti-de Sitter
  spacetimes}},}\ }\href {\doibase 10.1007/JHEP07(2014)086} {\bibfield
  {journal} {\bibinfo  {journal} {JHEP}\ }\textbf {\bibinfo {volume} {07}},\
  \bibinfo {pages} {086} (\bibinfo {year} {2014})},\ \Eprint
  {http://arxiv.org/abs/1309.1439} {arXiv:1309.1439 [hep-th]} \BibitemShut
  {NoStop}%
\bibitem [{\citenamefont {Horowitz}(2014)}]{Horowitz:2014awa}%
  \BibitemOpen
  \bibfield  {author} {\bibinfo {author} {\bibfnamefont {G.T.}\ \bibnamefont
  {Horowitz}},\ }\bibfield  {title} {\enquote {\bibinfo {title} {{Using general
  relativity to study superconductivity}},}\ }\href {\doibase
  10.1088/1742-6596/484/1/012002} {\bibfield  {journal} {\bibinfo  {journal}
  {J. Phys. Conf. Ser.}\ }\textbf {\bibinfo {volume} {484}},\ \bibinfo {pages}
  {012002} (\bibinfo {year} {2014})}\BibitemShut {NoStop}%
\bibitem [{\citenamefont {Hartnoll}(2009)}]{Hartnoll:2009sz}%
  \BibitemOpen
  \bibfield  {author} {\bibinfo {author} {\bibfnamefont {S.~A.}\ \bibnamefont
  {Hartnoll}},\ }\bibfield  {title} {\enquote {\bibinfo {title} {{Lectures on
  holographic methods for condensed matter physics}},}\ }\bibfield  {booktitle}
  {\emph {\bibinfo {booktitle} {{Strings, Supergravity and Gauge Theories.
  Proceedings, CERN Winter School, CERN, Geneva, Switzerland, February 9-13
  2009}}},\ }\href {\doibase 10.1088/0264-9381/26/22/224002} {\bibfield
  {journal} {\bibinfo  {journal} {Class. Quant. Grav.}\ }\textbf {\bibinfo
  {volume} {26}},\ \bibinfo {pages} {224002} (\bibinfo {year} {2009})},\
  \Eprint {http://arxiv.org/abs/0903.3246} {arXiv:0903.3246 [hep-th]}
  \BibitemShut {NoStop}%
\bibitem [{\citenamefont {Cardoso}\ \emph {et~al.}(2012)\citenamefont
  {Cardoso}, \citenamefont {Gualtieri}, \citenamefont {Herdeiro}, \citenamefont
  {Sperhake}, \citenamefont {Chesler} \emph {et~al.}}]{Cardoso:2012qm}%
  \BibitemOpen
  \bibfield  {author} {\bibinfo {author} {\bibfnamefont {V.}~\bibnamefont
  {Cardoso}}, \bibinfo {author} {\bibfnamefont {L.}~\bibnamefont {Gualtieri}},
  \bibinfo {author} {\bibfnamefont {C.}~\bibnamefont {Herdeiro}}, \bibinfo
  {author} {\bibfnamefont {U.}~\bibnamefont {Sperhake}}, \bibinfo {author}
  {\bibfnamefont {P.~M.}\ \bibnamefont {Chesler}},  \emph {et~al.},\ }\bibfield
   {title} {\enquote {\bibinfo {title} {{NR/HEP: roadmap for the future}},}\
  }\href {\doibase 10.1088/0264-9381/29/24/244001} {\bibfield  {journal}
  {\bibinfo  {journal} {Class. Quant. Grav.}\ }\textbf {\bibinfo {volume}
  {29}},\ \bibinfo {pages} {244001} (\bibinfo {year} {2012})},\ \Eprint
  {http://arxiv.org/abs/1201.5118} {arXiv:1201.5118 [hep-th]} \BibitemShut
  {NoStop}%
\bibitem [{\citenamefont {Sperhake}(2013)}]{Sperhake:2013wva}%
  \BibitemOpen
  \bibfield  {author} {\bibinfo {author} {\bibfnamefont {U.}~\bibnamefont
  {Sperhake}},\ }\bibfield  {title} {\enquote {\bibinfo {title} {{Black Holes
  on Supercomputers: Numerical Relativity Applications to Astrophysics and
  High-energy Physics}},}\ }\href {\doibase 10.5506/APhysPolB.44.2463}
  {\bibfield  {journal} {\bibinfo  {journal} {Acta Phys. Polon.}\ }\textbf
  {\bibinfo {volume} {B44}},\ \bibinfo {pages} {2463--2536} (\bibinfo {year}
  {2013})}\BibitemShut {NoStop}%
\bibitem [{\citenamefont {Christodoulou}\ and\ \citenamefont
  {Klainerman}(1994)}]{Christodoulou:1994ck}%
  \BibitemOpen
  \bibfield  {author} {\bibinfo {author} {\bibfnamefont {D.}~\bibnamefont
  {Christodoulou}}\ and\ \bibinfo {author} {\bibfnamefont {S.}~\bibnamefont
  {Klainerman}},\ }\href@noop {} {\emph {\bibinfo {title} {The Global Nonlinear
  Stability of the Minkowski Space (PMS-41)}}}\ (\bibinfo  {publisher}
  {Princeton University Press},\ \bibinfo {address} {Princeton},\ \bibinfo
  {year} {1994})\BibitemShut {NoStop}%
\bibitem [{\citenamefont {Lindblad}\ and\ \citenamefont
  {Rodnianski}(2010)}]{Lindblad_theglobal}%
  \BibitemOpen
  \bibfield  {author} {\bibinfo {author} {\bibfnamefont {H.}~\bibnamefont
  {Lindblad}}\ and\ \bibinfo {author} {\bibfnamefont {I.}~\bibnamefont
  {Rodnianski}},\ }\bibfield  {title} {\enquote {\bibinfo {title} {The global
  stability of minkowski space-time in harmonic gauge},}\ }\href@noop {}
  {\bibfield  {journal} {\bibinfo  {journal} {Ann. of Math}\ }\textbf {\bibinfo
  {volume} {171}},\ \bibinfo {pages} {1401--1477} (\bibinfo {year}
  {2010})}\BibitemShut {NoStop}%
\bibitem [{\citenamefont {Friedrich}(1986)}]{Friedrich:1986cm}%
  \BibitemOpen
  \bibfield  {author} {\bibinfo {author} {\bibfnamefont {H.}~\bibnamefont
  {Friedrich}},\ }\bibfield  {title} {\enquote {\bibinfo {title} {{On the
  existence of n-geodesically complete or future complete solutions of
  Einsteins field equations with smooth asymptotic structure}},}\ }\href@noop
  {} {\bibfield  {journal} {\bibinfo  {journal} {Commun. Math. Phys.}\ }\textbf
  {\bibinfo {volume} {107}},\ \bibinfo {pages} {587--609} (\bibinfo {year}
  {1986})}\BibitemShut {NoStop}%
\bibitem [{\citenamefont {Ishibashi}\ and\ \citenamefont
  {Wald}(2004)}]{Ishibashi:2004wx}%
  \BibitemOpen
  \bibfield  {author} {\bibinfo {author} {\bibfnamefont {A.}~\bibnamefont
  {Ishibashi}}\ and\ \bibinfo {author} {\bibfnamefont {R.~M.}\ \bibnamefont
  {Wald}},\ }\bibfield  {title} {\enquote {\bibinfo {title} {{Dynamics in
  nonglobally hyperbolic static space-times. 3. Anti-de Sitter space-time}},}\
  }\href {\doibase 10.1088/0264-9381/21/12/012} {\bibfield  {journal} {\bibinfo
   {journal} {Class. Quant. Grav.}\ }\textbf {\bibinfo {volume} {21}},\
  \bibinfo {pages} {2981--3014} (\bibinfo {year} {2004})},\ \Eprint
  {http://arxiv.org/abs/hep-th/0402184} {arXiv:hep-th/0402184 [hep-th]}
  \BibitemShut {NoStop}%
\bibitem [{\citenamefont {Friedrich}(2014{\natexlab{a}})}]{Friedrich:2014raa}%
  \BibitemOpen
  \bibfield  {author} {\bibinfo {author} {\bibfnamefont {H.}~\bibnamefont
  {Friedrich}},\ }\bibfield  {title} {\enquote {\bibinfo {title} {{On the AdS
  stability problem}},}\ }\href {\doibase 10.1088/0264-9381/31/10/105001}
  {\bibfield  {journal} {\bibinfo  {journal} {Class. Quant. Grav.}\ }\textbf
  {\bibinfo {volume} {31}},\ \bibinfo {pages} {105001} (\bibinfo {year}
  {2014}{\natexlab{a}})},\ \Eprint {http://arxiv.org/abs/1401.7172}
  {arXiv:1401.7172 [gr-qc]} \BibitemShut {NoStop}%
\bibitem [{\citenamefont {Husain}\ and\ \citenamefont
  {Olivier}(2001)}]{Husain:2000vm}%
  \BibitemOpen
  \bibfield  {author} {\bibinfo {author} {\bibfnamefont {V.}~\bibnamefont
  {Husain}}\ and\ \bibinfo {author} {\bibfnamefont {M.}~\bibnamefont
  {Olivier}},\ }\bibfield  {title} {\enquote {\bibinfo {title} {{Scalar field
  collapse in three-dimensional AdS space-time}},}\ }\href {\doibase
  10.1088/0264-9381/18/2/101} {\bibfield  {journal} {\bibinfo  {journal}
  {Class. Quant. Grav.}\ }\textbf {\bibinfo {volume} {18}},\ \bibinfo {pages}
  {L1--L10} (\bibinfo {year} {2001})},\ \Eprint
  {http://arxiv.org/abs/gr-qc/0008060} {arXiv:gr-qc/0008060 [gr-qc]}
  \BibitemShut {NoStop}%
\bibitem [{\citenamefont {Pretorius}\ and\ \citenamefont
  {Choptuik}(2000)}]{Pretorius:2000yu}%
  \BibitemOpen
  \bibfield  {author} {\bibinfo {author} {\bibfnamefont {F.}~\bibnamefont
  {Pretorius}}\ and\ \bibinfo {author} {\bibfnamefont {M.~W.}\ \bibnamefont
  {Choptuik}},\ }\bibfield  {title} {\enquote {\bibinfo {title} {{Gravitational
  collapse in (2+1)-dimensional AdS space-time}},}\ }\href {\doibase
  10.1103/PhysRevD.62.124012} {\bibfield  {journal} {\bibinfo  {journal} {Phys.
  Rev.}\ }\textbf {\bibinfo {volume} {D62}},\ \bibinfo {pages} {124012}
  (\bibinfo {year} {2000})},\ \Eprint {http://arxiv.org/abs/gr-qc/0007008}
  {arXiv:gr-qc/0007008 [gr-qc]} \BibitemShut {NoStop}%
\bibitem [{\citenamefont {Garfinkle}(2001)}]{Garfinkle:2000br}%
  \BibitemOpen
  \bibfield  {author} {\bibinfo {author} {\bibfnamefont {D.}~\bibnamefont
  {Garfinkle}},\ }\bibfield  {title} {\enquote {\bibinfo {title} {{An Exact
  solution for 2+1 dimensional critical collapse}},}\ }\href {\doibase
  10.1103/PhysRevD.63.044007} {\bibfield  {journal} {\bibinfo  {journal} {Phys.
  Rev.}\ }\textbf {\bibinfo {volume} {D63}},\ \bibinfo {pages} {044007}
  (\bibinfo {year} {2001})},\ \Eprint {http://arxiv.org/abs/gr-qc/0008023}
  {arXiv:gr-qc/0008023 [gr-qc]} \BibitemShut {NoStop}%
\bibitem [{\citenamefont {Bizo\'n}\ and\ \citenamefont
  {Rostworowski}(2011)}]{Bizon:2011gg}%
  \BibitemOpen
  \bibfield  {author} {\bibinfo {author} {\bibfnamefont {P.}~\bibnamefont
  {Bizo\'n}}\ and\ \bibinfo {author} {\bibfnamefont {A.}~\bibnamefont
  {Rostworowski}},\ }\bibfield  {title} {\enquote {\bibinfo {title} {{On weakly
  turbulent instability of Anti-de Sitter space}},}\ }\href {\doibase
  10.1103/PhysRevLett.107.031102} {\bibfield  {journal} {\bibinfo  {journal}
  {Phys. Rev. Lett.}\ }\textbf {\bibinfo {volume} {107}},\ \bibinfo {pages}
  {031102} (\bibinfo {year} {2011})},\ \Eprint {http://arxiv.org/abs/1104.3702}
  {arXiv:1104.3702 [gr-qc]} \BibitemShut {NoStop}%
\bibitem [{\citenamefont {Ja\l{}mu\.zna}\ \emph {et~al.}(2011)\citenamefont
  {Ja\l{}mu\.zna}, \citenamefont {Rostworowski},\ and\ \citenamefont
  {Bizo\'n}}]{Jalmuzna:2011qw}%
  \BibitemOpen
  \bibfield  {author} {\bibinfo {author} {\bibfnamefont {J.}~\bibnamefont
  {Ja\l{}mu\.zna}}, \bibinfo {author} {\bibfnamefont {A.}~\bibnamefont
  {Rostworowski}}, \ and\ \bibinfo {author} {\bibfnamefont {P.}~\bibnamefont
  {Bizo\'n}},\ }\bibfield  {title} {\enquote {\bibinfo {title} {{A Comment on
  AdS collapse of a scalar field in higher dimensions}},}\ }\href {\doibase
  10.1103/PhysRevD.84.085021} {\bibfield  {journal} {\bibinfo  {journal} {Phys.
  Rev.}\ }\textbf {\bibinfo {volume} {D84}},\ \bibinfo {pages} {085021}
  (\bibinfo {year} {2011})},\ \Eprint {http://arxiv.org/abs/1108.4539}
  {arXiv:1108.4539 [gr-qc]} \BibitemShut {NoStop}%
\bibitem [{\citenamefont {Buchel}\ \emph {et~al.}(2012)\citenamefont {Buchel},
  \citenamefont {Lehner},\ and\ \citenamefont {Liebling}}]{Buchel:2012uh}%
  \BibitemOpen
  \bibfield  {author} {\bibinfo {author} {\bibfnamefont {A.}~\bibnamefont
  {Buchel}}, \bibinfo {author} {\bibfnamefont {L.}~\bibnamefont {Lehner}}, \
  and\ \bibinfo {author} {\bibfnamefont {S.~L.}\ \bibnamefont {Liebling}},\
  }\bibfield  {title} {\enquote {\bibinfo {title} {{Scalar Collapse in AdS}},}\
  }\href {\doibase 10.1103/PhysRevD.86.123011} {\bibfield  {journal} {\bibinfo
  {journal} {Phys. Rev.}\ }\textbf {\bibinfo {volume} {D86}},\ \bibinfo {pages}
  {123011} (\bibinfo {year} {2012})},\ \Eprint {http://arxiv.org/abs/1210.0890}
  {arXiv:1210.0890 [gr-qc]} \BibitemShut {NoStop}%
\bibitem [{\citenamefont {Maliborski}\ and\ \citenamefont
  {Rostworowski}(2013{\natexlab{a}})}]{Maliborski:2013via}%
  \BibitemOpen
  \bibfield  {author} {\bibinfo {author} {\bibfnamefont {M.}~\bibnamefont
  {Maliborski}}\ and\ \bibinfo {author} {\bibfnamefont {A.}~\bibnamefont
  {Rostworowski}},\ }\bibfield  {title} {\enquote {\bibinfo {title} {{Lecture
  Notes on Turbulent Instability of Anti-de Sitter Spacetime}},}\ }\bibfield
  {booktitle} {\emph {\bibinfo {booktitle} {{Proceedings, Spring School on
  Numerical Relativity and High Energy Physics (NR/HEP2)}}},\ }\href {\doibase
  10.1142/S0217751X13400204} {\bibfield  {journal} {\bibinfo  {journal} {Int.
  J. Mod. Phys.}\ }\textbf {\bibinfo {volume} {A28}},\ \bibinfo {pages}
  {1340020} (\bibinfo {year} {2013}{\natexlab{a}})},\ \Eprint
  {http://arxiv.org/abs/1308.1235} {arXiv:1308.1235 [gr-qc]} \BibitemShut
  {NoStop}%
\bibitem [{\citenamefont {Maliborski}(2012)}]{Maliborski:2012gx}%
  \BibitemOpen
  \bibfield  {author} {\bibinfo {author} {\bibfnamefont {M.}~\bibnamefont
  {Maliborski}},\ }\bibfield  {title} {\enquote {\bibinfo {title} {{Instability
  of Flat Space Enclosed in a Cavity}},}\ }\href {\doibase
  10.1103/PhysRevLett.109.221101} {\bibfield  {journal} {\bibinfo  {journal}
  {Phys. Rev. Lett.}\ }\textbf {\bibinfo {volume} {109}},\ \bibinfo {pages}
  {221101} (\bibinfo {year} {2012})},\ \Eprint {http://arxiv.org/abs/1208.2934}
  {arXiv:1208.2934 [gr-qc]} \BibitemShut {NoStop}%
\bibitem [{\citenamefont {Okawa}\ \emph
  {et~al.}(2014{\natexlab{a}})\citenamefont {Okawa}, \citenamefont {Cardoso},\
  and\ \citenamefont {Pani}}]{Okawa:2013jba}%
  \BibitemOpen
  \bibfield  {author} {\bibinfo {author} {\bibfnamefont {H.}~\bibnamefont
  {Okawa}}, \bibinfo {author} {\bibfnamefont {V.}~\bibnamefont {Cardoso}}, \
  and\ \bibinfo {author} {\bibfnamefont {P.}~\bibnamefont {Pani}},\ }\bibfield
  {title} {\enquote {\bibinfo {title} {{Collapse of self-interacting fields in
  asymptotically flat spacetimes: do self-interactions render Minkowski
  spacetime unstable?}}}\ }\href {\doibase 10.1103/PhysRevD.89.041502}
  {\bibfield  {journal} {\bibinfo  {journal} {Phys. Rev.}\ }\textbf {\bibinfo
  {volume} {D89}},\ \bibinfo {pages} {041502} (\bibinfo {year}
  {2014}{\natexlab{a}})},\ \Eprint {http://arxiv.org/abs/1311.1235}
  {arXiv:1311.1235 [gr-qc]} \BibitemShut {NoStop}%
\bibitem [{\citenamefont {Okawa}\ \emph
  {et~al.}(2014{\natexlab{b}})\citenamefont {Okawa}, \citenamefont {Cardoso},\
  and\ \citenamefont {Pani}}]{Okawa:2014nea}%
  \BibitemOpen
  \bibfield  {author} {\bibinfo {author} {\bibfnamefont {H.}~\bibnamefont
  {Okawa}}, \bibinfo {author} {\bibfnamefont {V.}~\bibnamefont {Cardoso}}, \
  and\ \bibinfo {author} {\bibfnamefont {P.}~\bibnamefont {Pani}},\ }\bibfield
  {title} {\enquote {\bibinfo {title} {{Study of the nonlinear instability of
  confined geometries}},}\ }\href {\doibase 10.1103/PhysRevD.90.104032}
  {\bibfield  {journal} {\bibinfo  {journal} {Phys. Rev.}\ }\textbf {\bibinfo
  {volume} {D90}},\ \bibinfo {pages} {104032} (\bibinfo {year}
  {2014}{\natexlab{b}})},\ \Eprint {http://arxiv.org/abs/1409.0533}
  {arXiv:1409.0533 [gr-qc]} \BibitemShut {NoStop}%
\bibitem [{\citenamefont {Cardoso}\ and\ \citenamefont
  {Rocha}(2016)}]{Cardoso:2016wcr}%
  \BibitemOpen
  \bibfield  {author} {\bibinfo {author} {\bibfnamefont {V.}~\bibnamefont
  {Cardoso}}\ and\ \bibinfo {author} {\bibfnamefont {J.~V.}\ \bibnamefont
  {Rocha}},\ }\bibfield  {title} {\enquote {\bibinfo {title} {{Collapsing
  shells, critical phenomena and black hole formation}},}\ }\href@noop {} {\
  (\bibinfo {year} {2016})},\ \Eprint {http://arxiv.org/abs/1601.07552}
  {arXiv:1601.07552 [gr-qc]} \BibitemShut {NoStop}%
\bibitem [{\citenamefont {Cai}\ and\ \citenamefont {Yang}(2016)}]{Cai:2016yxd}%
  \BibitemOpen
  \bibfield  {author} {\bibinfo {author} {\bibfnamefont {R.-G.}\ \bibnamefont
  {Cai}}\ and\ \bibinfo {author} {\bibfnamefont {R.-Q.}\ \bibnamefont {Yang}},\
  }\bibfield  {title} {\enquote {\bibinfo {title} {{Multiple critical
  gravitational collapse of charged scalar with reflecting wall}},}\
  }\href@noop {} {\  (\bibinfo {year} {2016})},\ \Eprint
  {http://arxiv.org/abs/1602.00112} {arXiv:1602.00112 [gr-qc]} \BibitemShut
  {NoStop}%
\bibitem [{\citenamefont {Husain}\ \emph {et~al.}(2003)\citenamefont {Husain},
  \citenamefont {Kunstatter}, \citenamefont {Preston},\ and\ \citenamefont
  {Birukou}}]{Husain:2002nk}%
  \BibitemOpen
  \bibfield  {author} {\bibinfo {author} {\bibfnamefont {V.}~\bibnamefont
  {Husain}}, \bibinfo {author} {\bibfnamefont {G.}~\bibnamefont {Kunstatter}},
  \bibinfo {author} {\bibfnamefont {B.}~\bibnamefont {Preston}}, \ and\
  \bibinfo {author} {\bibfnamefont {M.}~\bibnamefont {Birukou}},\ }\bibfield
  {title} {\enquote {\bibinfo {title} {{Anti-de Sitter gravitational
  collapse}},}\ }\href {\doibase 10.1088/0264-9381/20/4/101} {\bibfield
  {journal} {\bibinfo  {journal} {Class. Quant. Grav.}\ }\textbf {\bibinfo
  {volume} {20}},\ \bibinfo {pages} {L23--L30} (\bibinfo {year} {2003})},\
  \Eprint {http://arxiv.org/abs/gr-qc/0210011} {arXiv:gr-qc/0210011 [gr-qc]}
  \BibitemShut {NoStop}%
\bibitem [{\citenamefont {Buchel}\ \emph {et~al.}(2013)\citenamefont {Buchel},
  \citenamefont {Liebling},\ and\ \citenamefont {Lehner}}]{Buchel:2013uba}%
  \BibitemOpen
  \bibfield  {author} {\bibinfo {author} {\bibfnamefont {A.}~\bibnamefont
  {Buchel}}, \bibinfo {author} {\bibfnamefont {S.~L.}\ \bibnamefont
  {Liebling}}, \ and\ \bibinfo {author} {\bibfnamefont {L.}~\bibnamefont
  {Lehner}},\ }\bibfield  {title} {\enquote {\bibinfo {title} {{Boson stars in
  AdS spacetime}},}\ }\href {\doibase 10.1103/PhysRevD.87.123006} {\bibfield
  {journal} {\bibinfo  {journal} {Phys. Rev.}\ }\textbf {\bibinfo {volume}
  {D87}},\ \bibinfo {pages} {123006} (\bibinfo {year} {2013})},\ \Eprint
  {http://arxiv.org/abs/1304.4166} {arXiv:1304.4166 [gr-qc]} \BibitemShut
  {NoStop}%
\bibitem [{\citenamefont {Maliborski}\ and\ \citenamefont
  {Rostworowski}(2013{\natexlab{b}})}]{Maliborski:2013ula}%
  \BibitemOpen
  \bibfield  {author} {\bibinfo {author} {\bibfnamefont {M.}~\bibnamefont
  {Maliborski}}\ and\ \bibinfo {author} {\bibfnamefont {A.}~\bibnamefont
  {Rostworowski}},\ }\bibfield  {title} {\enquote {\bibinfo {title} {{A comment
  on "Boson stars in AdS"}},}\ }\href@noop {} {\  (\bibinfo {year}
  {2013}{\natexlab{b}})},\ \Eprint {http://arxiv.org/abs/1307.2875}
  {arXiv:1307.2875} \BibitemShut {NoStop}%
\bibitem [{\citenamefont {Fodor}\ \emph {et~al.}(2014)\citenamefont {Fodor},
  \citenamefont {Forg\'acs},\ and\ \citenamefont
  {Grandcl\'ement}}]{Fodor:2013lza}%
  \BibitemOpen
  \bibfield  {author} {\bibinfo {author} {\bibfnamefont {G.}~\bibnamefont
  {Fodor}}, \bibinfo {author} {\bibfnamefont {P.}~\bibnamefont {Forg\'acs}}, \
  and\ \bibinfo {author} {\bibfnamefont {P.}~\bibnamefont {Grandcl\'ement}},\
  }\bibfield  {title} {\enquote {\bibinfo {title} {{Scalar field breathers on
  Anti-de Sitter background}},}\ }\href {\doibase 10.1103/PhysRevD.89.065027}
  {\bibfield  {journal} {\bibinfo  {journal} {Phys. Rev.}\ }\textbf {\bibinfo
  {volume} {D89}},\ \bibinfo {pages} {065027} (\bibinfo {year} {2014})},\
  \Eprint {http://arxiv.org/abs/1312.7562} {arXiv:1312.7562 [hep-th]}
  \BibitemShut {NoStop}%
\bibitem [{\citenamefont {Maliborski}\ and\ \citenamefont
  {Rostworowski}(2014)}]{Maliborski:2014rma}%
  \BibitemOpen
  \bibfield  {author} {\bibinfo {author} {\bibfnamefont {M.}~\bibnamefont
  {Maliborski}}\ and\ \bibinfo {author} {\bibfnamefont {A.}~\bibnamefont
  {Rostworowski}},\ }\bibfield  {title} {\enquote {\bibinfo {title} {{What
  drives AdS unstable?}}}\ }\href {\doibase 10.1103/PhysRevD.89.124006}
  {\bibfield  {journal} {\bibinfo  {journal} {Phys. Rev.}\ }\textbf {\bibinfo
  {volume} {D89}},\ \bibinfo {pages} {124006} (\bibinfo {year} {2014})},\
  \Eprint {http://arxiv.org/abs/1403.5434} {arXiv:1403.5434 [gr-qc]}
  \BibitemShut {NoStop}%
\bibitem [{\citenamefont {Balasubramanian}\ \emph {et~al.}(2014)\citenamefont
  {Balasubramanian}, \citenamefont {Buchel}, \citenamefont {Green},
  \citenamefont {Lehner},\ and\ \citenamefont
  {Liebling}}]{Balasubramanian:2014cja}%
  \BibitemOpen
  \bibfield  {author} {\bibinfo {author} {\bibfnamefont {V.}~\bibnamefont
  {Balasubramanian}}, \bibinfo {author} {\bibfnamefont {A.}~\bibnamefont
  {Buchel}}, \bibinfo {author} {\bibfnamefont {S.~R.}\ \bibnamefont {Green}},
  \bibinfo {author} {\bibfnamefont {L.}~\bibnamefont {Lehner}}, \ and\ \bibinfo
  {author} {\bibfnamefont {S.~L.}\ \bibnamefont {Liebling}},\ }\bibfield
  {title} {\enquote {\bibinfo {title} {{Holographic Thermalization, Stability
  of Anti-de Sitter Space, and the Fermi-Pasta-Ulam Paradox}},}\ }\href
  {\doibase 10.1103/PhysRevLett.113.071601} {\bibfield  {journal} {\bibinfo
  {journal} {Phys. Rev. Lett.}\ }\textbf {\bibinfo {volume} {113}},\ \bibinfo
  {pages} {071601} (\bibinfo {year} {2014})},\ \Eprint
  {http://arxiv.org/abs/1403.6471} {arXiv:1403.6471 [hep-th]} \BibitemShut
  {NoStop}%
\bibitem [{\citenamefont {Craps}\ \emph {et~al.}(2014)\citenamefont {Craps},
  \citenamefont {Evnin},\ and\ \citenamefont {Vanhoof}}]{Craps:2014vaa}%
  \BibitemOpen
  \bibfield  {author} {\bibinfo {author} {\bibfnamefont {B.}~\bibnamefont
  {Craps}}, \bibinfo {author} {\bibfnamefont {O.}~\bibnamefont {Evnin}}, \ and\
  \bibinfo {author} {\bibfnamefont {J.}~\bibnamefont {Vanhoof}},\ }\bibfield
  {title} {\enquote {\bibinfo {title} {{Renormalization group, secular term
  resummation and AdS (in)stability}},}\ }\href {\doibase
  10.1007/JHEP10(2014)048} {\bibfield  {journal} {\bibinfo  {journal} {JHEP}\
  }\textbf {\bibinfo {volume} {10}},\ \bibinfo {pages} {48} (\bibinfo {year}
  {2014})},\ \Eprint {http://arxiv.org/abs/1407.6273} {arXiv:1407.6273 [gr-qc]}
  \BibitemShut {NoStop}%
\bibitem [{\citenamefont {Dimitrakopoulos}\ \emph {et~al.}(2014)\citenamefont
  {Dimitrakopoulos}, \citenamefont {Freivogel}, \citenamefont {Lippert},\ and\
  \citenamefont {Yang}}]{Dimitrakopoulos:2014ada}%
  \BibitemOpen
  \bibfield  {author} {\bibinfo {author} {\bibfnamefont {F.~V.}\ \bibnamefont
  {Dimitrakopoulos}}, \bibinfo {author} {\bibfnamefont {B.}~\bibnamefont
  {Freivogel}}, \bibinfo {author} {\bibfnamefont {M.}~\bibnamefont {Lippert}},
  \ and\ \bibinfo {author} {\bibfnamefont {I.-S.}\ \bibnamefont {Yang}},\
  }\bibfield  {title} {\enquote {\bibinfo {title} {{Instability corners in AdS
  space}},}\ }\href@noop {} {\  (\bibinfo {year} {2014})},\ \Eprint
  {http://arxiv.org/abs/1410.1880} {arXiv:1410.1880 [hep-th]} \BibitemShut
  {NoStop}%
\bibitem [{\citenamefont {Bizo\'n}\ and\ \citenamefont
  {Rostworowski}(2015)}]{Bizon:2014bya}%
  \BibitemOpen
  \bibfield  {author} {\bibinfo {author} {\bibfnamefont {P.}~\bibnamefont
  {Bizo\'n}}\ and\ \bibinfo {author} {\bibfnamefont {A.}~\bibnamefont
  {Rostworowski}},\ }\bibfield  {title} {\enquote {\bibinfo {title} {{Comment
  on ``Holographic Thermalization, Stability of Anti-de Sitter Space, and the
  Fermi-Pasta-Ulam Paradox''}},}\ }\href {\doibase
  10.1103/PhysRevLett.115.049101} {\bibfield  {journal} {\bibinfo  {journal}
  {Phys. Rev. Lett.}\ }\textbf {\bibinfo {volume} {115}},\ \bibinfo {pages}
  {049101} (\bibinfo {year} {2015})},\ \Eprint {http://arxiv.org/abs/1410.2631}
  {arXiv:1410.2631 [gr-qc]} \BibitemShut {NoStop}%
\bibitem [{\citenamefont {Craps}\ \emph
  {et~al.}(2015{\natexlab{a}})\citenamefont {Craps}, \citenamefont {Evnin},\
  and\ \citenamefont {Vanhoof}}]{Craps:2014jwa}%
  \BibitemOpen
  \bibfield  {author} {\bibinfo {author} {\bibfnamefont {B.}~\bibnamefont
  {Craps}}, \bibinfo {author} {\bibfnamefont {O.}~\bibnamefont {Evnin}}, \ and\
  \bibinfo {author} {\bibfnamefont {J.}~\bibnamefont {Vanhoof}},\ }\bibfield
  {title} {\enquote {\bibinfo {title} {{Renormalization, averaging,
  conservation laws and AdS (in)stability}},}\ }\href {\doibase
  10.1007/JHEP01(2015)108} {\bibfield  {journal} {\bibinfo  {journal} {JHEP}\
  }\textbf {\bibinfo {volume} {01}},\ \bibinfo {pages} {108} (\bibinfo {year}
  {2015}{\natexlab{a}})},\ \Eprint {http://arxiv.org/abs/1412.3249}
  {arXiv:1412.3249 [gr-qc]} \BibitemShut {NoStop}%
\bibitem [{\citenamefont {Buchel}\ \emph {et~al.}(2015)\citenamefont {Buchel},
  \citenamefont {Green}, \citenamefont {Lehner},\ and\ \citenamefont
  {Liebling}}]{Buchel:2014xwa}%
  \BibitemOpen
  \bibfield  {author} {\bibinfo {author} {\bibfnamefont {A.}~\bibnamefont
  {Buchel}}, \bibinfo {author} {\bibfnamefont {S.~R.}\ \bibnamefont {Green}},
  \bibinfo {author} {\bibfnamefont {L.}~\bibnamefont {Lehner}}, \ and\ \bibinfo
  {author} {\bibfnamefont {S.~L.}\ \bibnamefont {Liebling}},\ }\bibfield
  {title} {\enquote {\bibinfo {title} {{Conserved quantities and dual turbulent
  cascades in Anti-de Sitter spacetime}},}\ }\href {\doibase
  10.1103/PhysRevD.91.064026} {\bibfield  {journal} {\bibinfo  {journal} {Phys.
  Rev.}\ }\textbf {\bibinfo {volume} {D91}},\ \bibinfo {pages} {064026}
  (\bibinfo {year} {2015})},\ \Eprint {http://arxiv.org/abs/1412.4761}
  {arXiv:1412.4761 [gr-qc]} \BibitemShut {NoStop}%
\bibitem [{\citenamefont {Bizo\'n}\ \emph {et~al.}(2015)\citenamefont
  {Bizo\'n}, \citenamefont {Maliborski},\ and\ \citenamefont
  {Rostworowski}}]{Bizon:2015pfa}%
  \BibitemOpen
  \bibfield  {author} {\bibinfo {author} {\bibfnamefont {P.}~\bibnamefont
  {Bizo\'n}}, \bibinfo {author} {\bibfnamefont {M.}~\bibnamefont {Maliborski}},
  \ and\ \bibinfo {author} {\bibfnamefont {A.}~\bibnamefont {Rostworowski}},\
  }\bibfield  {title} {\enquote {\bibinfo {title} {{Resonant dynamics and the
  instability of Anti-de Sitter spacetime}},}\ }\href@noop {} {\  (\bibinfo
  {year} {2015})},\ \Eprint {http://arxiv.org/abs/1506.03519} {arXiv:1506.03519
  [gr-qc]} \BibitemShut {NoStop}%
\bibitem [{\citenamefont {Balasubramanian}\ \emph {et~al.}(2015)\citenamefont
  {Balasubramanian}, \citenamefont {Buchel}, \citenamefont {Green},
  \citenamefont {Lehner},\ and\ \citenamefont
  {Liebling}}]{Balasubramanian:2015uua}%
  \BibitemOpen
  \bibfield  {author} {\bibinfo {author} {\bibfnamefont {V.}~\bibnamefont
  {Balasubramanian}}, \bibinfo {author} {\bibfnamefont {A.}~\bibnamefont
  {Buchel}}, \bibinfo {author} {\bibfnamefont {S.~R.}\ \bibnamefont {Green}},
  \bibinfo {author} {\bibfnamefont {L.}~\bibnamefont {Lehner}}, \ and\ \bibinfo
  {author} {\bibfnamefont {S.~L.}\ \bibnamefont {Liebling}},\ }\bibfield
  {title} {\enquote {\bibinfo {title} {{Reply to Comment on ``Holographic
  Thermalization, Stability of Anti-de Sitter Space, and the Fermi-Pasta-Ulam
  Paradox''}},}\ }\href {\doibase 10.1103/PhysRevLett.115.049102} {\bibfield
  {journal} {\bibinfo  {journal} {Phys. Rev. Lett.}\ }\textbf {\bibinfo
  {volume} {115}},\ \bibinfo {pages} {049102} (\bibinfo {year} {2015})},\
  \Eprint {http://arxiv.org/abs/1506.07907} {arXiv:1506.07907 [gr-qc]}
  \BibitemShut {NoStop}%
\bibitem [{\citenamefont {Dimitrakopoulos}\ and\ \citenamefont
  {Yang}(2015)}]{Dimitrakopoulos:2015pwa}%
  \BibitemOpen
  \bibfield  {author} {\bibinfo {author} {\bibfnamefont {F.}~\bibnamefont
  {Dimitrakopoulos}}\ and\ \bibinfo {author} {\bibfnamefont {I.-S.}\
  \bibnamefont {Yang}},\ }\bibfield  {title} {\enquote {\bibinfo {title}
  {{Occasionally Extended Validity of Perturbation Theory: Persistence of AdS
  Stability Islands}},}\ }\href@noop {} {\  (\bibinfo {year} {2015})},\ \Eprint
  {http://arxiv.org/abs/1507.02684} {arXiv:1507.02684 [hep-th]} \BibitemShut
  {NoStop}%
\bibitem [{\citenamefont {Green}\ \emph {et~al.}(2015)\citenamefont {Green},
  \citenamefont {Maillard}, \citenamefont {Lehner},\ and\ \citenamefont
  {Liebling}}]{Green:2015dsa}%
  \BibitemOpen
  \bibfield  {author} {\bibinfo {author} {\bibfnamefont {S.~R.}\ \bibnamefont
  {Green}}, \bibinfo {author} {\bibfnamefont {A.}~\bibnamefont {Maillard}},
  \bibinfo {author} {\bibfnamefont {L.}~\bibnamefont {Lehner}}, \ and\ \bibinfo
  {author} {\bibfnamefont {S.~L.}\ \bibnamefont {Liebling}},\ }\bibfield
  {title} {\enquote {\bibinfo {title} {{Islands of stability and recurrence
  times in AdS}},}\ }\href@noop {} {\  (\bibinfo {year} {2015})},\ \Eprint
  {http://arxiv.org/abs/1507.08261} {arXiv:1507.08261 [gr-qc]} \BibitemShut
  {NoStop}%
\bibitem [{\citenamefont {Deppe}\ and\ \citenamefont
  {Frey}(2015)}]{Deppe:2015qsa}%
  \BibitemOpen
  \bibfield  {author} {\bibinfo {author} {\bibfnamefont {N.}~\bibnamefont
  {Deppe}}\ and\ \bibinfo {author} {\bibfnamefont {A.~R.}\ \bibnamefont
  {Frey}},\ }\bibfield  {title} {\enquote {\bibinfo {title} {{Classes of Stable
  Initial Data for Massless and Massive Scalars in Anti-de Sitter
  Spacetime}},}\ }\href@noop {} {\  (\bibinfo {year} {2015})},\ \Eprint
  {http://arxiv.org/abs/1508.02709} {arXiv:1508.02709 [hep-th]} \BibitemShut
  {NoStop}%
\bibitem [{\citenamefont {Craps}\ \emph
  {et~al.}(2015{\natexlab{b}})\citenamefont {Craps}, \citenamefont {Evnin},\
  and\ \citenamefont {Vanhoof}}]{Craps:2015iia}%
  \BibitemOpen
  \bibfield  {author} {\bibinfo {author} {\bibfnamefont {B.}~\bibnamefont
  {Craps}}, \bibinfo {author} {\bibfnamefont {O.}~\bibnamefont {Evnin}}, \ and\
  \bibinfo {author} {\bibfnamefont {J.}~\bibnamefont {Vanhoof}},\ }\bibfield
  {title} {\enquote {\bibinfo {title} {{Ultraviolet asymptotics and singular
  dynamics of AdS perturbations}},}\ }\href@noop {} {\  (\bibinfo {year}
  {2015}{\natexlab{b}})},\ \Eprint {http://arxiv.org/abs/1508.04943}
  {arXiv:1508.04943 [gr-qc]} \BibitemShut {NoStop}%
\bibitem [{\citenamefont {Craps}\ \emph
  {et~al.}(2015{\natexlab{c}})\citenamefont {Craps}, \citenamefont {Evnin},
  \citenamefont {Jai-akson},\ and\ \citenamefont {Vanhoof}}]{Craps:2015xya}%
  \BibitemOpen
  \bibfield  {author} {\bibinfo {author} {\bibfnamefont {B.}~\bibnamefont
  {Craps}}, \bibinfo {author} {\bibfnamefont {O.}~\bibnamefont {Evnin}},
  \bibinfo {author} {\bibfnamefont {P.}~\bibnamefont {Jai-akson}}, \ and\
  \bibinfo {author} {\bibfnamefont {J.}~\bibnamefont {Vanhoof}},\ }\bibfield
  {title} {\enquote {\bibinfo {title} {{Ultraviolet asymptotics for
  quasiperiodic AdS\_4 perturbations}},}\ }\href@noop {} {\  (\bibinfo {year}
  {2015}{\natexlab{c}})},\ \Eprint {http://arxiv.org/abs/1508.05474}
  {arXiv:1508.05474 [gr-qc]} \BibitemShut {NoStop}%
\bibitem [{\citenamefont {Menon}\ and\ \citenamefont
  {Suneeta}(2015)}]{Menon:2015oda}%
  \BibitemOpen
  \bibfield  {author} {\bibinfo {author} {\bibfnamefont {D.~S.}\ \bibnamefont
  {Menon}}\ and\ \bibinfo {author} {\bibfnamefont {V.}~\bibnamefont
  {Suneeta}},\ }\bibfield  {title} {\enquote {\bibinfo {title} {{Necessary
  conditions for an AdS-type instability}},}\ }\href@noop {} {\  (\bibinfo
  {year} {2015})},\ \Eprint {http://arxiv.org/abs/1509.00232} {arXiv:1509.00232
  [gr-qc]} \BibitemShut {NoStop}%
\bibitem [{\citenamefont {Abajo-Arrastia}\ \emph {et~al.}(2014)\citenamefont
  {Abajo-Arrastia}, \citenamefont {da~Silva}, \citenamefont {Lopez},
  \citenamefont {Mas},\ and\ \citenamefont
  {Serantes}}]{Abajo-Arrastia:2014fma}%
  \BibitemOpen
  \bibfield  {author} {\bibinfo {author} {\bibfnamefont {J.}~\bibnamefont
  {Abajo-Arrastia}}, \bibinfo {author} {\bibfnamefont {E.}~\bibnamefont
  {da~Silva}}, \bibinfo {author} {\bibfnamefont {E.}~\bibnamefont {Lopez}},
  \bibinfo {author} {\bibfnamefont {J.}~\bibnamefont {Mas}}, \ and\ \bibinfo
  {author} {\bibfnamefont {A.}~\bibnamefont {Serantes}},\ }\bibfield  {title}
  {\enquote {\bibinfo {title} {{Holographic Relaxation of Finite Size Isolated
  Quantum Systems}},}\ }\href {\doibase 10.1007/JHEP05(2014)126} {\bibfield
  {journal} {\bibinfo  {journal} {JHEP}\ }\textbf {\bibinfo {volume} {05}},\
  \bibinfo {pages} {126} (\bibinfo {year} {2014})},\ \Eprint
  {http://arxiv.org/abs/1403.2632} {arXiv:1403.2632 [hep-th]} \BibitemShut
  {NoStop}%
\bibitem [{\citenamefont {da~Silva}\ \emph {et~al.}(2015)\citenamefont
  {da~Silva}, \citenamefont {Lopez}, \citenamefont {Mas},\ and\ \citenamefont
  {Serantes}}]{daSilva:2014zva}%
  \BibitemOpen
  \bibfield  {author} {\bibinfo {author} {\bibfnamefont {E.}~\bibnamefont
  {da~Silva}}, \bibinfo {author} {\bibfnamefont {E.}~\bibnamefont {Lopez}},
  \bibinfo {author} {\bibfnamefont {J.}~\bibnamefont {Mas}}, \ and\ \bibinfo
  {author} {\bibfnamefont {A.}~\bibnamefont {Serantes}},\ }\bibfield  {title}
  {\enquote {\bibinfo {title} {{Collapse and Revival in Holographic
  Quenches}},}\ }\href {\doibase 10.1007/JHEP04(2015)038} {\bibfield  {journal}
  {\bibinfo  {journal} {JHEP}\ }\textbf {\bibinfo {volume} {04}},\ \bibinfo
  {pages} {038} (\bibinfo {year} {2015})},\ \Eprint
  {http://arxiv.org/abs/1412.6002} {arXiv:1412.6002 [hep-th]} \BibitemShut
  {NoStop}%
\bibitem [{\citenamefont {Ishibashi}\ and\ \citenamefont
  {Maeda}(2012)}]{Ishibashi:2012xk}%
  \BibitemOpen
  \bibfield  {author} {\bibinfo {author} {\bibfnamefont {A.}~\bibnamefont
  {Ishibashi}}\ and\ \bibinfo {author} {\bibfnamefont {K.}~\bibnamefont
  {Maeda}},\ }\bibfield  {title} {\enquote {\bibinfo {title} {{Singularities in
  asymptotically Anti-de Sitter spacetimes}},}\ }\href {\doibase
  10.1103/PhysRevD.86.104012} {\bibfield  {journal} {\bibinfo  {journal} {Phys.
  Rev.}\ }\textbf {\bibinfo {volume} {D86}},\ \bibinfo {pages} {104012}
  (\bibinfo {year} {2012})},\ \Eprint {http://arxiv.org/abs/1208.1563}
  {arXiv:1208.1563 [hep-th]} \BibitemShut {NoStop}%
\bibitem [{\citenamefont {Friedrich}(2014{\natexlab{b}})}]{Friedrich:2014rpa}%
  \BibitemOpen
  \bibfield  {author} {\bibinfo {author} {\bibfnamefont {H.}~\bibnamefont
  {Friedrich}},\ }\bibfield  {title} {\enquote {\bibinfo {title} {{Geometric
  Asymptotics and Beyond}},}\ }\href@noop {} {\  (\bibinfo {year}
  {2014}{\natexlab{b}})},\ \Eprint {http://arxiv.org/abs/1411.3854}
  {arXiv:1411.3854 [gr-qc]} \BibitemShut {NoStop}%
\bibitem [{\citenamefont {Holzegel}\ \emph {et~al.}(2015)\citenamefont
  {Holzegel}, \citenamefont {Luk}, \citenamefont {Smulevici},\ and\
  \citenamefont {Warnick}}]{Holzegel:2015swa}%
  \BibitemOpen
  \bibfield  {author} {\bibinfo {author} {\bibfnamefont {G.}~\bibnamefont
  {Holzegel}}, \bibinfo {author} {\bibfnamefont {J.}~\bibnamefont {Luk}},
  \bibinfo {author} {\bibfnamefont {J.}~\bibnamefont {Smulevici}}, \ and\
  \bibinfo {author} {\bibfnamefont {C.}~\bibnamefont {Warnick}},\ }\bibfield
  {title} {\enquote {\bibinfo {title} {{Asymptotic properties of linear field
  equations in Anti-de Sitter space}},}\ }\href@noop {} {\  (\bibinfo {year}
  {2015})},\ \Eprint {http://arxiv.org/abs/1502.04965} {arXiv:1502.04965
  [gr-qc]} \BibitemShut {NoStop}%
\bibitem [{\citenamefont {Holzegel}\ and\ \citenamefont
  {Shao}(2015)}]{Holzegel:2015bna}%
  \BibitemOpen
  \bibfield  {author} {\bibinfo {author} {\bibfnamefont {G.}~\bibnamefont
  {Holzegel}}\ and\ \bibinfo {author} {\bibfnamefont {A.}~\bibnamefont
  {Shao}},\ }\bibfield  {title} {\enquote {\bibinfo {title} {{Unique
  continuation from infinity in asymptotically Anti-de Sitter spacetimes}},}\
  }\href@noop {} {\  (\bibinfo {year} {2015})},\ \Eprint
  {http://arxiv.org/abs/1508.03820} {arXiv:1508.03820 [gr-qc]} \BibitemShut
  {NoStop}%
\bibitem [{\citenamefont {Kim}(2015)}]{Kim:2014ida}%
  \BibitemOpen
  \bibfield  {author} {\bibinfo {author} {\bibfnamefont {N.}~\bibnamefont
  {Kim}},\ }\bibfield  {title} {\enquote {\bibinfo {title} {{Time-periodic
  solutions of massive scalar fields in dynamical AdS background: Perturbative
  constructions}},}\ }\href {\doibase 10.1016/j.physletb.2015.01.045}
  {\bibfield  {journal} {\bibinfo  {journal} {Phys. Lett.}\ }\textbf {\bibinfo
  {volume} {B742}},\ \bibinfo {pages} {274--278} (\bibinfo {year} {2015})},\
  \Eprint {http://arxiv.org/abs/1411.1633} {arXiv:1411.1633 [hep-th]}
  \BibitemShut {NoStop}%
\bibitem [{\citenamefont {Okawa}\ \emph {et~al.}(2015)\citenamefont {Okawa},
  \citenamefont {Lopes},\ and\ \citenamefont {Cardoso}}]{Okawa:2015xma}%
  \BibitemOpen
  \bibfield  {author} {\bibinfo {author} {\bibfnamefont {H.}~\bibnamefont
  {Okawa}}, \bibinfo {author} {\bibfnamefont {J.~C.}\ \bibnamefont {Lopes}}, \
  and\ \bibinfo {author} {\bibfnamefont {V.}~\bibnamefont {Cardoso}},\
  }\bibfield  {title} {\enquote {\bibinfo {title} {{Collapse of massive fields
  in Anti-de Sitter spacetime}},}\ }\href@noop {} {\  (\bibinfo {year}
  {2015})},\ \Eprint {http://arxiv.org/abs/1504.05203} {arXiv:1504.05203
  [gr-qc]} \BibitemShut {NoStop}%
\bibitem [{\citenamefont {Yang}(2015)}]{Yang:2015jha}%
  \BibitemOpen
  \bibfield  {author} {\bibinfo {author} {\bibfnamefont {I.-S.}\ \bibnamefont
  {Yang}},\ }\bibfield  {title} {\enquote {\bibinfo {title} {{Missing top of
  the AdS resonance structure}},}\ }\href {\doibase 10.1103/PhysRevD.91.065011}
  {\bibfield  {journal} {\bibinfo  {journal} {Phys. Rev.}\ }\textbf {\bibinfo
  {volume} {D91}},\ \bibinfo {pages} {065011} (\bibinfo {year} {2015})},\
  \Eprint {http://arxiv.org/abs/1501.00998} {arXiv:1501.00998 [hep-th]}
  \BibitemShut {NoStop}%
\bibitem [{\citenamefont {Basu}\ \emph {et~al.}(2015)\citenamefont {Basu},
  \citenamefont {Krishnan},\ and\ \citenamefont
  {Bala~Subramanian}}]{Basu:2015efa}%
  \BibitemOpen
  \bibfield  {author} {\bibinfo {author} {\bibfnamefont {P.}~\bibnamefont
  {Basu}}, \bibinfo {author} {\bibfnamefont {C.}~\bibnamefont {Krishnan}}, \
  and\ \bibinfo {author} {\bibfnamefont {P.~N.}\ \bibnamefont
  {Bala~Subramanian}},\ }\bibfield  {title} {\enquote {\bibinfo {title} {{AdS
  (In)stability: Lessons From The Scalar Field}},}\ }\href {\doibase
  10.1016/j.physletb.2015.05.009} {\bibfield  {journal} {\bibinfo  {journal}
  {Phys. Lett.}\ }\textbf {\bibinfo {volume} {B746}},\ \bibinfo {pages}
  {261--265} (\bibinfo {year} {2015})},\ \Eprint
  {http://arxiv.org/abs/1501.07499} {arXiv:1501.07499 [hep-th]} \BibitemShut
  {NoStop}%
\bibitem [{\citenamefont {Evnin}\ and\ \citenamefont
  {Krishnan}(2015)}]{Evnin:2015gma}%
  \BibitemOpen
  \bibfield  {author} {\bibinfo {author} {\bibfnamefont {O.}~\bibnamefont
  {Evnin}}\ and\ \bibinfo {author} {\bibfnamefont {C.}~\bibnamefont
  {Krishnan}},\ }\bibfield  {title} {\enquote {\bibinfo {title} {{A Hidden
  Symmetry of AdS Resonances}},}\ }\href {\doibase 10.1103/PhysRevD.91.126010}
  {\bibfield  {journal} {\bibinfo  {journal} {Phys. Rev.}\ }\textbf {\bibinfo
  {volume} {D91}},\ \bibinfo {pages} {126010} (\bibinfo {year} {2015})},\
  \Eprint {http://arxiv.org/abs/1502.03749} {arXiv:1502.03749 [hep-th]}
  \BibitemShut {NoStop}%
\bibitem [{\citenamefont {Cai}\ \emph {et~al.}(2015)\citenamefont {Cai},
  \citenamefont {Ji},\ and\ \citenamefont {Yang}}]{Cai:2015jbs}%
  \BibitemOpen
  \bibfield  {author} {\bibinfo {author} {\bibfnamefont {R.-G.}\ \bibnamefont
  {Cai}}, \bibinfo {author} {\bibfnamefont {L.-W.}\ \bibnamefont {Ji}}, \ and\
  \bibinfo {author} {\bibfnamefont {R.-Q.}\ \bibnamefont {Yang}},\ }\bibfield
  {title} {\enquote {\bibinfo {title} {{Collapse of self-interacting scalar
  field in Anti-de Sitter space}},}\ }\href@noop {} {\  (\bibinfo {year}
  {2015})},\ \Eprint {http://arxiv.org/abs/1511.00868} {arXiv:1511.00868
  [gr-qc]} \BibitemShut {NoStop}%
\bibitem [{\citenamefont {Deppe}\ \emph {et~al.}(2015)\citenamefont {Deppe},
  \citenamefont {Kolly}, \citenamefont {Frey},\ and\ \citenamefont
  {Kunstatter}}]{Deppe:2014oua}%
  \BibitemOpen
  \bibfield  {author} {\bibinfo {author} {\bibfnamefont {N.}~\bibnamefont
  {Deppe}}, \bibinfo {author} {\bibfnamefont {A.}~\bibnamefont {Kolly}},
  \bibinfo {author} {\bibfnamefont {A.}~\bibnamefont {Frey}}, \ and\ \bibinfo
  {author} {\bibfnamefont {G.}~\bibnamefont {Kunstatter}},\ }\bibfield  {title}
  {\enquote {\bibinfo {title} {{Stability of AdS in Einstein Gauss Bonnet
  Gravity}},}\ }\href {\doibase 10.1103/PhysRevLett.114.071102} {\bibfield
  {journal} {\bibinfo  {journal} {Phys. Rev. Lett.}\ }\textbf {\bibinfo
  {volume} {114}},\ \bibinfo {pages} {071102} (\bibinfo {year} {2015})},\
  \Eprint {http://arxiv.org/abs/1410.1869} {arXiv:1410.1869 [hep-th]}
  \BibitemShut {NoStop}%
\bibitem [{\citenamefont {Golod}\ and\ \citenamefont
  {Piran}(2012)}]{Golod:2012yt}%
  \BibitemOpen
  \bibfield  {author} {\bibinfo {author} {\bibfnamefont {S.}~\bibnamefont
  {Golod}}\ and\ \bibinfo {author} {\bibfnamefont {T.}~\bibnamefont {Piran}},\
  }\bibfield  {title} {\enquote {\bibinfo {title} {{Choptuik's Critical
  Phenomenon in Einstein-Gauss-Bonnet Gravity}},}\ }\href {\doibase
  10.1103/PhysRevD.85.104015} {\bibfield  {journal} {\bibinfo  {journal} {Phys.
  Rev.}\ }\textbf {\bibinfo {volume} {D85}},\ \bibinfo {pages} {104015}
  (\bibinfo {year} {2012})},\ \Eprint {http://arxiv.org/abs/1201.6384}
  {arXiv:1201.6384 [gr-qc]} \BibitemShut {NoStop}%
\bibitem [{\citenamefont {Dias}\ \emph
  {et~al.}(2012{\natexlab{a}})\citenamefont {Dias}, \citenamefont {Horowitz},\
  and\ \citenamefont {Santos}}]{Dias:2011ss}%
  \BibitemOpen
  \bibfield  {author} {\bibinfo {author} {\bibfnamefont {O.~J.~C.}\
  \bibnamefont {Dias}}, \bibinfo {author} {\bibfnamefont {G.~T.}\ \bibnamefont
  {Horowitz}}, \ and\ \bibinfo {author} {\bibfnamefont {J.~E.}\ \bibnamefont
  {Santos}},\ }\bibfield  {title} {\enquote {\bibinfo {title} {{Gravitational
  Turbulent Instability of Anti-de Sitter Space}},}\ }\href {\doibase
  10.1088/0264-9381/29/19/194002} {\bibfield  {journal} {\bibinfo  {journal}
  {Class. Quant. Grav.}\ }\textbf {\bibinfo {volume} {29}},\ \bibinfo {pages}
  {194002} (\bibinfo {year} {2012}{\natexlab{a}})},\ \Eprint
  {http://arxiv.org/abs/1109.1825} {arXiv:1109.1825 [hep-th]} \BibitemShut
  {NoStop}%
\bibitem [{\citenamefont {Dias}\ \emph
  {et~al.}(2012{\natexlab{b}})\citenamefont {Dias}, \citenamefont {Horowitz},
  \citenamefont {Marolf},\ and\ \citenamefont {Santos}}]{Dias:2012tq}%
  \BibitemOpen
  \bibfield  {author} {\bibinfo {author} {\bibfnamefont {O.~J.~C.}\
  \bibnamefont {Dias}}, \bibinfo {author} {\bibfnamefont {G.~T.}\ \bibnamefont
  {Horowitz}}, \bibinfo {author} {\bibfnamefont {D.}~\bibnamefont {Marolf}}, \
  and\ \bibinfo {author} {\bibfnamefont {J.~E.}\ \bibnamefont {Santos}},\
  }\bibfield  {title} {\enquote {\bibinfo {title} {{On the Nonlinear Stability
  of Asymptotically Anti-de Sitter Solutions}},}\ }\href {\doibase
  10.1088/0264-9381/29/23/235019} {\bibfield  {journal} {\bibinfo  {journal}
  {Class. Quant. Grav.}\ }\textbf {\bibinfo {volume} {29}},\ \bibinfo {pages}
  {235019} (\bibinfo {year} {2012}{\natexlab{b}})},\ \Eprint
  {http://arxiv.org/abs/1208.5772} {arXiv:1208.5772 [gr-qc]} \BibitemShut
  {NoStop}%
\bibitem [{\citenamefont {Horowitz}\ and\ \citenamefont
  {Santos}(2014)}]{Horowitz:2014hja}%
  \BibitemOpen
  \bibfield  {author} {\bibinfo {author} {\bibfnamefont {G.~T.}\ \bibnamefont
  {Horowitz}}\ and\ \bibinfo {author} {\bibfnamefont {J.~E.}\ \bibnamefont
  {Santos}},\ }\bibfield  {title} {\enquote {\bibinfo {title} {{Geons and the
  Instability of Anti-de Sitter Spacetime}},}\ }\href@noop {} {\  (\bibinfo
  {year} {2014})},\ \Eprint {http://arxiv.org/abs/1408.5906} {arXiv:1408.5906
  [gr-qc]} \BibitemShut {NoStop}%
\bibitem [{\citenamefont {Bizo\'n}(2014)}]{Bizon:2013gxa}%
  \BibitemOpen
  \bibfield  {author} {\bibinfo {author} {\bibfnamefont {P.}~\bibnamefont
  {Bizo\'n}},\ }\bibfield  {title} {\enquote {\bibinfo {title} {{Is AdS
  stable?}}}\ }\bibfield  {booktitle} {\emph {\bibinfo {booktitle}
  {{Proceedings, 20th International Conference on General Relativity and
  Gravitation and 10th Amaldi Conference on Gravitational Waves (GR20 /
  Amaldi10): The First Century of General Relativity}}},\ }\href {\doibase
  10.1007/s10714-014-1724-0} {\bibfield  {journal} {\bibinfo  {journal} {Gen.
  Rel. Grav.}\ }\textbf {\bibinfo {volume} {46}},\ \bibinfo {pages} {1724}
  (\bibinfo {year} {2014})},\ \Eprint {http://arxiv.org/abs/1312.5544}
  {arXiv:1312.5544 [gr-qc]} \BibitemShut {NoStop}%
\bibitem [{\citenamefont {Santos-Oliv\'an}\ and\ \citenamefont
  {Sopuerta}(2016)}]{SantosOlivan:2015fmy}%
  \BibitemOpen
  \bibfield  {author} {\bibinfo {author} {\bibfnamefont {D.}~\bibnamefont
  {Santos-Oliv\'an}}\ and\ \bibinfo {author} {\bibfnamefont {C.~F.}\
  \bibnamefont {Sopuerta}},\ }\bibfield  {title} {\enquote {\bibinfo {title}
  {{New Features of Gravitational Collapse in Anti-de Sitter Spacetimes}},}\
  }\href {\doibase 10.1103/PhysRevLett.116.041101} {\bibfield  {journal}
  {\bibinfo  {journal} {Phys. Rev. Lett.}\ }\textbf {\bibinfo {volume} {116}},\
  \bibinfo {pages} {041101} (\bibinfo {year} {2016})},\ \Eprint
  {http://arxiv.org/abs/1511.04344} {arXiv:1511.04344 [gr-qc]} \BibitemShut
  {NoStop}%
\bibitem [{\citenamefont {Courant}\ and\ \citenamefont
  {Hilbert}(1989)}]{Courant:1989aa}%
  \BibitemOpen
  \bibfield  {author} {\bibinfo {author} {\bibfnamefont {R.}~\bibnamefont
  {Courant}}\ and\ \bibinfo {author} {\bibfnamefont {D.}~\bibnamefont
  {Hilbert}},\ }\href@noop {} {\emph {\bibinfo {title} {{Methods of
  Mathematical Physics Volume II}}}}\ (\bibinfo  {publisher} {John Wiley and
  Sons},\ \bibinfo {year} {1989})\BibitemShut {NoStop}%
\bibitem [{\citenamefont {Arnowitt}\ \emph {et~al.}(1962)\citenamefont
  {Arnowitt}, \citenamefont {Deser},\ and\ \citenamefont
  {Misner}}]{adm:1962ok}%
  \BibitemOpen
  \bibfield  {author} {\bibinfo {author} {\bibfnamefont {R.}~\bibnamefont
  {Arnowitt}}, \bibinfo {author} {\bibfnamefont {S.}~\bibnamefont {Deser}}, \
  and\ \bibinfo {author} {\bibfnamefont {C.~W.}\ \bibnamefont {Misner}},\
  }\bibfield  {title} {\enquote {\bibinfo {title} {The dynamics of general
  relativity},}\ }in\ \href@noop {} {\emph {\bibinfo {booktitle} {Gravitation:
  An introduction to current research}}},\ \bibinfo {editor} {edited by\
  \bibinfo {editor} {\bibfnamefont {L.}~\bibnamefont {Witten}}}\ (\bibinfo
  {publisher} {Wiley},\ \bibinfo {address} {New York},\ \bibinfo {year}
  {1962})\ pp.\ \bibinfo {pages} {227--265}\BibitemShut {NoStop}%
\bibitem [{\citenamefont {Bland}\ \emph {et~al.}(2005)\citenamefont {Bland},
  \citenamefont {Preston}, \citenamefont {Becker}, \citenamefont {Kunstatter},\
  and\ \citenamefont {Husain}}]{Bland:2005kk}%
  \BibitemOpen
  \bibfield  {author} {\bibinfo {author} {\bibfnamefont {J.}~\bibnamefont
  {Bland}}, \bibinfo {author} {\bibfnamefont {B.}~\bibnamefont {Preston}},
  \bibinfo {author} {\bibfnamefont {M.}~\bibnamefont {Becker}}, \bibinfo
  {author} {\bibfnamefont {G.}~\bibnamefont {Kunstatter}}, \ and\ \bibinfo
  {author} {\bibfnamefont {V.}~\bibnamefont {Husain}},\ }\bibfield  {title}
  {\enquote {\bibinfo {title} {{Dimension dependence of the critical exponent
  in spherically symmetric gravitational collapse}},}\ }\href {\doibase
  10.1088/0264-9381/22/24/009} {\bibfield  {journal} {\bibinfo  {journal}
  {Class. Quant. Grav.}\ }\textbf {\bibinfo {volume} {22}},\ \bibinfo {pages}
  {5355--5364} (\bibinfo {year} {2005})}\BibitemShut {NoStop}%
\bibitem [{\citenamefont {Bland}\ and\ \citenamefont
  {Kunstatter}(2007)}]{Bland:2007sg}%
  \BibitemOpen
  \bibfield  {author} {\bibinfo {author} {\bibfnamefont {J.}~\bibnamefont
  {Bland}}\ and\ \bibinfo {author} {\bibfnamefont {G.}~\bibnamefont
  {Kunstatter}},\ }\bibfield  {title} {\enquote {\bibinfo {title} {{The 5-D
  Choptuik critical exponent and holography}},}\ }\href {\doibase
  10.1103/PhysRevD.75.101501} {\bibfield  {journal} {\bibinfo  {journal} {Phys.
  Rev.}\ }\textbf {\bibinfo {volume} {D75}},\ \bibinfo {pages} {101501}
  (\bibinfo {year} {2007})},\ \Eprint {http://arxiv.org/abs/hep-th/0702226}
  {arXiv:hep-th/0702226 [hep-th]} \BibitemShut {NoStop}%
\bibitem [{\citenamefont {John}(1991)}]{John:1991fj}%
  \BibitemOpen
  \bibfield  {author} {\bibinfo {author} {\bibfnamefont {F.}~\bibnamefont
  {John}},\ }\href@noop {} {\emph {\bibinfo {title} {Partial Differential
  Equations}}}\ (\bibinfo  {publisher} {Springer Verlag New York Inc.},\
  \bibinfo {address} {New York},\ \bibinfo {year} {1991})\BibitemShut {NoStop}%
\bibitem [{\citenamefont {Canizares}\ and\ \citenamefont
  {Sopuerta}(2009)}]{Canizares:2009ay}%
  \BibitemOpen
  \bibfield  {author} {\bibinfo {author} {\bibfnamefont {P.}~\bibnamefont
  {Canizares}}\ and\ \bibinfo {author} {\bibfnamefont {C.~F.}\ \bibnamefont
  {Sopuerta}},\ }\bibfield  {title} {\enquote {\bibinfo {title} {{An Efficient
  Pseudospectral Method for the Computation of the Self-force on a Charged
  Particle: Circular Geodesics around a Schwarzschild Black Hole}},}\ }\href
  {\doibase 10.1103/PhysRevD.79.084020} {\bibfield  {journal} {\bibinfo
  {journal} {Phys. Rev.}\ }\textbf {\bibinfo {volume} {D79}},\ \bibinfo {pages}
  {084020} (\bibinfo {year} {2009})},\ \Eprint {http://arxiv.org/abs/0903.0505}
  {arXiv:0903.0505 [gr-qc]} \BibitemShut {NoStop}%
\bibitem [{\citenamefont {Canizares}\ \emph {et~al.}(2010)\citenamefont
  {Canizares}, \citenamefont {Sopuerta},\ and\ \citenamefont
  {Jaramillo}}]{Canizares:2010yx}%
  \BibitemOpen
  \bibfield  {author} {\bibinfo {author} {\bibfnamefont {P.}~\bibnamefont
  {Canizares}}, \bibinfo {author} {\bibfnamefont {C.~F.}\ \bibnamefont
  {Sopuerta}}, \ and\ \bibinfo {author} {\bibfnamefont {J.~L.}\ \bibnamefont
  {Jaramillo}},\ }\bibfield  {title} {\enquote {\bibinfo {title}
  {{Pseudospectral Collocation Methods for the Computation of the Self-Force on
  a Charged Particle: Generic Orbits around a Schwarzschild Black Hole}},}\
  }\href {\doibase 10.1103/PhysRevD.82.044023} {\bibfield  {journal} {\bibinfo
  {journal} {Phys. Rev.}\ }\textbf {\bibinfo {volume} {D82}},\ \bibinfo {pages}
  {044023} (\bibinfo {year} {2010})},\ \Eprint {http://arxiv.org/abs/1006.3201}
  {arXiv:1006.3201 [gr-qc]} \BibitemShut {NoStop}%
\bibitem [{\citenamefont {Canizares}\ and\ \citenamefont
  {Sopuerta}(2011)}]{Canizares:2011kw}%
  \BibitemOpen
  \bibfield  {author} {\bibinfo {author} {\bibfnamefont {P.}~\bibnamefont
  {Canizares}}\ and\ \bibinfo {author} {\bibfnamefont {C.~F.}\ \bibnamefont
  {Sopuerta}},\ }\bibfield  {title} {\enquote {\bibinfo {title} {{Tuning
  Time-Domain Pseudospectral Computations of the Self-Force on a Charged Scalar
  Particle}},}\ }\href {\doibase 10.1088/0264-9381/28/13/134011} {\bibfield
  {journal} {\bibinfo  {journal} {Class. Quant. Grav.}\ }\textbf {\bibinfo
  {volume} {28}},\ \bibinfo {pages} {134011} (\bibinfo {year} {2011})},\
  \Eprint {http://arxiv.org/abs/1101.2526} {arXiv:1101.2526 [gr-qc]}
  \BibitemShut {NoStop}%
\bibitem [{\citenamefont {Jaramillo}\ \emph {et~al.}(2011)\citenamefont
  {Jaramillo}, \citenamefont {Sopuerta},\ and\ \citenamefont
  {Canizares}}]{Jaramillo:2011gu}%
  \BibitemOpen
  \bibfield  {author} {\bibinfo {author} {\bibfnamefont {J.~L.}\ \bibnamefont
  {Jaramillo}}, \bibinfo {author} {\bibfnamefont {C.~F.}\ \bibnamefont
  {Sopuerta}}, \ and\ \bibinfo {author} {\bibfnamefont {P.}~\bibnamefont
  {Canizares}},\ }\bibfield  {title} {\enquote {\bibinfo {title} {{Are
  Time-Domain Self-Force Calculations Contaminated by Jost Solutions?}}}\
  }\href {\doibase 10.1103/PhysRevD.83.061503} {\bibfield  {journal} {\bibinfo
  {journal} {Phys. Rev.}\ }\textbf {\bibinfo {volume} {D83}},\ \bibinfo {pages}
  {061503} (\bibinfo {year} {2011})},\ \Eprint {http://arxiv.org/abs/1101.2324}
  {arXiv:1101.2324 [gr-qc]} \BibitemShut {NoStop}%
\bibitem [{\citenamefont {Boyd}(2001)}]{Boyd}%
  \BibitemOpen
  \bibfield  {author} {\bibinfo {author} {\bibfnamefont {J.~P.}\ \bibnamefont
  {Boyd}},\ }\href@noop {} {\emph {\bibinfo {title} {Chebyshev and Fourier
  Spectral Methods}}},\ \bibinfo {edition} {2nd}\ ed.\ (\bibinfo  {publisher}
  {Dover},\ \bibinfo {address} {New York},\ \bibinfo {year} {2001})\BibitemShut
  {NoStop}%
\bibitem [{\citenamefont {Fornberg}(1996)}]{Fornberg:1996psc}%
  \BibitemOpen
  \bibfield  {author} {\bibinfo {author} {\bibfnamefont {B.}~\bibnamefont
  {Fornberg}},\ }\href@noop {} {\emph {\bibinfo {title} {A Practical Guide to
  Pseudospectral Methods}}}\ (\bibinfo  {publisher} {Cambridge University
  Press},\ \bibinfo {address} {Cambridge},\ \bibinfo {year} {1996})\BibitemShut
  {NoStop}%
\bibitem [{\citenamefont {Canuto}\ \emph {et~al.}(2006)\citenamefont {Canuto},
  \citenamefont {Quarteroni}, \citenamefont {Hussaini},\ and\ \citenamefont
  {Zang}}]{Canutoetal:2006sm1}%
  \BibitemOpen
  \bibfield  {author} {\bibinfo {author} {\bibfnamefont {C.}~\bibnamefont
  {Canuto}}, \bibinfo {author} {\bibfnamefont {A.}~\bibnamefont {Quarteroni}},
  \bibinfo {author} {\bibfnamefont {M.~Y.}\ \bibnamefont {Hussaini}}, \ and\
  \bibinfo {author} {\bibfnamefont {T.~A.}\ \bibnamefont {Zang}},\ }\href@noop
  {} {\emph {\bibinfo {title} {Spectral Methods. Fundamentals in Single
  Domains}}}\ (\bibinfo  {publisher} {Springer-Verlag},\ \bibinfo {address}
  {Berlin Heidelberg},\ \bibinfo {year} {2006})\BibitemShut {NoStop}%
\bibitem [{\citenamefont {Butcher}(2003)}]{Butcher:2003jcb}%
  \BibitemOpen
  \bibfield  {author} {\bibinfo {author} {\bibfnamefont {J.~C.}\ \bibnamefont
  {Butcher}},\ }\href@noop {} {\emph {\bibinfo {title} {Numerical Methods for
  Ordinary Differential Equations}}}\ (\bibinfo  {publisher} {{John Wiley \&
  Sons}},\ \bibinfo {address} {Chichester},\ \bibinfo {year}
  {2003})\BibitemShut {NoStop}%
\bibitem [{\citenamefont {Press}\ \emph {et~al.}(1992)\citenamefont {Press},
  \citenamefont {Flannery}, \citenamefont {Teukolsky},\ and\ \citenamefont
  {Vetterling}}]{Press:1992nr}%
  \BibitemOpen
  \bibfield  {author} {\bibinfo {author} {\bibfnamefont {W.~H.}\ \bibnamefont
  {Press}}, \bibinfo {author} {\bibfnamefont {B.~P.}\ \bibnamefont {Flannery}},
  \bibinfo {author} {\bibfnamefont {S.~A.}\ \bibnamefont {Teukolsky}}, \ and\
  \bibinfo {author} {\bibfnamefont {W.~T.}\ \bibnamefont {Vetterling}},\
  }\href@noop {} {\emph {\bibinfo {title} {Numerical Recipes: The Art of
  Scientific Computing}}}\ (\bibinfo  {publisher} {Cambridge University
  Press},\ \bibinfo {address} {Cambridge (UK) and New York},\ \bibinfo {year}
  {1992})\BibitemShut {NoStop}%
\bibitem [{\citenamefont {Gustafsson}\ \emph {et~al.}(1995)\citenamefont
  {Gustafsson}, \citenamefont {Kreiss},\ and\ \citenamefont
  {Oliger}}]{Gustafsson:1995tb}%
  \BibitemOpen
  \bibfield  {author} {\bibinfo {author} {\bibfnamefont {B.}~\bibnamefont
  {Gustafsson}}, \bibinfo {author} {\bibfnamefont {H.}~\bibnamefont {Kreiss}},
  \ and\ \bibinfo {author} {\bibfnamefont {J.}~\bibnamefont {Oliger}},\
  }\href@noop {} {\emph {\bibinfo {title} {Time dependent problems}}}\
  (\bibinfo  {publisher} {John Wiley \& Sons},\ \bibinfo {address} {New York},\
  \bibinfo {year} {1995})\BibitemShut {NoStop}%
\bibitem [{\citenamefont {Gough}(2009)}]{Gough:2009:GSL:1538674}%
  \BibitemOpen
  \bibfield  {author} {\bibinfo {author} {\bibfnamefont {B.}~\bibnamefont
  {Gough}},\ }\href@noop {} {\emph {\bibinfo {title} {GNU Scientific Library
  Reference Manual - Third Edition}}},\ \bibinfo {edition} {3rd}\ ed.\
  (\bibinfo  {publisher} {Network Theory Ltd.},\ \bibinfo {year}
  {2009})\BibitemShut {NoStop}%
\bibitem [{\citenamefont {Garfinkle}\ and\ \citenamefont
  {Duncan}(1998)}]{Garfinkle:1998va}%
  \BibitemOpen
  \bibfield  {author} {\bibinfo {author} {\bibfnamefont {D.}~\bibnamefont
  {Garfinkle}}\ and\ \bibinfo {author} {\bibfnamefont {G.~C.}\ \bibnamefont
  {Duncan}},\ }\bibfield  {title} {\enquote {\bibinfo {title} {{Scaling of
  curvature in subcritical gravitational collapse}},}\ }\href {\doibase
  10.1103/PhysRevD.58.064024} {\bibfield  {journal} {\bibinfo  {journal} {Phys.
  Rev.}\ }\textbf {\bibinfo {volume} {D58}},\ \bibinfo {pages} {064024}
  (\bibinfo {year} {1998})},\ \Eprint {http://arxiv.org/abs/gr-qc/9802061}
  {arXiv:gr-qc/9802061 [gr-qc]} \BibitemShut {NoStop}%
\bibitem [{\citenamefont {Hod}\ and\ \citenamefont {Piran}(1997)}]{Hod:1996az}%
  \BibitemOpen
  \bibfield  {author} {\bibinfo {author} {\bibfnamefont {S.}~\bibnamefont
  {Hod}}\ and\ \bibinfo {author} {\bibfnamefont {T.}~\bibnamefont {Piran}},\
  }\bibfield  {title} {\enquote {\bibinfo {title} {{Fine structure of
  Choptuik's mass scaling relation}},}\ }\href {\doibase
  10.1103/PhysRevD.55.440} {\bibfield  {journal} {\bibinfo  {journal} {Phys.
  Rev.}\ }\textbf {\bibinfo {volume} {D55}},\ \bibinfo {pages} {440--442}
  (\bibinfo {year} {1997})},\ \Eprint {http://arxiv.org/abs/gr-qc/9606087}
  {arXiv:gr-qc/9606087 [gr-qc]} \BibitemShut {NoStop}%
\bibitem [{\citenamefont {Frigo}\ and\ \citenamefont
  {Johnson}(2005)}]{fftw:2005}%
  \BibitemOpen
  \bibfield  {author} {\bibinfo {author} {\bibfnamefont {M.}~\bibnamefont
  {Frigo}}\ and\ \bibinfo {author} {\bibfnamefont {S.~G.}\ \bibnamefont
  {Johnson}},\ }\bibfield  {title} {\enquote {\bibinfo {title} {The design and
  implementation of {FFTW3}},}\ }\href@noop {} {\bibfield  {journal} {\bibinfo
  {journal} {Proceedings of the IEEE}\ }\textbf {\bibinfo {volume} {93}},\
  \bibinfo {pages} {216} (\bibinfo {year} {2005})},\ \bibinfo {note} {special
  issue on "Program Generation, Optimization, and Platform
  Adaptation"}\BibitemShut {NoStop}%
\end{thebibliography}

%


\end{document}